\documentclass[useAMS,usenatbib]{mn2e}
\usepackage{epsfig,natbib}
\usepackage{eufrak}
\usepackage{amsmath}

\newcommand{\Msun}{M_\odot}
\newcommand{\bmf}[1]{\mathbf {#1}}
\newcommand{\kaco}[1]{\langle {#1}\rangle}

\newcounter{remcount}

\newcommand{\simgt}{\lower.5ex\hbox{$\; \buildrel > \over \sim \;$}}
\newcommand{\simlt}{\lower.5ex\hbox{$\; \buildrel < \over \sim \;$}}

\newcommand{\pdd}{P_{\delta\delta}}
\newcommand{\pdv}{P_{\delta\theta}}
\newcommand{\pvv}{P_{\theta\theta}}

\newcommand{\pshotnoise}{P_{\rm sn}}
\newcommand{\hompc}{h^{-1}{\rm Mpc}}

\title[Likelihood reconstruction 
of density-velocity power spectra]{Likelihood reconstruction method of
real-space density and velocity power spectra from a redshift galaxy survey
}
\author[J.~Tang, I.~Kayo, M.~Takada]
{Jiayu Tang,$^{1,2}$\thanks{Email: jiayu.tang@ipmu.jp} Issha Kayo,$^{1,3}$
  Masahiro Takada$^1$\\
$^1$Institute for the Physics and Mathematics of the Universe (IPMU), 
 The University of Tokyo, Chiba 277-8582, Japan\\
 $^2$Department of Physics, the Chinese University of Hong Kong, Shatin, New Territories, Hong Kong, China\\
$^3$Department of Physics, Toho University, Funabashi, Chiba 274-8510, Japan}

\pagerange{\pageref{firstpage}--\pageref{lastpage}} \pubyear{2010}
\begin{document}
\setcounter{remcount}{0}
\maketitle \label{firstpage}
\begin{abstract}
We develop a maximum likelihood based method of
 reconstructing band powers of the density and velocity power spectra at
 each wavenumber bins from the measured clustering features of galaxies
 in redshift space, including marginalization over uncertainties
 inherent in the small-scale, nonlinear redshift distortion, the
 Fingers-of-God (FoG) effect. The reconstruction can be done assuming
 that the density and velocity power spectra depend on the
 redshift-space power spectrum having different angular modulations of
 $\mu$ with $\mu^{2n}$ ($n=0,1,2$) and that the model FoG effect is
 given as a multiplicative function in the redshift-space spectrum.

By using N-body simulations and the halo catalogs, we test our method by
comparing the reconstructed power spectra with the spectra directly
measured from the simulations. For the spectrum of $\mu^0$ or
equivalently the density power spectrum $\pdd(k)$, our method recovers
the amplitudes to a few percent accuracies up to $k\simeq 0.3~h{\rm
Mpc}^{-1}$ for both dark matter and halos.
For the power spectrum of
$\mu^2$, 
which is equivalent to the density-velocity power spectrum
$\pdv(k)$ in the linear regime,
our method can recover, within the statistical errors, the input power
spectrum for dark matter up to $k\simeq 0.2~h{\rm Mpc}^{-1}$ and at
both redshifts $z=0$ and $1$, if the adequate FoG model being
marginalized over is employed.  However, for the halo spectrum that is
least affected by the FoG effect, the reconstructed spectrum shows
greater amplitudes than the spectrum $\pdv(k)$ inferred from the
simulations over a range of wavenumbers $0.05\le k \le
0.3~h$Mpc$^{-1}$. We argue that the disagreement is ascribed to
nonlinearity effect that arises from the cross-bispectra of density and
velocity perturbations. Using the perturbation theory and assuming
Einstein gravity as in simulations, we derive the nonlinear
correction term to the redshift-space spectrum, and find that the
leading-order correction term is proportional to $\mu^2$ and increases
the $\mu^2$-power spectrum amplitudes more significantly at larger $k$,
at lower redshifts and for more massive halos.  We find that adding the
nonlinearity correction term to the simulation $\pdv(k)$ can fairly well
reproduce the reconstructed $\pdv(k)$ for halos up to $k\simeq
0.2~h$Mpc$^{-1}$. 
\end{abstract}
\begin{keywords}
cosmology: theory -- galaxy clustering -- dark energy -- gravity test
\end{keywords}
\section{Introduction}
\label{sec:intro}

Cosmic accelerating expansion is the most tantalizing problem in modern
cosmology and physics.  Within the framework of Einstein's general
relativity (GR), the cosmic acceleration requires that
roughly 70\% of total energy of the present-day universe is in the form
of unknown, mysterious energy component having negative pressure, dubbed
as dark energy.  An alternative explanation is the so-called
modified gravity scenario, where the cosmic acceleration is conjectured
as a result of 
 breakdown of Einstein's gravity on cosmological scales.  There are
growing attempts in the community trying to develop a consistent model
of modified gravity that can explain the cosmic acceleration on
cosmological scales, yet recovering GR on small scales
such as solar system scales, without the need of dark energy \citep[e.g. see][for a
review]{JainKhoury:10}.

There are various methods capable of addressing the nature of the cosmic
acceleration: type-Ia supernovae, cluster experiments,
galaxy clustering, and weak gravitational lensing. These methods are
sensitive to cosmic expansion and structure formation histories, in
a complementary way, over different length scales and/or different ranges
of redshifts. In particular, an essential approach to discriminate the
dark energy and modified gravity scenarios is exploring both the cosmic
expansion history and the growth rate of structure formation by
combining more than two different methods above
\citep{DETF:06,EsoFundamental:06,JainZhang:08,Guziketal:10}.

In this paper we focus on cosmological observables derivable from a
wide-field galaxy redshift survey. A robust method feasible with a
galaxy redshift survey is the baryon acoustic oscillation (BAO)
experiment, which allows us to infer the angular diameter distance as
well as the Hubble expansion rate from the measured pattern of galaxy
clustering \citep{Eisenstein:05,Cole:05,BlakeetalBAO:11}. 
There are many ongoing and planned galaxy redshift surveys aimed at
achieving the BAO experiments at higher precisions: the Baryon
Oscillation Spectroscopic Survey
(BOSS)\footnote{http://cosmology.lbl.gov/BOSS/}, the BigBOSS project
\citep{BigBoSS}, the HETDEX survey\footnote{http://hetdex.org/}, and the
Subaru Prime Focus Spectrograph (PFS)
project\footnote{http://sumire.ipmu.jp/en/}.

Adding the redshift distortion measurement 
can further improve the cosmological power of a galaxy
redshift survey
\citep{Peacocketal:01,Guzzoetal:08,Whiteetal:10,Yamamotoetal:10,Blakeetal:11,songsabiu:11}.
In real space
galaxy clustering is statistically isotropic in a statistically
homogeneous and isotropic universe.  However, in redshift space the
line-of-sight component of galaxies' peculiar velocities induces an
angular anisotropic modulation in the clustering pattern.  In a
structure formation scenario the peculiar velocities of galaxies are
caused by gravitational attracting force in large-scale structure and
the gravitational field can be inferred from the observed galaxy
distribution or directly probed by weak gravitational lensing. 

More precisely there are two kinds of redshift distortion effects. 
One is caused by large-scale coherent velocities or bulk motions of
halos which are associated with large-scale structure of large length
scales, $\simgt 1~{\rm Mpc}$. This large-scale redshift distortion in
the linear regime is
called Kaiser effect \citep{Kaiser:87}. This effect amplifies clustering
amplitudes of galaxies in redshift space. It is now becoming recognized
that, even at length scales of $100~h^{-1}{\rm Mpc}$ relevant for
BAO experiments, the Kaiser effect ceases to be
accurate, and nonlinearity effects needs to be included for a level of
precision ongoing/upcoming survey can achieve. Encouragingly, however,
the refined, accurate modeling has been developed based on perturbation
theory and/or simulations \citep[][and see references
therein]{Scoccimarro04,Matsubara:08,Taruyaetal:09,Taruyaetal:10}.  In
this sense the Kaiser effect contains a cleaner cosmological
information. Hence, if Einstein GR is a priori assumed, adding the
large-scale velocity information to the BAO constraints or more
generally the density clustering information allows us to 
significantly improve geometrical constraints
\citep[e.g.][]{AP,MatsubaraSuto:96,Ballingeretal:96} as well as
cosmological parameter estimation
\citep[e.g.][]{Eisensteinetal:99,Takadaetal:06,Takada:06,Saitoetal:08}.


Probably more interestingly, if the density and velocity power spectra
can be reconstructed from the measured redshift-space clustering of
galaxies without assuming any gravity theory, we can now open up a
window of exploring properties of gravity on cosmological scales in a
model-independent way, by comparing the reconstructed density and
velocity power spectra, because the density and velocity fields are
related via gravity theory
\citep{Linder:05,Zhangetal:07,Zhangetal:08,Guzzoetal:08,Wang:08,Yamamotoetal:08,
Whiteetal:09,PercivalWhite:09,
SimpsonPeacock:10,Song:10,SongKayo:10,Yamamotoetal:10,Reyesetal:10,Shapiroetal:10}. For
example, Einstein gravity or a concordance $\Lambda$CDM model gives us
specific predictions on how these two spectra are related to each other:
the two spectra have a constant overall offset in the amplitudes in the
linear regime.  Hence, if any scale-dependent differences in the
amplitudes are found from data, it is a signature of failure of Einstein
gravity.

However, a viable reconstruction method needs to be not much influenced
by uncertainties arising from small-scale, nonlinear redshift distortion
effect due to
internal virial motions
of galaxies within halos, the so-called Fingers-of-God (FoG) effect
\citep[e.g. see][]{Jackson:72,Peacock:book99,Hamilton:98,Scoccimarro04}. 
This effect causes a significant suppression
in redshift-space clustering amplitudes along the line-of-sight
direction.
How does the small-scale velocity field affect the BAO-scale clustering
of galaxies? Here is a rough estimate on the physics. Recall that virial
velocity dispersion for massive halos of $10^{15}M_\odot$ can have
velocities of a few $ 10^3$~km s$^{-1}$. This causes a redshift
modulation given as $\Delta z\simeq v_{\parallel} \simeq 10^{-2}$ (in
units of speed-of-light $c=1$), which in turn causes an apparent
displacement in the position space as $\Delta r_\parallel=\Delta
z/H(z)\simeq 30~h^{-1}{\rm Mpc}$. This corresponds to Fourier modes of
$k=2\pi/\lambda\simeq 0.2~h{\rm Mpc}^{-1}$, which are indeed relevant
for the BAO scales. Thus the real-space galaxy clustering within halos
at scales smaller than a few Mpc blows up to large scales up to $\sim
$50Mpc in redshift space.  Since the FoG effect arises from highly
nonlinear regime and is affected by baryonic and astrophysical effects,
it is still very challenging to have a sufficiently accurate model
needed for precision cosmology. In fact, the FoG effect is one of the
major systematic errors in galaxy clustering observables.


Hence the purpose of this paper is to develop a method that allows us to
unbiasedly reconstruct the real-space density and velocity power spectra of
large length scales from the measured redshift-space clustering of
galaxies, removing the contamination of FoG effect \citep[also see][for a similar study]{SongKayo:10}.  This can be done by
developing a maximum likelihood based method of reconstructing band
powers of the real-space power spectra
at each wavenumber bins, including marginalization over
uncertainties in parameters to model the FoG effect.  In this method the
real-space power spectra at each wavenumber bins are estimated such that
the likelihood of redshift-space power spectrum is maximized, assuming
that the original density perturbation field is a Gaussian field.  The
{\em real-space} velocity power spectra on large scales include only the
information on the large-scale redshift distortion effect, because the
spectra arise from the density and velocity fields at physical scales
corresponding to the wavenumbers without the FoG effect contamination.  Our
method is analogous to the cosmic microwave background (CMB) power
spectrum reconstruction \citep{Verde03}.
 By using N-body simulations of 70 realizations and the halo
catalogs, we will carefully test the method by studying whether or not
the reconstructed real-space power spectra can recover the input spectra
in the simulations.

The structure of this paper is as follows. In
\S~\ref{sec:redshift_distortion} we review how the redshift distortion
effect due to peculiar velocities causes an angular modulation in
redshift-space power spectrum, after briefly describing how the peculiar
velocity field is related to metric scalar perturbations. In
\S~\ref{sec:method} we develop a maximum likelihood method of
reconstructing the real-space density and velocity power spectra from
the redshift-space power spectrum. After describing the N-body
simulations and halo catalogs in \S~\ref{sec:dm_catalog}, we will show
in \S~\ref{sec:results} the main results of this paper; by applying the
method to the N-body simulations and halo catalogs, we assess accuracies
of reconstructing the real-space power spectra with the
method. \S~\ref{sec:summary} is devoted to summary and discussion.

\section{Preliminaries}\label{sec:redshift_distortion}

\subsection{Metric perturbations}
\label{sec:metric}

In the Newtonian gauge the perturbed Friedmann-Robertson-Walker metric
that has scalar perturbations can be fully specified by the form of
\begin{equation}
ds^2=-(1+2\Psi)dt^2+a^2(1-2\Phi)d\bmf{x}^2,
\label{eq:metric}
\end{equation}
where $a(t)$ is the expansion scale factor. Note that we here assumed a
flat universe for simplicity.  The metric form (\ref{eq:metric}) is
fully general for any metric theory of gravity, as long as the vector
and tensor perturbations are negligible. Redshift $z$ is the most
important observable in astronomy, and it is given as $1+z=1/a(t_e)$,
where $a(t_e)$ is the scale factor at the epoch when an object of
interest, e.g. galaxy, emitted the photon to be observed by an
observer. We use the convention $a(t_0)=1$ at present. $\Psi$
corresponds to the Newtonian potential that describes the acceleration
of particles, while $\Phi$ denotes the curvature perturbation.

The expansion history of the universe is specified by the function of
$a(t)$ or the Hubble function $H(t)=\dot{a}/a$, where $\dot{}$ denotes
the derivative with respect to time $t$. Given gravity theory, the time
evolution of $a(t)$ or $H(t)$ is specified once the energy content of
the universe is specified, as in the case of Einstein gravity.

\subsection{The case of Einstein gravity}

Although the rest of this paper does not assume any theory of gravity,
it would be instructive to discuss the case of Einstein gravity. This
subsection also gives a background motivation of our work.

Theory of gravity relates the metric perturbations in
Eq.~(\ref{eq:metric}) to matter variables. In the matter dominated
era, if assuming the Einstein gravity, the Einstein equations yield, for
example, the Poisson equation on sub-horizon scales, which relates the
metric perturbation $\Phi$ to the density perturbation field of total
matter as
\begin{equation}
-k^2\Phi=4\pi G a^2\delta. 
\end{equation}
Note that the Poisson equation here is given in the Fourier space,
yielding the factor $k^2$ on the l.h.s.
The matter distribution can be inferred from galaxy surveys or weak
lensing surveys.

In a case that the anisotropic energy stress is negligible as in a 
cold dark matter (CDM) dominated structure formation 
model, the two metric perturbations are equivalent to each other 
on sub-horizon
scales:
\begin{equation}
\Psi\simeq\Phi.
\end{equation}
Thus the two metric perturbations have only one degree of freedom, which
corresponds to the density field $\delta$ in the matter sector.

The geodesic equation for a test particle is given by
\begin{equation}
\frac{dp^\alpha}{d\lambda}+\Gamma^{\alpha}_{\mu\nu}p^\mu p^\nu=0, 
\end{equation}
where $\Gamma$ is the Christophel symbols. Let's consider a test
particle which only slowly move with respect to the comoving coordinates
i.e. a non-relativistic particle. Dark matter and galaxies are such
particles.  The equation of motion for
such a test particle is given in the linear regime as
%
\begin{equation}
\frac{d\bmf{v}}{dt}+H\bmf{v}=-\frac{1}{a}\nabla\Psi,
\end{equation}
where $\bmf{v}$ is the comoving peculiar velocity defined as
$\bmf{v}\equiv d\bmf{x}/dt$. 
Thus the velocity field follows the gravitational potential. For this
reason the peculiar velocity field of galaxies is expected as a powerful
tool for probing the gravitational potential field. 

Another important observable is gravitational lensing. Solving the
geodesic equation for a photon, which is a relativistic particle, leads
the lensing deflection angle to be given as
\begin{equation}
\bmf{\alpha}=\int\!d\chi~W_{\rm GL}(\chi)\nabla_\perp
\left(\Psi+\Phi\right),
\end{equation}
where $W_{\rm GL}(\chi)$ is the lensing geometrical kernel that depends
on the background metric quantity, i.e the scale factor
\cite[e.g.][]{Guziketal:10}.
Thus lensing depends on a combination of the two metric perturbations,
$\Psi+\Phi$.

Therefore combining different observables such as the galaxy
distribution, the peculiar velocity and weak lensing in principle allow
to test the consistency relation $\Psi=\Phi$ or more generally explore
properties of gravity on cosmological scales
\citep[e.g.][]{JainZhang:08}.  
However in this paper we address
we can use the measured clustering features of galaxies in redshift space
to reconstruct the power spectrum of the peculiar velocity field
$\bmf{v}$, independently from the density power spectrum. Hence, our
method allows us to use the redshift-space clustering to test gravity
theory on cosmological scales, by comparing the reconstructed power
spectra of density and velocity fields.

\subsection{Redshift-space power spectrum}\label{sec:FOG_theory}

What we can measure from a spectroscopic survey of galaxies is angular
positions and redshifts of the galaxies. However, the observed redshift
of a given galaxy, $\hat{z}$, is modulated from the true redshift, $z$,
due to its peculiar velocity as well as the metric perturbations 
 -- the so-called redshift-space
distortion. According to the metric theory of gravity, the observed
redshift is given \citep[e.g., see][]{Sasaki:87} as
\begin{eqnarray}
1+\hat{z}&\simeq& (1+z)\left\{
1+
\left[\Phi+
v_z\right]^{\rm e}_{\rm o}
\right\}\nonumber\\
&\simeq &(1+z)\left[1+\left.v_\chi\right|^{\rm e}_{\rm o}\right],
\end{eqnarray}
where $\Phi$ is 
the gravitational potential perturbation (Eq.~[\ref{eq:metric}]), $v_z$
denotes the line-of-sight component of the comoving peculiar velocity of
tracer considered, and the notation $\left.\cdots\right|^{\rm e}_{\rm
o}$ denotes the difference between quantities at observer's and galaxy's
positions. In the second line on the r.h.s. of the equation above, we
assumed that the effect of peculiar velocity $v_\chi$, which is the
order of $10^{-3}$ (corresponding to $300$km s$^{-1}$) for the
large-scale coherent peculiar velocities, is much larger than the
potential amplitude $\Phi\sim O(10^{-5})$ for $\Lambda$CDM-like
cosmologies. In other words, while we will later focus on the density
perturbations of matter or galaxies in large-scale structure, the effect
of the metric perturbation $\Phi$ is safely negligible compared to the
density perturbations on relevant length scales. In addition the
perturbation contributions at an observer's position ($O$) only
contribute to the monopole offset (e.g. a shift in the overall
normalization of galaxy number density at a given redshift), therefore
we can ignore these contributions in the following.

Via the redshift-distance relation $\chi(z)$, the apparent radial
distance to a galaxy at redshift $z$, $\hat{\chi}(z)$, is modulated from
the true position $\chi(z)$ due to the peculiar velocity as
\begin{eqnarray}
\hat{\chi}&\equiv& \chi(\hat{z})\nonumber\\
&\simeq& \chi(z)+(1+z)\frac{d\chi}{dz}v_\chi\nonumber\\
&=& \chi(z)+\frac{(1+z)}{H(z)}v_\chi\nonumber\\
&=& \chi(z)+u_\chi,
\label{eq:u_chi}
\end{eqnarray}
where $u_\chi$ is the normalized peculiar velocity field defined as
$u_\chi\equiv (1+z)v_\chi/H(z)$.

The mass conservation, or the number conservation of galaxies, tells
that the density perturbation in redshift space, $\delta_s$, is related
to the real-space density perturbation as
\begin{eqnarray}
1+\delta_s&=&(1+\delta)
\left(1+\frac{\partial u_\chi}{\partial \chi}\right)^{-1}\nonumber\\
&\simeq& 1+\delta - \frac{\partial u_\chi}{\partial \chi}+
O\!\left(\delta u, u^2\right),
\label{eq:delta_s}
\end{eqnarray}
where in the second equality of the equation above we have used the
Taylor expansion of $(1+\partial u/\partial \chi)^{-1}$, and we have
ignored the higher-order terms of the perturbations (see below for
further discussion).
Exactly
speaking the Jacobian transformation above breaks down when the particle
motions have shell crossing or multi-streamings at a single spatial
position, which can occur in the nonlinear stage such as a region within
a virialized halo. In other words, the equation above is valid only at
large length scales greater than a size of halos, which is validated on
scales $k\simlt 0.3~\hompc$ we are interested in.

Fourier-transforming the equation above yields
\begin{equation}
\tilde{\delta}_s(\bmf{k})
\simeq \tilde{\delta}(\bmf{k}) + \mu^2 \tilde{\theta}(\bmf{k}) 
+O(\delta \theta, \theta^2),
\label{eq:delta_s_app}
\end{equation}
where $\mu$ is the cosine between the wavevector
$\bmf{k}$ and the line-of-sight direction. 
Here we have assumed that the peculiar velocity is irrotational,
therefore is given in terms of the scalar velocity potential, and the
quantity, $\tilde{\theta}$, denotes the Fourier-transformed coefficient
of the divergence of peculiar velocity field, $\theta\equiv
-\nabla\cdot\bmf{u}$.
Also notice that in the equation above we have employed a distant
observer approximation and ignored the curvature of the sky, where one
axis of the coordinate system can be chosen to be along the
line-of-sight direction.  
We again ignored the higher-order terms of the
perturbations such
as $O(\delta \theta, \theta^2)$.
Thus the redshift distortion induces angle-dependent
modulations, given by $\mu^{2n}$ ($n=1,2,\cdots$), in the redshift-space density field.

Motivated by the discussion above and the previous works
\citep{Kaiser:87,Scoccimarro04}, we {\em assume} that the redshift-space
power spectrum of galaxies (dark matter or halos)
is given by the
following functional form:
\begin{eqnarray}
&&\hspace{-1em}\langle\tilde{\delta}_s(\bmf{k})
\tilde{\delta}^\ast_s(\bmf{k}')
\rangle\equiv (2\pi)^3\pdd^{s}(k,\mu)\delta_D^3(\bmf{k}-\bmf{k}')\nonumber\\
&&\hspace{-1em}\rightarrow 
  \pdd^{s}(k,\mu)=
  \left[P_{\delta\delta}(k)+2\mu^2
P_{\delta\theta}(k)+\mu^4
P_{\theta\theta}(k)\right]
  F(k,\mu),\nonumber\\
\label{eq:Pgg}
\end{eqnarray}
where $P_{\delta\delta}$ and $P_{\theta\theta}$ are the power spectra of
density perturbation and velocity divergence and $P_{\delta\theta}$ is
the cross power spectrum:
\begin{eqnarray}
\langle\tilde{\delta}(\bmf{k})
\tilde{\delta}^\ast(\bmf{k}')\rangle&\equiv&
(2\pi)^3P_{\delta\delta}(k)\delta^3_D(\bmf{k}-\bmf{k}'),\nonumber \\
\langle\tilde{\delta}(\bmf{k})
 \tilde{\theta}^\ast(\bmf{k}')\rangle&\equiv &
(2\pi)^3P_{\delta\theta}(k)\delta^3_D(\bmf{k}-\bmf{k}'),\nonumber \\
\langle\tilde{\theta}(\bmf{k})
\tilde{\theta}^\ast(\bmf{k}')\rangle&\equiv &
(2\pi)^3P_{\theta\theta}(k)\delta^3_D(\bmf{k}-\bmf{k}'). 
\label{eq:ps_def}
\end{eqnarray}
The form of Eq.~(\ref{eq:Pgg}) is often assumed in the literature
\citep[e.g.][and references
therein]{Hamilton:98,Taruyaetal:09}. However, as can be found from
Eqs.~(\ref{eq:delta_s_app}), (\ref{eq:Pgg}) and (\ref{eq:ps_def}), we
ignored the contributions of higher-order perturbations to the
redshift-space power spectrum. In fact we will discuss later that the
higher-order terms of Eq.~(\ref{eq:delta_s_app}) 
can be important for the redshift-space power
spectrum, especially for massive halos.  To be more precise, if we
recall that the velocity perturbation is smaller than the density
perturbation at relevant low redshifts, 
the leading-order correction to the Kaiser formula is
found in \cite{Taruyaetal:10} to be
\begin{equation}
\delta P_s(k,\mu)\leftarrow
k_\parallel\left\langle\tilde{\delta}(\bmf{k}')
\int\!\frac{d^3\bmf{q}}{(2\pi)^3}\frac{q_\parallel}{q^2}\tilde{\theta}(\bmf{q})
\tilde{\delta}(\bmf{k}-\bmf{q})
\right
\rangle.
\label{eq:higher_Ps}
\end{equation}
In Appendix~\ref{sec:higher_Kaiser} we derive the correction terms for
dark matter and halos based on the perturbation theory, and will use the
results for the following discussion.  
Meanwhile we will assume Eq.~(\ref{eq:Pgg}) for simplicity. 

The function in the square bracket on the r.h.s. of Eq.~(\ref{eq:Pgg})
denotes the Kaiser formula for redshift-space power spectrum, which is
valid only at large length scales in the linear regime. The function
$F(k,\mu)$ was introduced so as to take into account the nonlinear
distortion effect, the so-called Fingers-of-God (FoG) effect, which
causes a smearing of redshift-space clustering due to random virial
motions of dark matter particles or galaxies within halos.  Thus the
assumption we employed in Eq.~(\ref{eq:Pgg}) is the FoG redshift
distortion and the Kaiser formula are separable functions in the
redshift-space power spectrum.  This does not necessarily hold, although
an empirical model based on the halo model gives such a functional form
of redshift-space power spectrum \citep[][also see Hikage et al. in
preparation]{White:01,Seljak:01}.  Hence the validity of
Eq.~(\ref{eq:Pgg}) needs to be further tested in combination with
simulations.  The recent study done in \cite{Taruyaetal:09} gives a
possible verification on this treatment, where it was shown that the
form (\ref{eq:Pgg}) can well reproduce the simulation results in the
weakly nonlinear regime down to $k\simeq 0.2~ h{\rm Mpc}^{-1}$ if an
appropriate function $F(k,\mu)$ is employed.

A theoretical understanding of the FoG effect is still lacking
due to complicated physics involved in the nonlinear clustering
regime. In this paper we rather employ an empirical approach: we will
consider the following functional forms of $F(k,\mu)$ in order to study
how the results change with the different FoG models: 
\begin{equation}
F(k,\mu)=
\left\{
\begin{array}{l}
\exp[-\sigma^2k^2\mu^2],\\
{\displaystyle \frac{1}{1+\sigma^2k^2\mu^2}},\\
{\displaystyle 1-\sigma^2k^2\mu^2+\frac{1}{2}\tau^4 k^4\mu^4}.\\
\end{array}
\right.
\label{eq:FOG}
\end{equation}
All the models have a limit of $F\rightarrow 1$ when $k\rightarrow 0$.
The first and second forms correspond to the Gaussian and Lorentzian FoG
models that are sometimes employed in the literature
\citep[e.g. see][for a review]{Hamilton:98}. We will treat $\sigma$
appearing in the forms as a free parameter in the following analyses,
motivated by the results in \cite{Taruyaetal:09}. The third form can be
considered as a more general form, in analogy to the Taylor expansion of
the FoG function in terms of $k\mu$, and this includes the Gaussian and
Lorentzian models in the range $k\mu\ll 1$. Similarly we will treat
$\sigma$ and $\tau$ as free parameters in the model fitting.  We will
refer to these models as Gaussian, Lorentzian, Taylor-$\sigma$
and Taylor-($\sigma+\tau$) models, respectively.

Besides the assumed form of redshift-space power spectrum and the
FoG function (see Eqs.~[\ref{eq:Pgg}] and [\ref{eq:FOG}]), in the
following we will explore a model-independent reconstruction of the
density and velocity power spectra $P_{\delta\delta}$,
$P_{\delta\theta}$ and $P_{\theta\theta}$ at each $k$ bins, from the
measured galaxy distribution in redshift space. 
More exactly speaking,
since the reconstructed power spectra are not necessarily same as the
density and velocity spectra, 
our method recovers the real-space power spectra that are proportional
to $\mu^{2n}$ ($n=0,1,2$) in the redshift-space power spectrum, being
marginalized over uncertainties of the FoG effect.
Then we will assess the
performance of this reconstruction method by comparing the reconstructed
spectra with the spectra directly measured from simulations. This
reconstruction problem is not a linear problem, because the FoG function
is non-linearly coupled with the density and velocity spectra. Hence,
the reconstructed band powers at different $k$-bins become correlated
with each other even if the underlying density and velocity fields are
Gaussian.

Finally we remark on the Einstein gravity case, where the two metric
perturbations are equivalent: $\Psi=\Phi$ as discussed in
\S~\ref{sec:metric}. In this case the density and velocity power spectra
are related to each other in the linear regime 
as
$P_{\delta\theta}\simeq \beta P_{\delta\delta}$ and
$P_{\theta\theta}\simeq \beta^2P_{\delta\delta}$, where $\beta=(1/b)d\ln
D/d\ln a$ with $D$ and $b$ being the linear growth rate and the linear
bias parameter, respectively. 
Note that the possible nonlinear correction terms (see
Eq.~[\ref{eq:higher_Ps}]) can be also accurately computed based on
perturbation theory and/or simulations for a given cosmological model
\citep[e.g.][]{Taruyaetal:10,Jenningsetal:11}. 
Thus, if the Einstein gravity is {\em a
priori} assumed, measuring the density and velocity power spectra helps
to break parameter degeneracies, especially the degeneracy between the
galaxy bias and the power spectrum amplitudes, which in turn helps to
significantly improve parameter constraints \citep[e.g.][]{Takadaetal:06}.

\section{A maximum likelihood reconstruction method of 
redshift-space power spectra}
\label{sec:method}

In this section we develop a method for reconstructing the real-space
power spectra from the galaxy distribution in redshift space, based on
a maximum likelihood method. 

We start with assuming that the mass density fluctuation field
$\delta_m(\bmf{x})$ in redshift space obeys the Gaussian likelihood
function:
\begin{eqnarray}
&&{\cal L}[\delta_s\!(\bmf{x})]
\propto \frac{1}{\sqrt{\rm det(\bmf{C})}}
\int\!\frac{d^3\bmf{x}_i}{V_s}
\int\!\frac{d^3\bmf{x}_j}{V_s}\nonumber\\
&&\hspace{5em}\times
\exp\left[
-\frac{1}{2}\delta_s(\bmf{x}_i)
(\bmf{C}^{-1})_{ij}\delta_s(\bmf{x}_j)
\right],
\label{eq:like_real}
\end{eqnarray}
where $V_s$ is the survey volume, $\bmf{C}(\bmf{x}_i-\bmf{x}_j)$ is
defined as $\bmf{C}(\bmf{x}_i-\bmf{x}_j)\equiv
\langle\delta_s(\bmf{x}_i)\delta_s(\bmf{x}_j) \rangle$, the two-point
correlation function between the density fields $\delta_s(\bmf{x}_i)$
and $\delta_s(\bmf{x}_j)$ in redshift space, and $\bmf{C}^{-1}$ is its
inverse matrix.
 
Analogously to the likelihood of the CMB temperature power spectrum
\citep[e.g.,][]{Verde03}, by converting the likelihood function to
Fourier space, we can derive the log-likelihood function for the
redshift-space power spectrum (see Appendix~\ref{sec:like} for the
detailed derivation; also see \cite{Percival:2005yq} for the similar
discussion):
\begin{equation}
-2\ln{\cal L}=\sum_{k_i, \mu_a}N(k_i,\mu_a)\left[ \frac{\hat{P^s}(k_i,\mu_a)}{P^s(k_i,\mu_a)} +\ln \frac{P^s(k_i,\mu_a)}{\hat{P^s}(k_i,\mu_a)}-1
\right], \label{eq:2dmethod}
\end{equation}
where $\hat{P}_s(k_i,\mu_a)$ is the power spectrum estimated at the bin
($k_i, \mu_a$):
\begin{equation}
\hat{P}^s(k_i,\mu_a)\equiv \frac{1}{N(k_i,\mu_a)}\sum_{\bmf{k}\in
 (k_i,\mu_a)} \left|
\tilde{\delta}_s(\bmf{k})
\right|^2.
\label{eq:measuredPk}
\end{equation}
The quantity $N(k_i,\mu_a)$ is the number of independent Fourier modes
confined within the bin ($k_i, \mu_a$):
$
N(k_i,\mu_a)\equiv \sum_{\bmf{k}\in (k_i,\mu_a)} 1. 
$
If a surveyed volume has a cubic geometry with side length $L$,
i.e. $V_s=L^3$, the fundamental mode to discriminate different Fourier
modes has the length given by $k_f=2\pi/L$. Hence the number of
independent Fourier modes, for the bin $(k_i,\mu_a)$, is approximately
given as $N(k_i,\mu_a)\approx 2\pi k_i^2\Delta k\Delta\mu/(2\pi/L)^3$
in the limit $k_i\gg k_f$, where $\Delta k$ and $\Delta \mu$ are the bin
widths. In the equation above, we ignored observational effects due to
survey geometry and masking of the surveyed region for simplicity. For
actual data we need to include these effects. 

In Eq.~(\ref{eq:2dmethod}) $P^s(k_i,\mu_a)$ is the underlying true
redshift-space power spectrum at the bin $(k_i,\mu_a)$. We assume the
form given by Eq.~(\ref{eq:Pgg}) for $P^s(k_i,\mu_a)$, which is given by
the model power spectra $P_{\delta\delta}(k)$, $P_{\delta\theta}(k)$ and
$P_{\theta\theta}(k)$ and the parameters to model the FoG effect (see
Eq.~[\ref{eq:FOG}]). Hence, given the measured redshift-space power
spectrum $\hat{P}^s(k_i,\mu_a)$
(Eq.~[\ref{eq:measuredPk}]), we can estimate the best-fit
power spectra $P_{\delta\delta}(k)$, $P_{\delta\theta}(k)$, and
$P_{\theta\theta}(k)$ at each $k_i$-bin, including marginalization
 over the band
powers at different $k$-bins and the FoG effect parameters, in such a
way that the log-likelihood (\ref{eq:2dmethod}) is maximized. This is
the maximum likelihood method for reconstructing the real-space
spectra. 

We will demonstrate how the method above allows a reconstruction of the
real-space power spectra using simulations. To do this, 
we will use the Markov-Chain Monte Carlo (MCMC) sampling method
\citep[e.g.][]{cosmomc}, more specifically Metropolis-Hasting algorithm in
our work.
The chain convergence is diagnosed by using the criteria given in
\citet{dunkley_mcmc}. The free parameters are: the band powers at each
$k$ 
bin, $\pdd(k_i)$, $\pdv(k_i)$ and
$\pvv(k_i)$, and the parameters to model the FoG effect given by
Eq.~(\ref{eq:FOG}), where $k_i$ denotes the $i$-th wavenumber bin and
the index $i$ runs over the number of bins. If we employ $N_{\rm bin}$
for the bin number over $k_{\rm min}\le k\le k_{\rm max}$, the total
number of model parameters are $3\times N_{\rm bin}$ plus the number of
the FoG parameters, 1 or 2 ($\sigma$ or $\sigma$ and $\tau$,
respectively), depending on which FoG model to use. 
In the MCMC parameter search, we adopted
the following priors on model parameters: 
$\pdd>\pdv>\pvv>0$ and the FoG function $0<F(k,\mu)\le 1$.  
Note that the latter prior, $0<F(k,\mu)\le 1$ is automatically satisfied by the
Gaussian and Lorentzian FoG models in Eq.~(\ref{eq:FOG}). 

Another assumption we employed in the log-likelihood function is the Gaussian
field assumption. This Gaussian assumption breaks down in the weakly
nonlinear regime, indeed over a range of wavenumbers relevant for the
method above. However, \cite{Takahashietal:11} showed that, using 5000
N-body simulation realizations, the non-Gaussianity of the density field
does not cause any large impact on parameter estimation in the weakly
nonlinear regime. Hence we do not think that the non-Gaussianity
affects the following results.

\section{N-body simulations and halo catalogs}
\label{sec:dm_catalog}

To test the performance of the power spectrum reconstruction method
described in the preceding section, we will 
implement a hypothetical
experiment: we will apply the method to mock data from N-body
simulations, and then compare the reconstructed spectra with the input
spectra directly measured from simulations.  
In this section we describe
some details of N-body simulations and the halo catalogs we will use in
the following sections.

\subsection{N-body Simulations}
The N-body simulations are generated by running the GADGET
\citep{Springel:05} assuming a flat universe; the matter density
$\Omega_{\rm m}=0.238$, the baryon content $\Omega_{\rm b}=0.041$, the
Hubble constant $H_0=73.2 {\rm km s^{-1} Mpc^{-1}}$, the spectral index
$n_s=0.958$,  
and the amplitude of
the linear power spectrum $\sigma_8=0.76$ \citep{Spergel:2007fk}. The
transfer function is calculated by the CAMB \citep{Lewis:1999bs}.  We
include $512^{3}$ N-body particles in a box of 
1${ h^{-3}{\rm
Gpc}}^3$ volume.  We started the simulations from the initial redshift $z=30$,
and set the initial conditions of N-body particles using the Zel'dovich
approximation. In this paper we use the outputs of $z=0$ and 1. 
Our initial redshift may not be sufficiently early to accurately
compute the nonlinear clustering of N-body particles as discussed in 
e.g. \citet[][]{Crocce:2006uq}.  
However, the main purpose of this paper is to study 
whether the reconstruction method of the power spectra $\pdd$ and $\pdv$
can reproduce the spectra directly measured from simulations, so
the accuracy of N-body simulations is not our concern.  
We will use 70 realizations in order to reduce the statistical scatters.

Using these N-body simulations, we also construct halo catalogs by
adopting the friend-of-friend (FOF) method with the linking length of
$b=0.2$ ($20\%$ of the mean separation).  The minimum number of member
particles is set to 20, which corresponds to the mass threshold of halos
 $9.8\times10^{12}\Msun$ for both the $z=0$ and $1$ outputs, and
the resulting number density of halos is $\bar{n}\simeq 3.8\times
10^{-4}~h^3{\rm Mpc}^{-3}$, which is comparable to the number density of
SDSS luminous red galaxies (LRGs) targeted for the ongoing BOSS survey
\citep{Whiteetal:10}.
We will use these
halo catalogs to compute the halo power spectrum, and then address
whether the power spectrum reconstruction method can also work for the
halo power spectrum.

\subsection{Power Spectrum Measurement from Simulations}\label{sec:ps_measurement}

From the simulation data above, we measure
the redshift-space power spectrum $P^s(k_i,\mu_a)$ in the
two-dimensional $(k_i,\mu_a)$ bins,
as well as the real-space spectra; the density-density power spectrum
$P_{\delta\delta}(k)$, the density-velocity power spectrum
$P_{\delta\theta}(k)$, and the velocity-velocity power spectrum
$P_{\theta\theta}(k)$ (see Eq.~[\ref{eq:ps_def}]).
In the following we describe how we measure these spectra from the
N-body simulations and the halo catalogs.

\subsubsection{Dark matter spectra}
 
\begin{figure*}
\begin{center}
\includegraphics[width=15.0cm,angle=0]{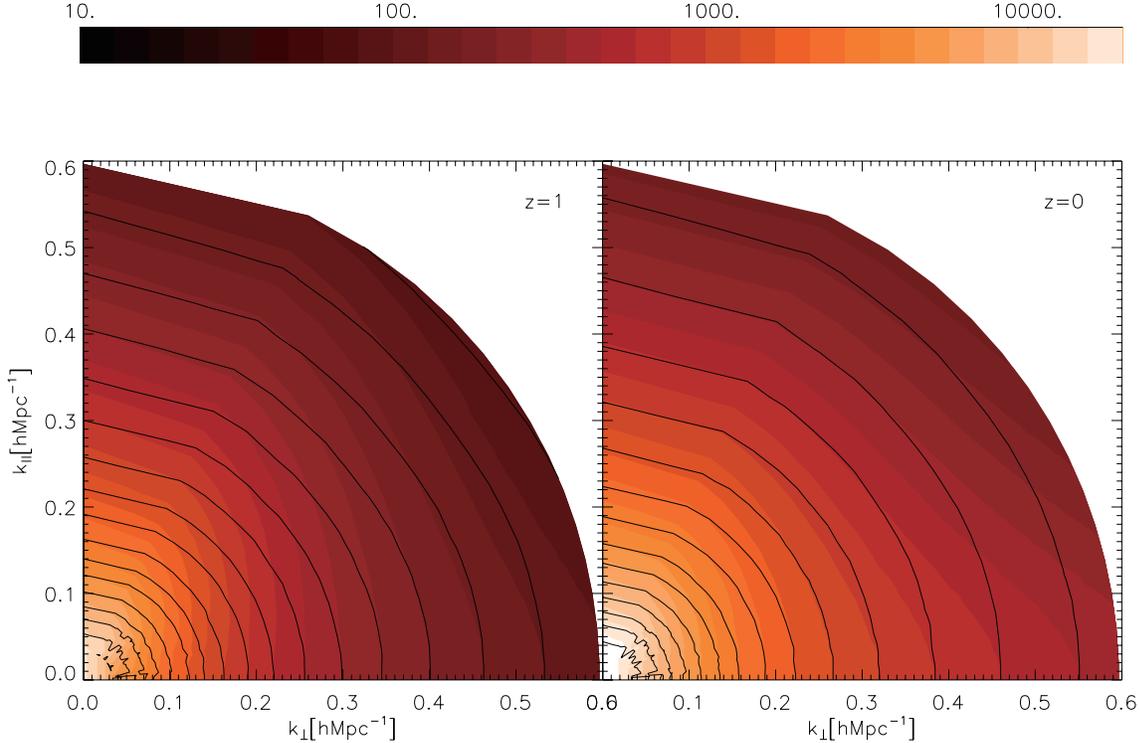}
\end{center}
\caption{Power spectra measured from N-body simulations at redshifts
 $z=0$ ({\em right panel}) and $z=1$ ({\em left}), respectively.  The
 color scales show the redshift-space power spectrum amplitudes as a
 function of $k_\perp$ and $k_{\parallel}$, where $k_{\perp}$ and
 $k_{\parallel}$ are wavelengths perpendicular and parallel to the
 line-of-sight direction (which is taken as the $z$-axis direction in
 simulations). Shown is the mean power spectrum among the spectra of
 70 realizations, each of which has a volume of $1~[h^{-1}{\rm
 Gpc}]^3$. The anisotropic modulations of the band powers are due to the
 redshift distortion effect due to the peculiar motions of N-body
 particles (see text for the details).  The spectra for $z=0$ shows a
 stronger FoG effect: a stronger squashed feature of the iso-contours
 along the $k_{\parallel}$ direction.  For comparison, the solid
 contours show the real-space spectra, which have isotropic
 contours. The contours are stepped by $\Delta \log P(k) = 0.11$.  } \label{fig:2dps_dm}
\end{figure*}

First let us discuss the spectra measured from the N-body simulations.
For redshift-space power spectrum, the distribution of N-body
particles is mapped into the redshift-space distribution taking into
account the modulation of their positions due to the redshift
distortion, where the line-of-sight direction is simply taken to be in
the $z$-axis direction in each simulation.
Note that we here adopted the distant observer approximation for simplicity.
The power spectrum of N-body particle distribution is measured using the
fast Fourier transform method (FFT). In doing this we first used the
``Cloud-in-Cell'' (CIC) interpolation method for assigning N-body
particles to the $512^{3}$ uniformly-distributed grids in order to
construct the grid-based density field. Then we implemented the FFT
method on the density field to obtain the Fourier-transformed
coefficients,
$\tilde{\delta}_s(\mathbf k)$.
Using Eq.~(\ref{eq:measuredPk}), we
estimate, in each simulation realization, the redshift-space power
spectrum, $\hat{P}^s(k_i,\mu_a)$, from the Fourier coefficients of the
density field. To reduce the statistical scatters, we will use the
averaged power spectrum of 70 realizations, and infer the 1$\sigma$
statistical errors from the scatters among the 70 realizations, which
correspond to the sampling variance for a volume of $1~h^{-3}{\rm Gpc}^3$.

Fig.~\ref{fig:2dps_dm} shows the redshift-space power spectrum (color
scales) for dark matter (N-body particles), measured from the
simulations of $z=0$ and 1 outputs. The redshift-space power spectrum
is compared with the real-space density power spectrum (contours). The
figure clearly shows redshift-space distortion effects. The Kaiser
effect due to large-scale bulk motions increases the redshift-space
power spectrum amplitudes along the line-of-sight direction or
equivalently $k_\parallel$, stretching the iso-contours towards the
larger $k_\parallel$. On the other hands, the FoG effect squashes the
iso-contours towards the smaller $k_\parallel$. Comparing the left- and
right-panels clarifies that the FoG effect is stronger at lower redshifts.

The real-space power spectra $\pdd(k), \pdv(k)$ and $\pvv(k)$ are
estimated from simulations as follows. The power spectrum $\pdd(k)$ is
just similar to the redshift-space spectrum as described above, but
skipping a step to compute redshift modulation due to the peculiar
velocities.  For $\pdv(k)$ and $\pvv(k)$, we first assign the velocity
components of each N-body particles to the $512^3$ uniformly-distributed
grids based on the CIC method, and then use the FFT method to generate
the Fourier coefficients of the velocity fields,
$\tilde{v}_i(\bmf{k})$. The velocity-divergence field is computed at
each Fourier grid as $\tilde{\theta}(\bmf{k})\propto
\bmf{k}\cdot\tilde{\bmf{v}}(\bmf{k})$.  If a larger number of grids than
$512^3$ (i.e. the smaller-size grid) is used, some grids may not contain
any N-body particle, which causes an ill-behaved spectrum $\pdv(k)$ at
small $k$ bins.  On the other hand, if we use a smaller number of
grids than $512^3$, the CIC interpolation causes a smoothing of the
velocity power spectrum amplitudes at large $k$ bins we are interested
in, as carefully studied in \cite{Pueblas:2009qy}. Hence we checked that
the 512$^3$ grids are rather close to an optimal choice of the grid
number in order to avoid these artificial effects over a range of scales
we are interested in. We again use the averaged power spectra, $\pdd,
\pdv$ and $\pvv$ from 70 realizations to reduce the statistical
scatters.

\begin{figure}
\begin{center}
\includegraphics[width=9.0cm,angle=0]{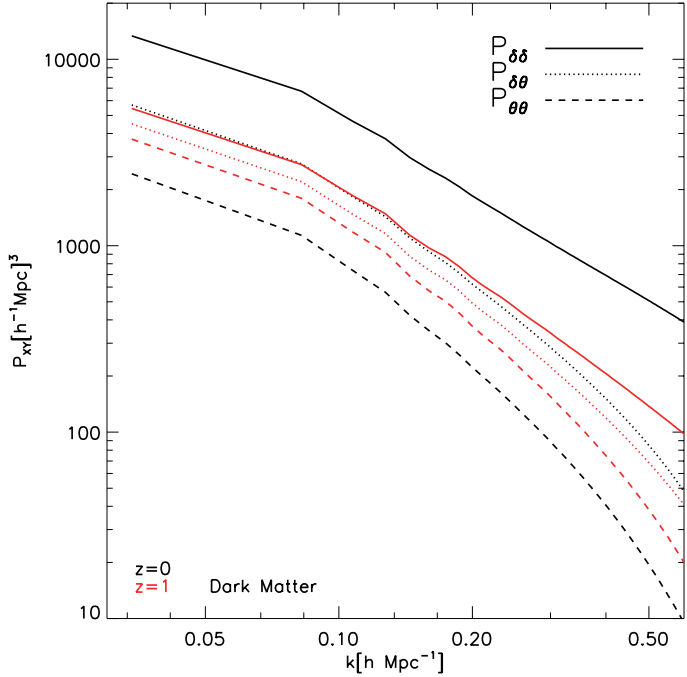}
\end{center}
\caption{The density-density ($\pdd$), density-velocity ($\pdv$) and
velocity-velocity ($\pvv$) power spectra at $z=0$ and $1$, respectively,
for N-body simulation particles. Similarly to the previous plot, shown
is the mean spectra of 70 realizations (see text for the details). The
statistical scatters around the mean spectra are sufficiently small, so
we do not show the scatters here (in other words, the average spectra
are well-converged). 
}
\label{fig:ps_dm}
\end{figure}
\begin{figure}
\includegraphics[width=8.0cm,angle=0]{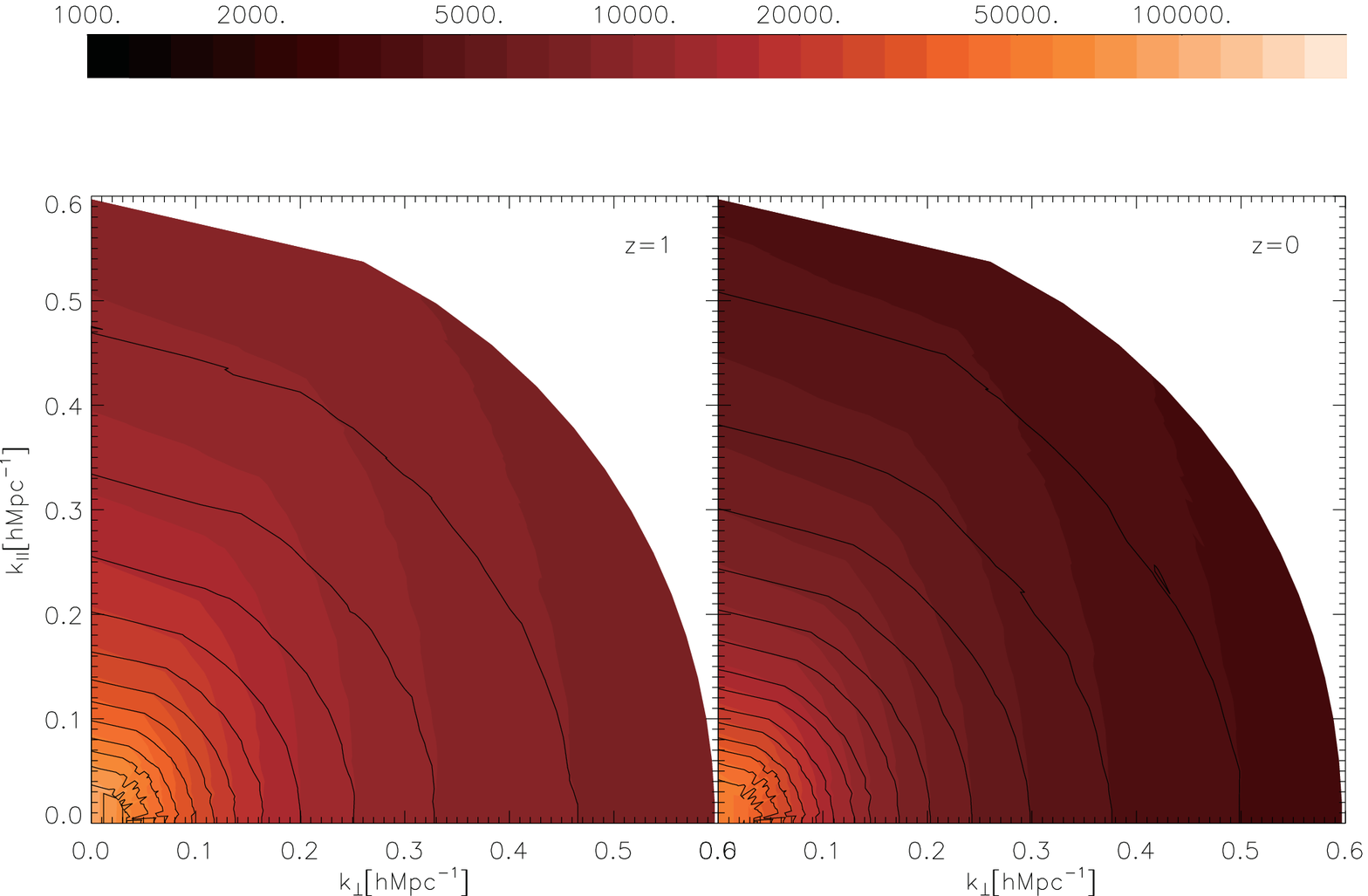}\\
\includegraphics[width=8.0cm,angle=0]{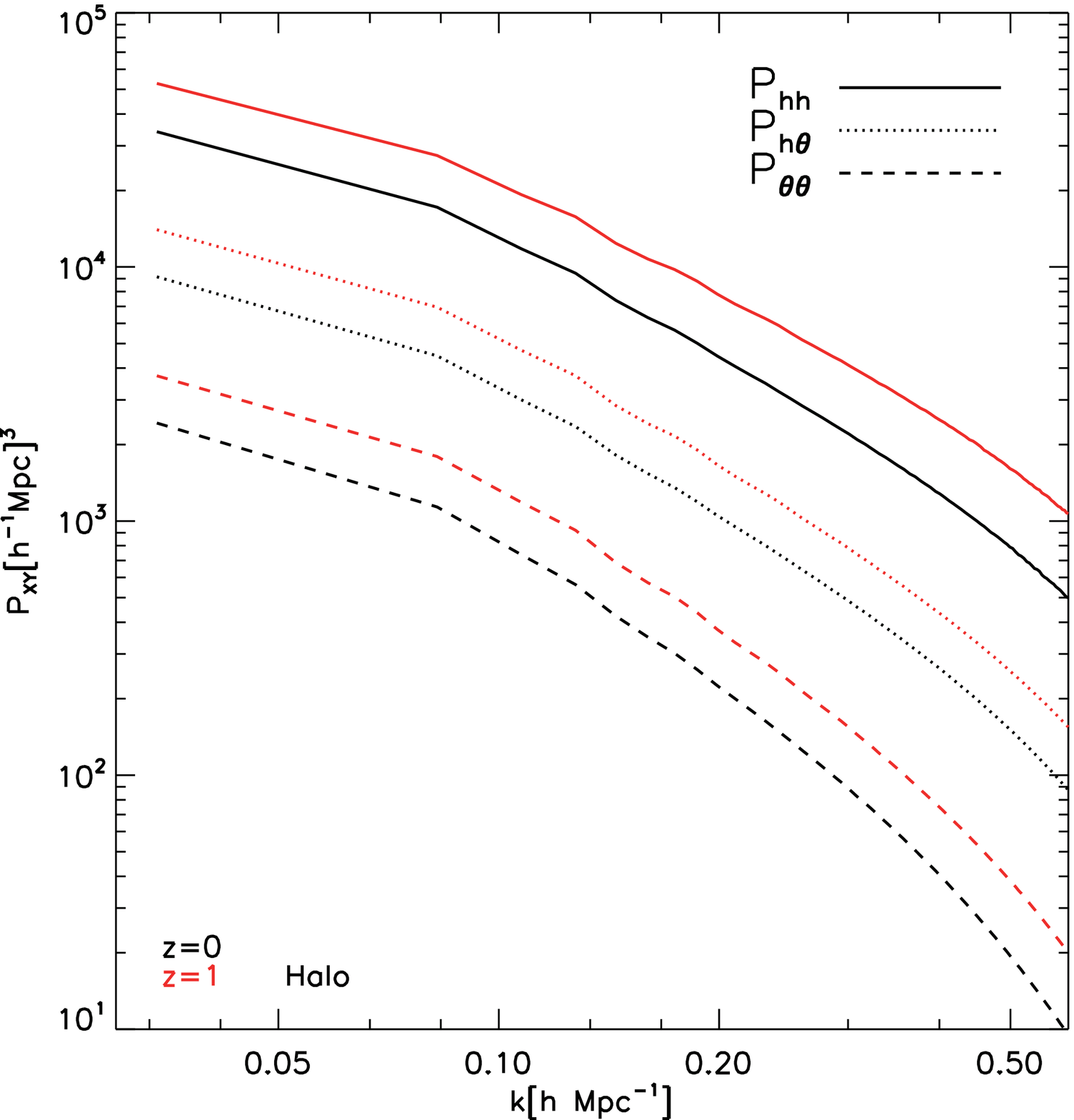}
\caption{ {\em Top panel}: The redshift-space power spectra of halos, 
similarly to Fig.~\ref{fig:2dps_dm}. We used halo catalogs
containing halos with masses greater than $\approx
10^{13}h^{-1}M_\odot$, roughly corresponding to halos hosting LRGs.
Slightly jaggy contours in the plot are due to the smaller number of
halos compared to the case of N-body particles. 
Compared to Fig.~\ref{fig:2dps_dm}, the halo spectra shows a less FoG
effect than in N-body particles, because halos have only bulk-velocity
contributions in large-scale structure.  
{\em Lower panel}: Similarly to
Fig.~\ref{fig:ps_dm}, the real-space power spectra 
for the halo
distribution. For the density-density power spectrum $\pdd$, the shot
 noise contamination $P_{\rm sn}=1/\bar{n}_{\rm halo}$ is subtracted. 
Note that the density power spectra (solid curves) have
greater amplitudes at $z=1$ than at $z=0$ because of the greater halo
biases, where we used the same mass threshold of $M_{\rm min}\approx
10^{13}h^{-1}M_\odot$ for both the two redshift outputs.  For the
velocity-related spectra we used the velocity field defined from N-body
particles, instead of the velocity field of halos, because we found the
difficulty of defining the velocity field for halos that are too
sparsely sampled (see text for the details). Hence the velocity-velocity
spectra $P_{\theta\theta}$ shown here are same as those in
Fig.~\ref{fig:ps_dm}.  \label{fig:ps_halo}}
\end{figure}


Fig.~\ref{fig:ps_dm} shows the real-space spectra of dark
matter (N-body particles): $\pdd(k)$, $\pdv(k)$ and $\pvv(k)$, for the two
redshift outputs of $z=0$ and 1. 
One can clearly find the relation $\pdd>\pdv>\pvv$,
and therefore the approximation given by Eq.~(\ref{eq:delta_s_app}) is
considered valid.

\subsubsection{Halo spectra}
\label{sec:halo_spectra}

Now let us move on to discussion on the the power spectra measured from
 the halo catalogs.
First we need to define the spatial position and the velocity for each
halo in the simulation.
We use the center-of-mass position, computed from N-body particles
contained within each halo, as the spatial position of the halo, while
we assign the mean of member N-body particles' velocities to the
velocity of the halo. Then, to get the density field for the
discrete halo distribution in each realization, we adopt the
Nearest-Grid-Point (NGP) method to assign the density field in $512^3$
uniformly-distributed grids, after the redshift modulation due to the
halo velocity field in redshift space are taken into account.  Similarly
to the cases for N-body particles, we computed the density power spectra
in redshift- and real-space, $P^{s}(k,\mu)$ and
$P_{\delta\delta}(k)$. 

On the other hand, however, the velocity related power spectra for the
halo distribution, $P_{\delta\theta}$ and $P_{\theta\theta}$, require
some caution, because it is not straightforward to define the
continuously-varying velocity field from the halo distribution that has
a much smaller number density (typically $\sim 10^{-4}~[h^{-1}{\rm
Mpc}]^{-3}$) than that of N-body particles.  We tried several
interpolation methods such as the CIC and the Delaunay triangulation
interpolation method for which we used the publicly available code from
the Computational Geometrical Algorithms Library (CGAL:
http://www.cgal.org/).  However, we could not find a reliable result for
the velocity power spectra at scales of interest in such a way that the
power spectra obtained become insensitive to the interpolation
method. Hence, instead of pursuing a more appropriate method to obtain
the halo velocity field, we use the grid-based velocity field of N-body
particles, {\em assuming} no velocity bias between the halo and N-body
particle (dark matter) distribution.
Thus we computed the power spectra, $P_{\delta\theta}$ and
$P_{\theta\theta}$, combining the halo density field and the N-body particle
velocity field. 
We will again use the mean halo power spectra from
70 realizations.

Fig.~\ref{fig:ps_halo} shows the redshift-space and real-space spectra
for the halo distribution, measured from the 70 realizations according
to the method we described above.  Compared to Fig.~\ref{fig:2dps_dm},
the redshift-space power spectrum of halos shows almost no FoG effect,
because the halo spectrum does not include contributions from the virial
motions of particles within each halo, and rather includes only the
contribution from the bulk motion of each halo.

\section{Results}
\label{sec:results}

\subsection{Reconstruction of matter power spectra}
\label{sec:dm_reconstruction}

\begin{figure}
\begin{center}
\includegraphics[width=8.5cm,angle=0]{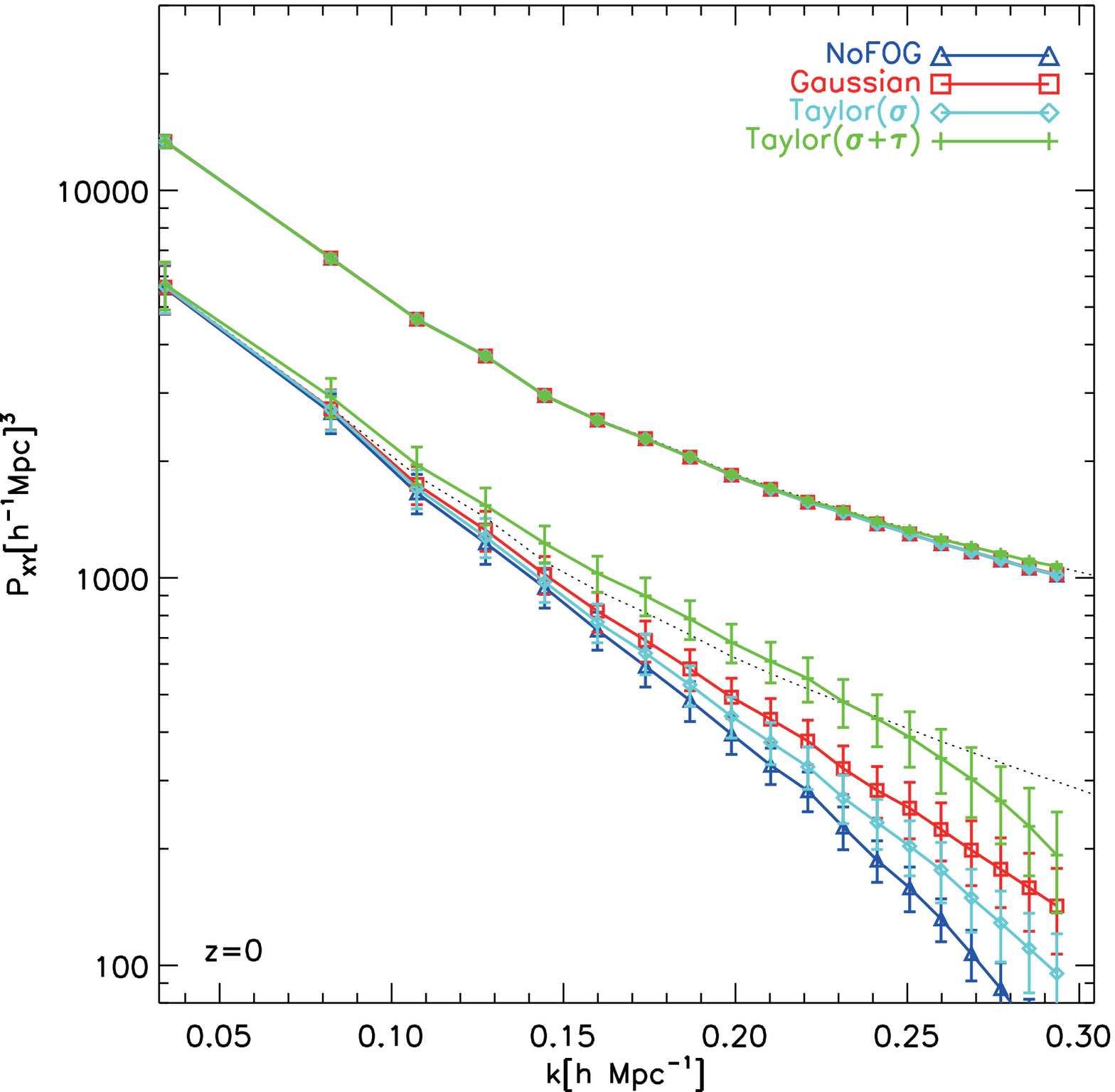}
\end{center}
\caption{The symbols show the reconstructed power spectra,
$P_{\delta\delta}(k)$ and $P_{\delta\theta}(k)$, based on the maximum
likelihood method (see Eq.~[\ref{eq:2dmethod}] and
\S~\ref{sec:method}), for dark matter. 
The different symbols are the results assuming the
different models of FoG effect: $F(k,\mu)$ (Eq.~[\ref{eq:FOG}]) in the
model redshift-space power spectrum (Eq.~[\ref{eq:Pgg}]). The triangle
symbols are the results ignoring the FoG effect $F=1$; the square shows
the results assuming the Gaussian $F(k,\mu)$, which has a single
parameter; the diamond and plus symbols show the results assuming the
Taylor expansion forms for $F(k,\mu)$ up to different orders of $(k\mu)
$, which are characterized by one ($\sigma$) and two ($\sigma, \tau$)
free parameters, respectively. The reconstructed band powers at each $k$
bins include marginalization over uncertainties in reconstructing band
powers of $P_{\delta\delta}$, $P_{\delta\theta}$ and $P_{\theta\theta}$
at different $k$ bins as well as the FoG parameter(s). The error bars
 around the symbols denote statistical uncertainties in the
 reconstruction for the volume $1~h^{-3}{\rm Gpc}^3$, computed from the
MCMC-based posterior distributions. 
Note that the
band powers at different $k$ bins are correlated.  For comparison, the
dotted curves show the power spectra, $\pdd(k)$ and $\pdv(k)$,
 directly measured from the
simulations. For $P_{\delta\delta}(k)$ the dotted curve and the symbols are
almost perfectly overlaid.  } \label{fig:re_dm}
\end{figure}

\begin{figure}
\begin{center}
\includegraphics[width=8.cm,angle=0]{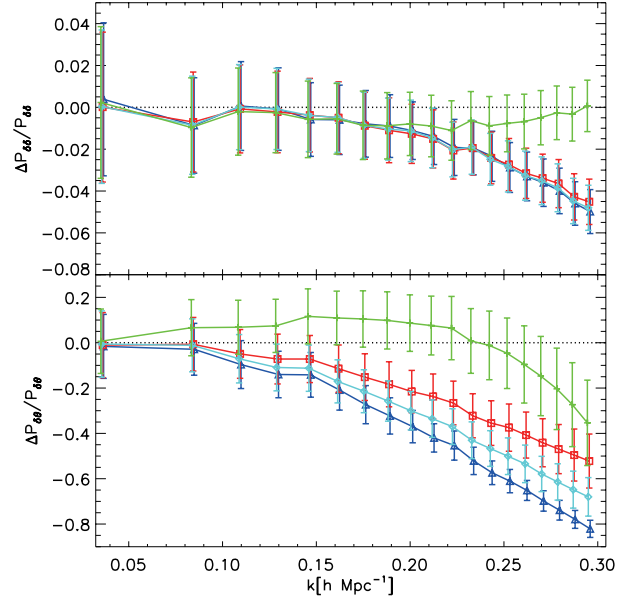}
\end{center}
\caption{Using the results in Fig.~\ref{fig:re_dm}, the plot shows
fractional differences between the reconstructed power spectra and the
spectra measured from the simulations for $P_{\delta\delta}$ (upper
panel) and $P_{\delta\theta}$ (lower), respectively. To be more
precise,  $\Delta P/P\equiv \left[ P({\rm reconst.})-P({\rm
input})\right]/P({\rm input})$, where $P_{\rm input}$ and $P_{\rm
 reconst.}$ are the input and reconstructed power spectra,
 respectively. 
The different
symbols are as in the previous plot. } \label{fig:re_diff_dm}
\end{figure}

\begin{figure*}
\begin{center}
\includegraphics[width=5.7cm,angle=0]{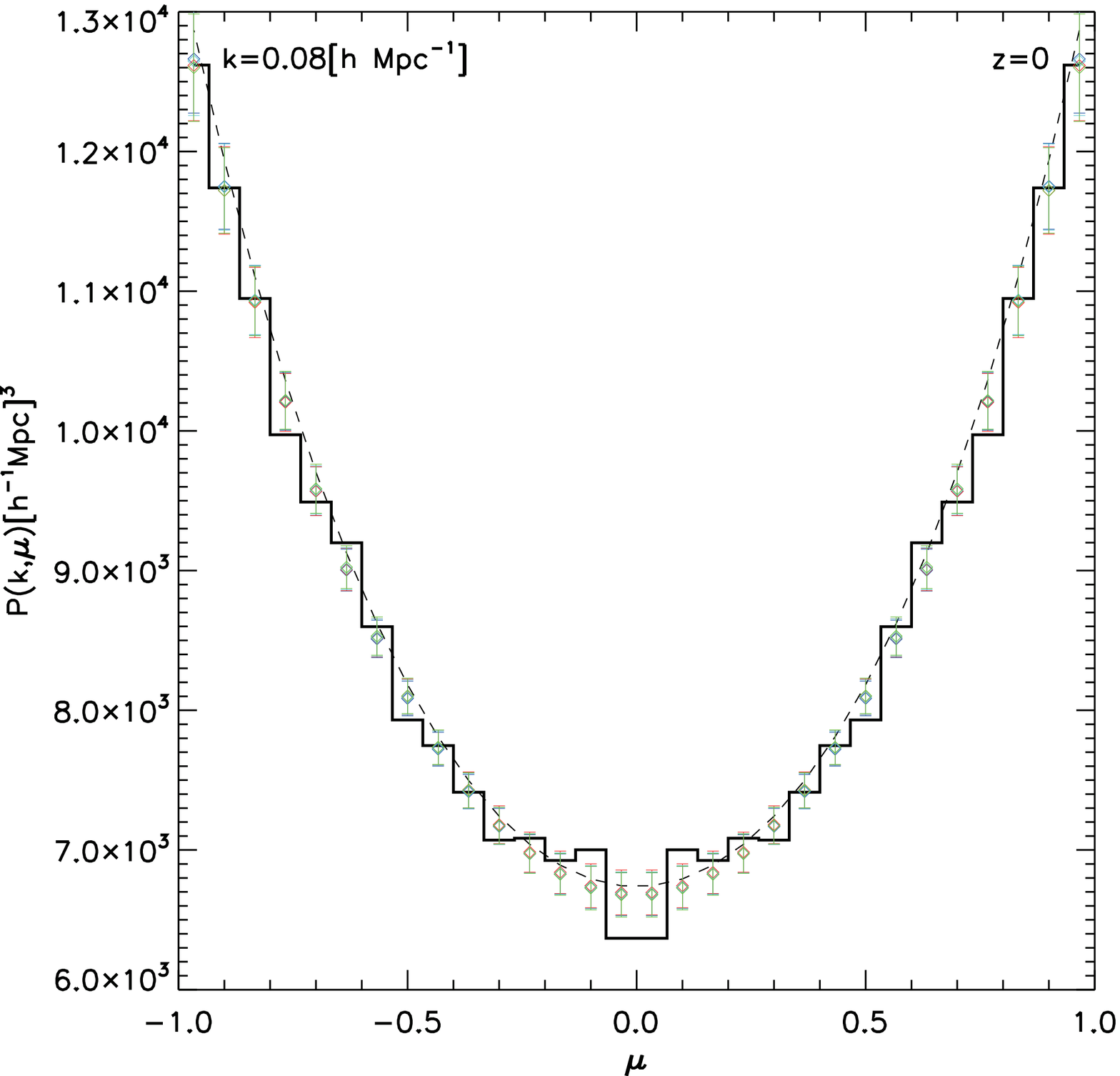}
\includegraphics[width=5.7cm,angle=0]{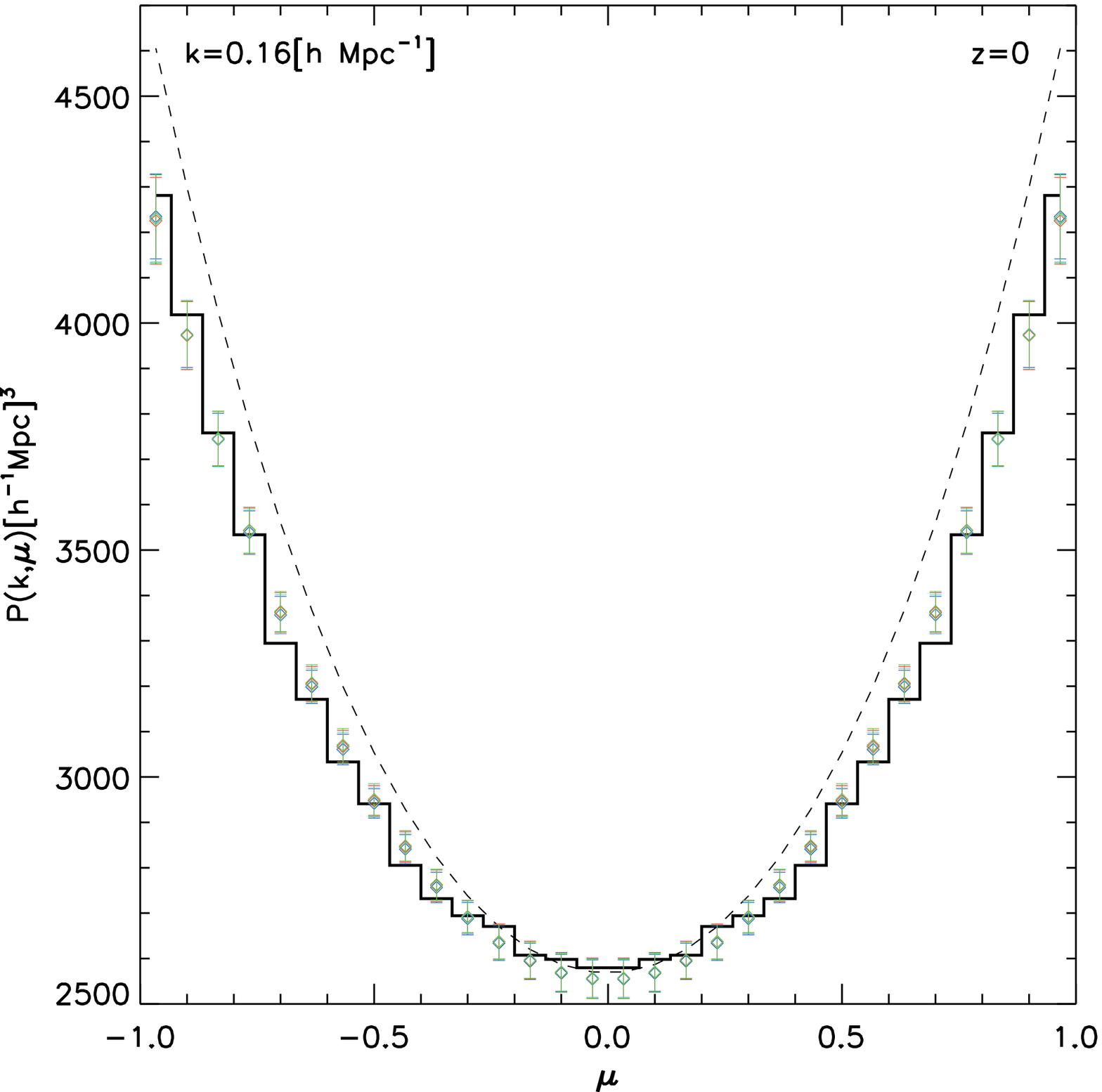}
\includegraphics[width=5.7cm,angle=0]{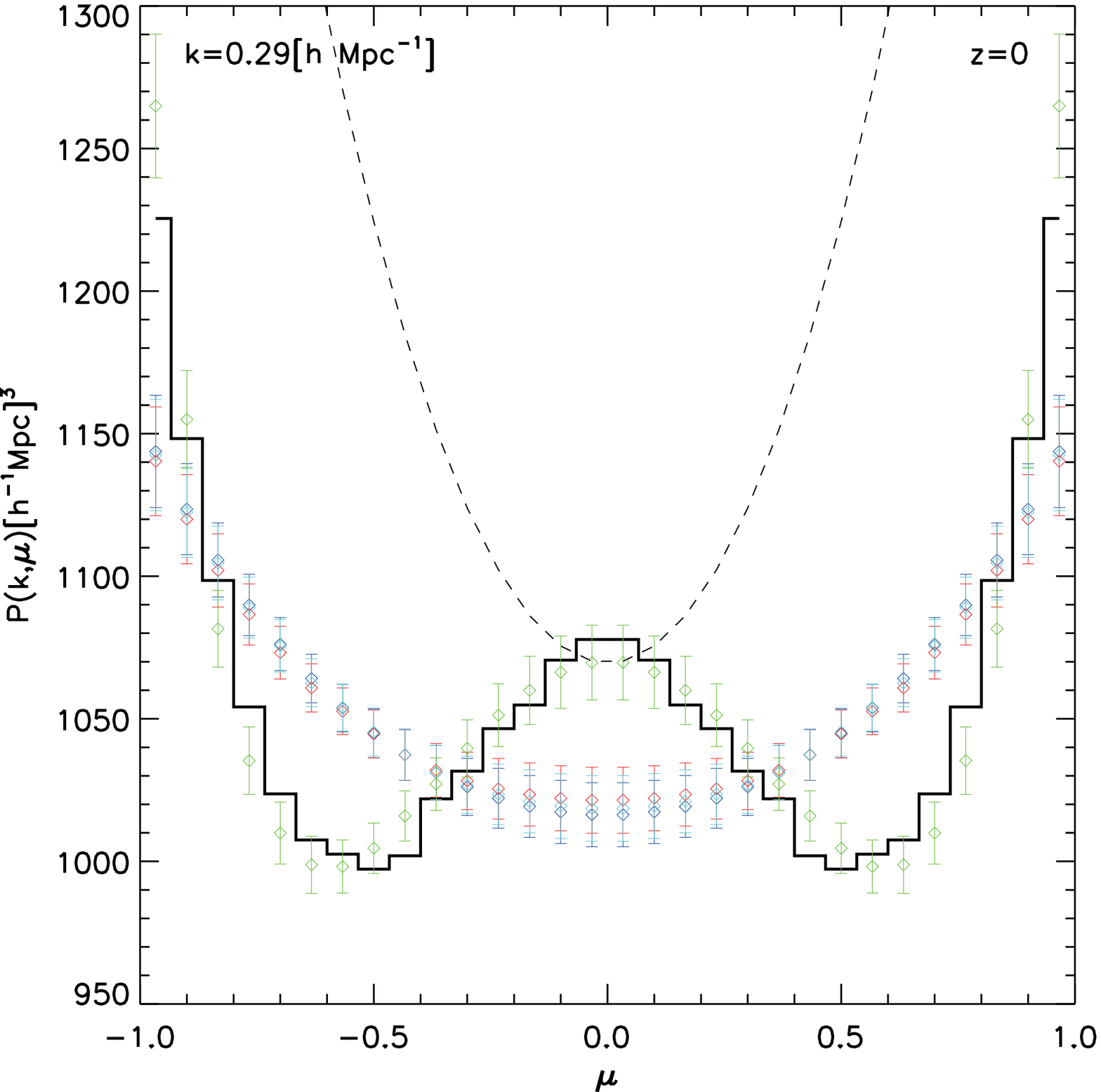}
\end{center}
\caption{Comparing the best-fit redshift-space power spectrum, based on
 the maximum likelihood method, with the redshift-space spectrum
 directly measured from simulations.  The solid histograms show slices
 of the redshift-space power spectrum amplitudes as a function of $\mu$, 
for a fixed $k$: $k=0.08$, $0.16$ and
 $0.29~ h{\rm Mpc}^{-1}$ from the left to right panels,
 respectively. Note $\mu$ denotes the cosine angle between the
 wavevector $\bmf{k}$ and the line-of-sight direction: $\mu\equiv
 \cos(\hat{\bmf{k}\cdot}\hat{\bmf{k}}_\parallel)$. The best-fit spectra,
 denoted by the different symbols, are computed by inserting the
 best-fit band powers of $\pdd,\pdv$ and $\pvv$ at each $k$ bin and the
 best-fit FoG parameters into Eq.~(\ref{eq:Pgg}), which are shown in
 Fig.~\ref{fig:re_dm}.  For comparison, the dashed curves in each panel
 show the spectra computed by inserting the simulation-measured spectra
 $P_{\delta\delta}, P_{\delta\theta}$ and $P_{\theta\theta}$ in
 Eq.~(\ref{eq:Pgg}), but ignoring the FoG effect, i.e. setting
 $F(k,\mu)=1$. Hence the differences between the dashed curves and, for
 example, the histograms are due to the FoG effect. The reconstructed
 power spectra obtained using the Taylor ($\sigma+\tau$) FoG model
 (denoted by the cross symbols) are found to well reproduce the
 redshift-space power spectrum. 
}  \label{fig:2d_slices_z=0}
\end{figure*}
\begin{figure}
\begin{center}
\includegraphics[width=4.1cm,angle=0]{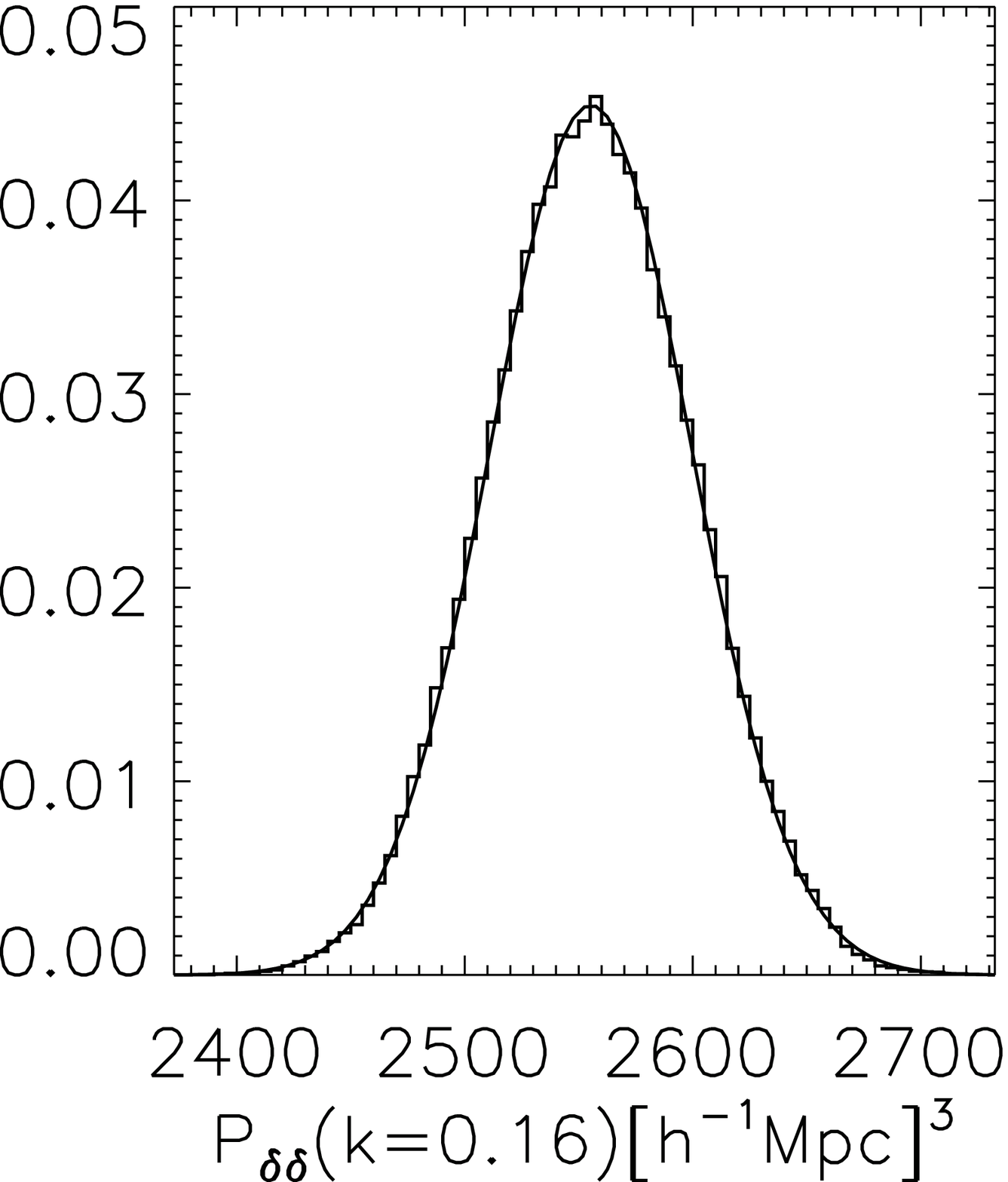}
\includegraphics[width=4.1cm,angle=0]{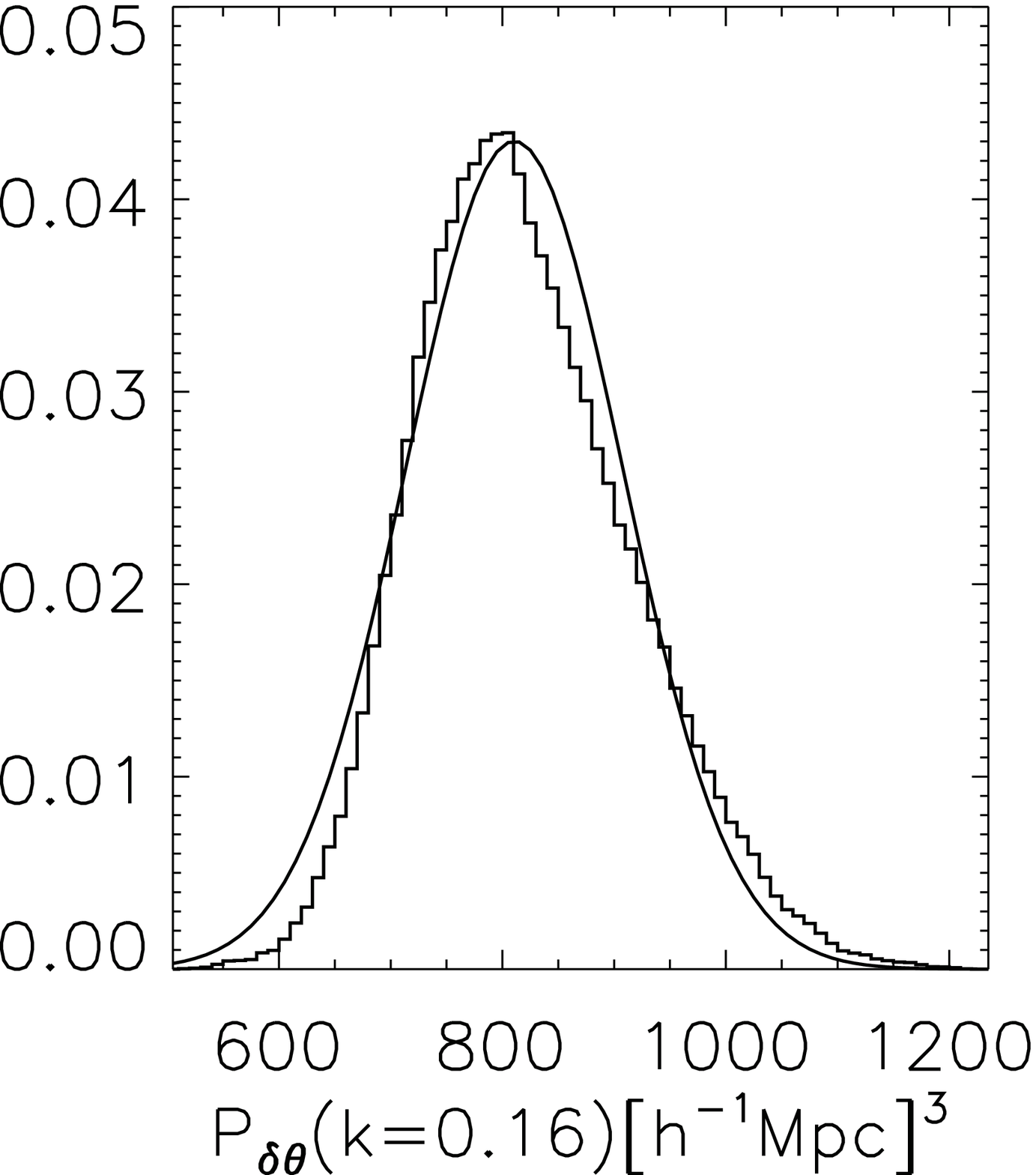}
\includegraphics[width=4.1cm,angle=0]{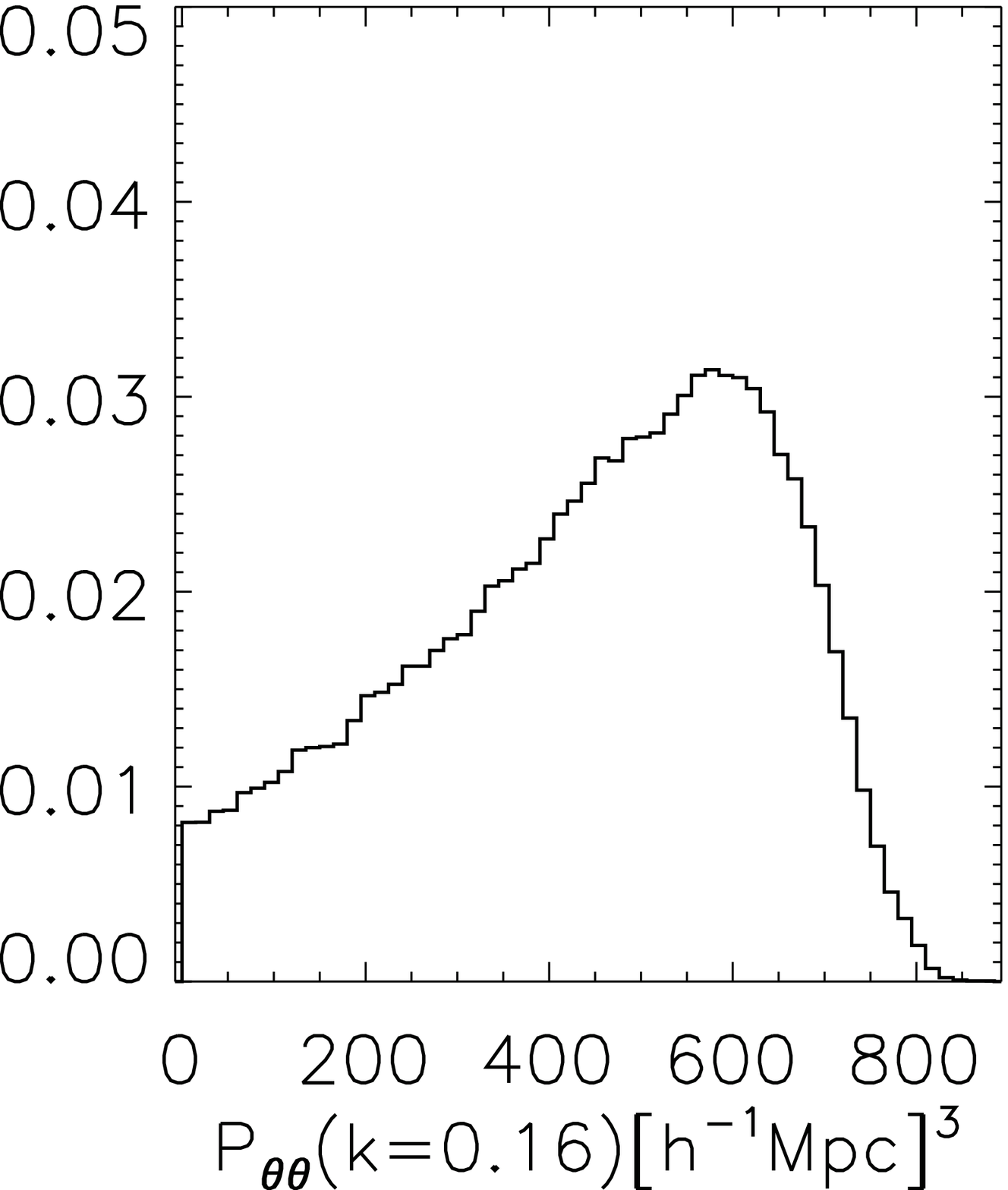}
\includegraphics[width=4.1cm,angle=0]{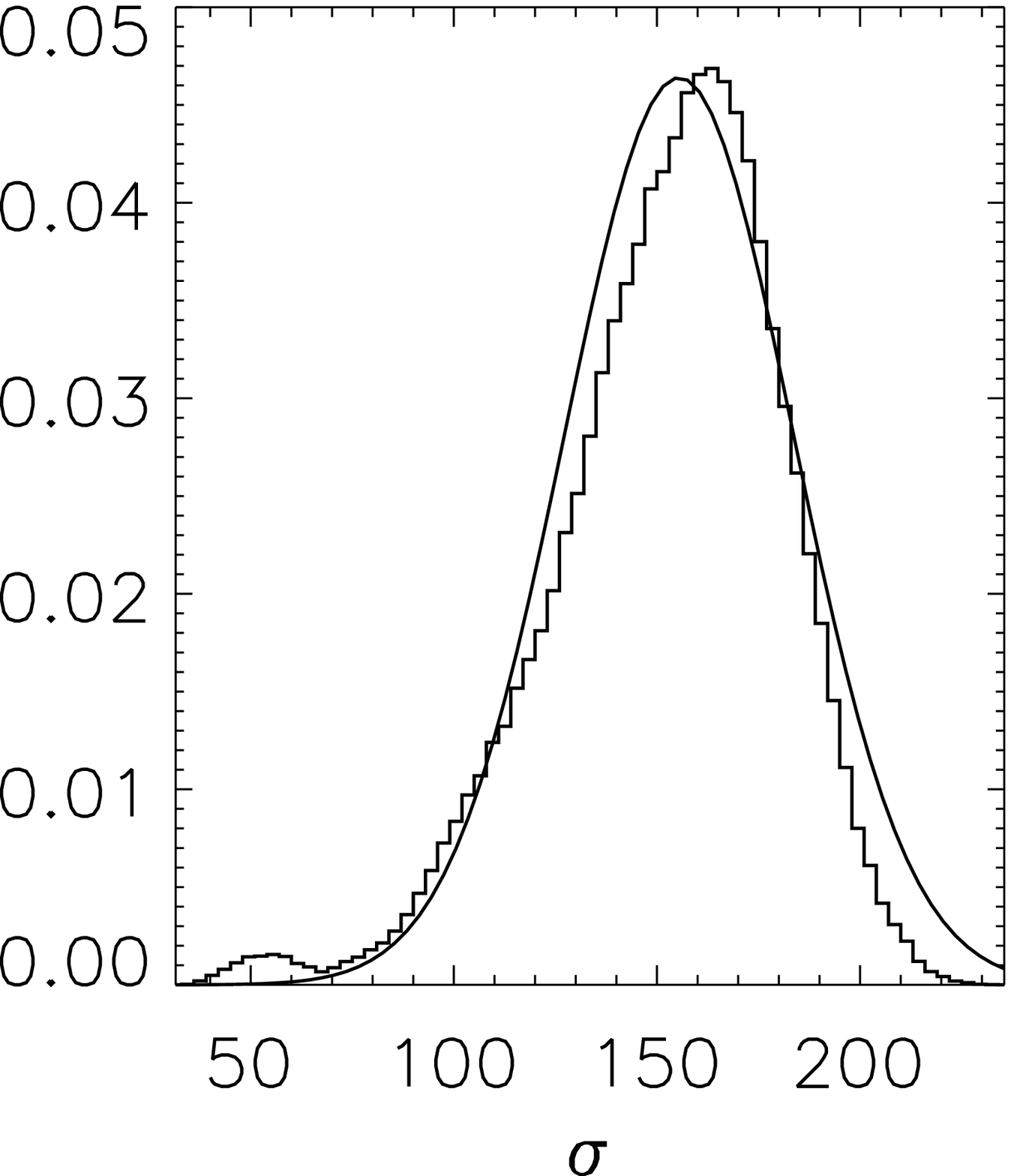}
\end{center}
\caption{
The posterior distributions of the band powers $\pdd$, $\pdv$ and $\pvv$
at $k=0.16\hompc$ and the FoG parameter $\sigma$, for the power spectrum
reconstruction of $z=0$ using the Gaussian FoG model as a demonstration
example. The histograms are computed from MCMC chains. The solid curve
in each panel represents a Gaussian distribution with the mean and
variance given by the MCMC chains. 
The distribution of $\pvv$ includes a range of $\pvv=0$, meaning that
 the band power is not well constrained. 
}  
\label{fig:exmp_pxx}
\end{figure}
\begin{figure}
\begin{center}
\includegraphics[width=5.8cm,angle=0]{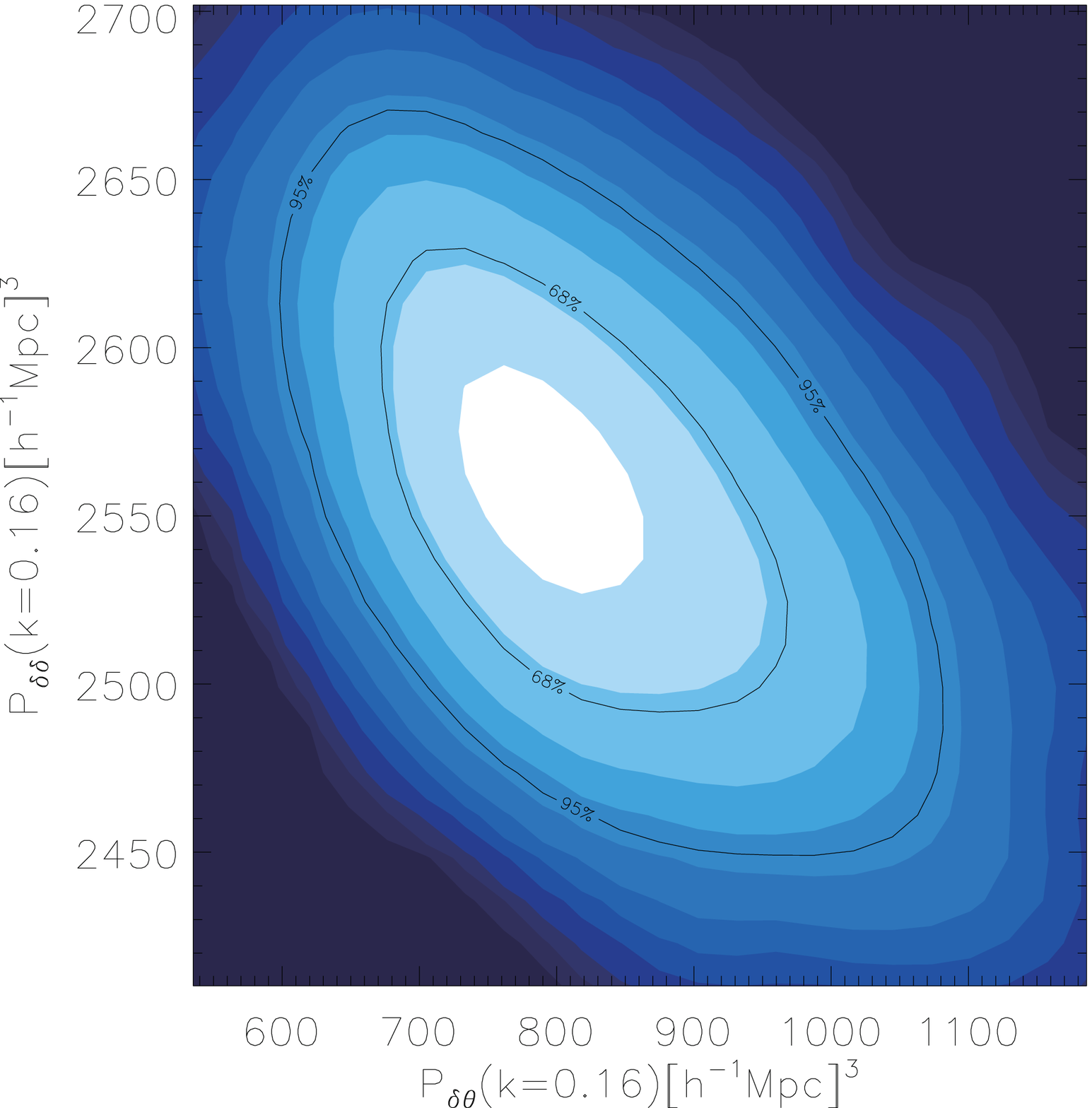}
\includegraphics[width=5.8cm,angle=0]{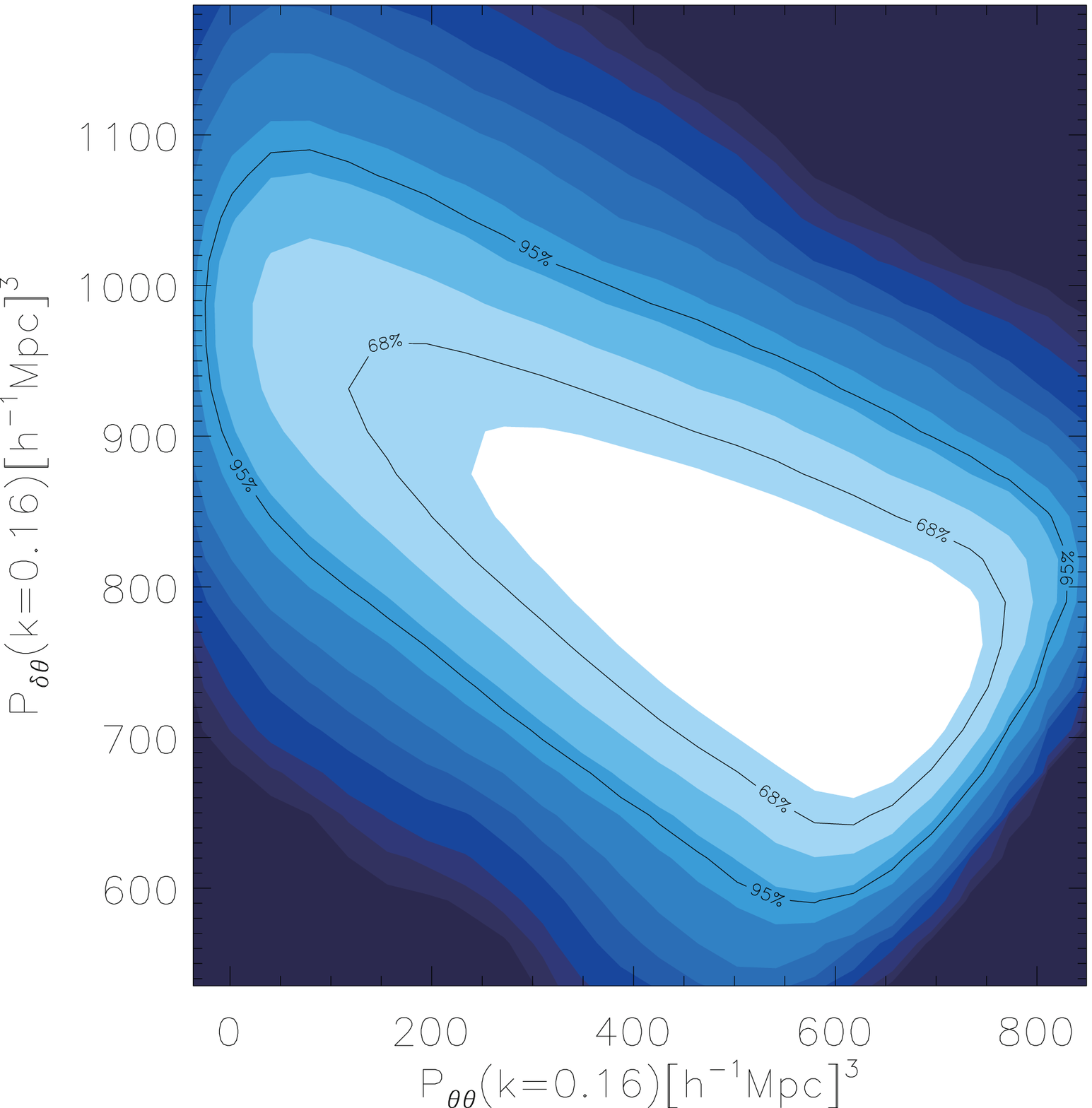}
\includegraphics[width=5.8cm,angle=0]{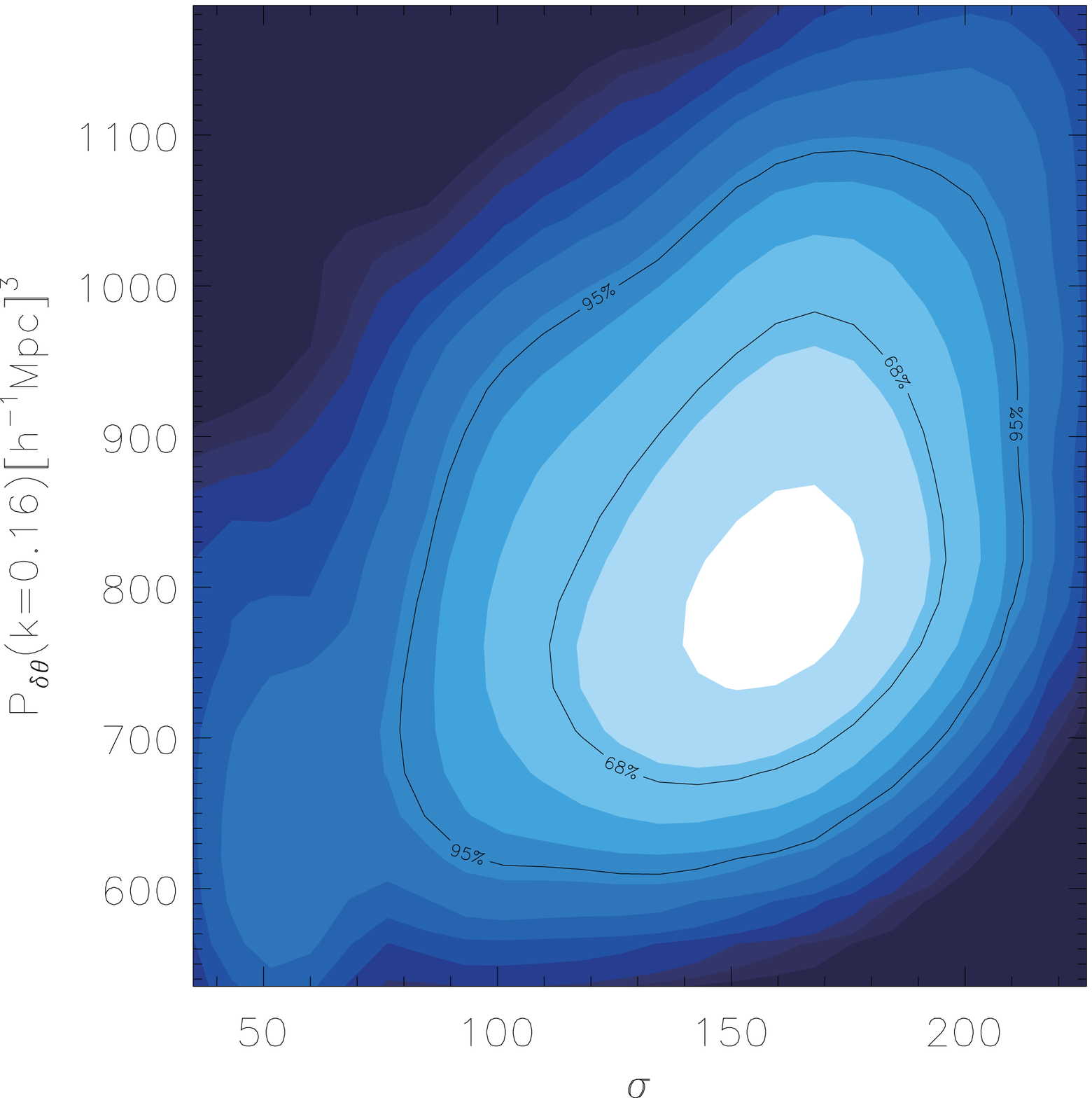}
\end{center}
\caption{The color scales represent the marginalized 2d probability
 distribution between the band powers ($\pdd$, $\pdv$, $\pvv$) and
 $\sigma$ as in the previous figure.
The two contours levels represent the confidence levels of 
68\% and 95\%,
 respectively.  
}
\label{fig:exmp_contour}
\end{figure}

We first assess performance of the maximum likelihood reconstruction
method developed in \S~\ref{sec:method} for matter spectra, by using
N-body simulations.  
We stress here again that the real-space density and velocity spectra
used to compare with the reconstructed power spectra, shown in figures
of this and following sections, are the spectra directly measured from
the simulations, and therefore include nonlinearity effects arising from
nonlinear clustering in structure formation.

To apply our method to N-body simulations, we need to compute the
likelihood function, given by Eq.~(\ref{eq:2dmethod}), for the
redshift-space density field. More precisely, in
Eq.~(\ref{eq:2dmethod}), we need to specify a survey volume $V_s$, which
determines the statistical uncertainties, and need to compute the
redshift-space power spectrum $P_s(k_i,\mu_a)$ at each $k$- and
$\mu$-bins from the simulations.  In the following we assume
$V_s=1h^{-3}{\rm Gpc}^3$ and use the spectrum $P_s(k_i,\mu_a)$ averaged
from 70 realizations to reduce the statistical scatters for illustrative
purpose. For the $k$-binning, we mainly use 19
wavenumber bins over $0.034\le k\le 0.3$~$h$Mpc$^{-1}$. We determined
the bin widths of $k$ and $\mu$ such that the area $\Delta k \times
k\Delta \mu$ in the two-dimensional Fourier space of $(k,\mu)$ is kept
about constant.
Therefore, instead of using the constant bin width, 
$\Delta k$ is taken to be 
large at small $k$ and gradually become smaller at larger $k$. 
Since we use the bin width $\Delta \mu =0.067$, we use $\Delta k\simeq 
0.02~h$Mpc$^{-1}$ 
at small $k$ bins, 
while we use $\Delta k=0.01$ around $k\simeq 0.2~{\rm Mpc}^{-1}$.
We will 
show below the reconstruction results for the density-density and
density-velocity power spectra, $P_{\delta\delta}$ and
$P_{\delta\theta}$, but not for the velocity-velocity spectrum
$P_{\theta\theta}$, because the reconstruction of $P_{\theta\theta}$ is
very noisy due to the smaller amplitudes, i.e. the small
signal-to-noise ratios, compared to $P_{\delta\delta}$ and
$P_{\delta\theta}$ \citep[also see][for the similar discussion]{Tegmarketal:04}.

Fig.~\ref{fig:re_dm} shows the results when using the simulation outputs
at $z=0$, for dark matter (N-body particle) distribution.  
The top-dotted curve shows the input density-density power
spectrum, $P_{\delta\delta}(k)$, directly measured from N-body
simulations (the average of 70 realizations). The three symbols around
the curve, although almost perfectly overlapped with each other, show
the reconstructed power spectra assuming different FoG models
(Eq.~[\ref{eq:FOG}]).  Note that we here show the results for the
Gaussian and Taylor-type FoG models, and do not and will not show the
results for the Lorentzian FoG model for illustrative purpose. The
results for the Lorentzian FoG model is very similar to the results of
the Gaussian and Taylor ($\sigma$) models; that is, we have found that 
all the results for
one-parameter FoG models are similar.  
The error bars around the symbols, although again
overlapped, show $1\sigma$ statistical uncertainties in the band power
reconstruction at each $k$-bin, including marginalization over
uncertainties in the reconstructed band powers at different $k$ bins and
for  
 different spectra $(P_{\delta\delta}, P_{\delta\theta},
P_{\theta\theta}$) as well as the FoG effect parameters. 
Encouragingly our method can well recover $P_{\delta\delta}(k)$, rather
irrespectively of the assumed FoG model.

 To be more precise the upper panel of Fig.~\ref{fig:re_diff_dm} shows
fractional differences between the input and reconstructed spectra:
$\Delta P/P\equiv \left[ P({\rm reconst.})-P({\rm input})\right]/P({\rm
sim.})$, where $P({\rm reconst.})$ and $P({\rm input})$ are the
reconstructed spectrum and the spectrum directly measured from
simulations, respectively.  Our method recovers the input power spectrum
within the statistical errors, achieving a few percent accuracy up to
$k\simeq 0.2h$Mpc$^{-1}$.  If we employ the Taylor-type FoG model
including the orders up to $(k\mu)^4$ (hereafter Taylor ``$\sigma+\tau$
model'') in Eq.~(\ref{eq:FOG}), our method can recover $P_{\delta\delta
}$ up to $k\simeq 0.3h$Mpc$^{-1}$, which is well in the nonlinear
regime.

The lower curves with different symbols in Fig.~\ref{fig:re_dm} show the
reconstruction results for the density-velocity power spectrum,
$P_{\delta\theta}(k)$, assuming different FoG models
(Eq.~[\ref{eq:FOG}]). The reconstruction of $P_{\delta\theta}$ is noisier
than in $P_{\delta \delta}$ due to the lower signal-to-noise ratios
\citep{Tegmarketal:04}. Also the reconstruction is sensitive to which
FoG model is assumed, reflecting that the redshift-space power spectrum
is affected by the FoG effect over a range of wavenumbers we
consider. Fig.~\ref{fig:re_dm} shows that the reconstructed
$P_{\delta\theta}$ is in closest agreement with the input spectrum, if
using the Taylor-($\sigma+\tau$) FoG model that is given by two free
parameters and has more degrees of freedom to describe a scale-dependent
FoG effect than other models (that respectively has only one free
parameter).

The lower panel of Fig.~\ref{fig:re_diff_dm} shows fractional
differences between the input and reconstructed spectra for
$P_{\delta\theta}$. Combined with the result for $\pdd$ shown in the
upper panel, one can notice that, although $\pdd$ and $\pdv$ are
unbiasedly recovered regardless of the FoG models at small $k$, the
results are substantially different at large $k$ depending on which FoG
model to use.  The FoG redshift distortion increasingly affects the
power spectrum with increasing $k$. 
As a result, the reconstructed band
powers at different $k$ bins are correlated 
with each other via the FoG effect being marginalized over, 
and therefore the correlations need to be properly taken into account 
(see below). 
Fig.~\ref{fig:re_diff_dm} shows that, among the different FoG
models, the performance of the Taylor ($\sigma+\tau$) model is of
promise; it can unbiasedly recover $\pdd$ over all the scales as well as
$\pdv$ within the statistical uncertainties up to $k\simeq 0.25~h{\rm
Mpc}^{-1}$, implying that the FoG model can nicely fit the strong FoG
effect in simulations.

To have more insights on the results in Figs.~\ref{fig:re_dm} and
\ref{fig:re_diff_dm}, Fig.~\ref{fig:2d_slices_z=0} show slices of the
redshift-space power spectrum amplitudes, $P_s(k,\mu)$, as a function of
the azimuthal angle $\mu$, for a fixed radius $k$: $k=0.08$, $0.16$ and
$0.29$~$h{\rm Mpc}^{-1}$ from the left to right panels, respectively.
The histograms in each panel show the band powers measured from
simulations, which are compared with the best-fit power spectra
(different symbols) obtained in Fig.~\ref{fig:re_dm}. The best-fit
spectra are computed by inserting the best-fit parameters (band powers
at each $k$ bins and the FoG parameters) into Eq.~(\ref{eq:Pgg}).  As in
Fig.~\ref{fig:re_dm}, the different symbols are computed for the
different FoG models.  For comparison, we also plot the spectra, by
dashed curves, which are computed by inserting the directly- measured
$P_{\delta\delta}$, $P_{\delta\theta}$ and $P_{\theta\theta}$ in
Eq.~(\ref{eq:Pgg}), without FoG term, i.e.  $F(k,\mu)=1$.  Hence, the
difference between the dashed curve and the solid histogram clarifies
how strongly the FoG affects the redshift-space power spectrum at each
$k$ bin.

The two extreme
cases are very distinctive.  At small $k$ ($k=0.08\hompc$) the FoG
effect is very small, and all the reconstructed power spectra can well
match the input spectra
 independently of the FoG models. 
One the other hand, at the largest $k$
($k=0.29\hompc$), one can clearly see that the FoG effect is so
strong that none of the Taylor ($\sigma$) or Gaussian models (also or
Lorentzian model)
can fit the $\mu$-dependence of redshift-space power spectrum. 
It
is essential to add an additional parameter in the FoG model, like
Taylor ($\sigma+\tau$) model, to
reproduce the simulation results. 
The middle panel shows the result at the
intermediate scale ($k=0.16\hompc$), where the FoG
effect is mild and the FoG models of one parameter work to
a good approximation.

Fig.~\ref{fig:exmp_pxx} shows the posterior, marginalized 
distributions of the band
powers, $\pdd$, $\pdv$, and $\pvv$ at $k=0.16~ \hompc$ and the $\sigma$
parameter of the Gaussian FoG model, 
which are obtained from the MCMC
chains. 
Here we show the reconstruction results 
assuming the Gaussian
FoG model, which well works at the scale of $k=0.16\hompc$ as shown in
Fig.~\ref{fig:2d_slices_z=0}. 
The figure shows that, while the distribution of $\pdd$ looks Gaussian,
the distributions of $\pdv$, $\pvv$ and $\sigma$ show skewed,
non-Gaussian distributions. In particular, the distribution of $\pvv$
has a wide distribution and includes a region around $\pvv=0$, showing
no constraint on the band power of $\pvv$. Given these results, we
conclude that it is very difficult to reliably reconstruct $\pvv$ based on the
method developed in this paper, at least for a survey with survey volume
$\sim 1$Gpc$^3$. 

The origin of the skewed distributions in Fig.~\ref{fig:exmp_pxx} is
explored in Fig.~\ref{fig:exmp_contour}, which shows the posterior
distributions in a two-parameter sub-space between the parameters in
Fig.~\ref{fig:exmp_pxx}.  The figure clearly shows that the different
parameters are correlated with each other after the nonlinear
reconstruction. In particular, the $\sigma$ parameter of Gaussian
FoG model, shown here as an example, shows a strong correlation with the
band power $\pdv$.  Thus the band powers of different spectra
($\pdd,\pdv,\pvv$) at different $k$ bins are correlated with each other,
and the correlation needs to be properly taken into account for the power
spectrum reconstruction.

 \begin{figure}
\centering
\includegraphics[width=8.8cm,angle=0]{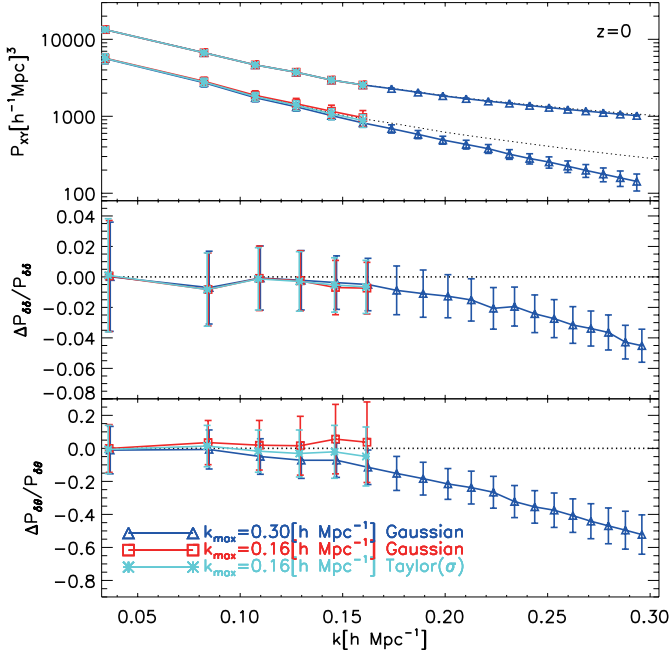}
\caption{ {\em Upper panel}: Sensitivity of the reconstructed power
  spectra $P_{\delta \delta}$ and $P_{\delta\theta}$ at $z=0$ to the
  maximum wavenumber $k_{\rm max}$, where the redshift-space power
  spectrum information up to $k_{\rm max}$ is used for the power
  spectrum reconstruction. The triangle and square symbols show the
  results for $k_{\rm max}=0.16$ and $0.30~\hompc$, respectively,
  assuming the Gaussian FoG effect as a working example.  From
  comparison, the star symbols show the results assuming $k_{\rm
  max}=0.16~\hompc$ and the Taylor ($\sigma$) FoG model.  {\em Middle
  and lower panels}: The fractional differences between the input and
  reconstructed power spectra as in Fig.~\ref{fig:re_diff_dm}.
\label{fig:2d_nk_cmp} 
}
\end{figure}

Given such strong correlations between the band powers, how sensitive is
the reconstruction of the power spectrum to a choice of the maximum
wavenumber $k_{\rm max}$? Fig.~\ref{fig:2d_nk_cmp} studies this
question.  With increasing $k_{\rm max}$, the redshift-space power
spectrum to use for the reconstruction is more affected by the FoG
effect, and in turn the reconstructed $\pdd$ and $\pdv$ become
increasingly affected by the FoG effect after marginalization.  The
figure shows the reconstructed $\pdd$ and $\pdv$ obtained when including
the redshift-space power spectrum information up to $k_{\rm max}=0.16$
and $0.3~h{\rm Mpc}^{-1}$, respectively.  For the triangle and square
symbols, we assumed the Gaussian FoG model for comparison. The results
for $\pdd$ agree and are only slightly different at scales around
$k_{\rm max}=0.16~h{\rm Mpc}^{-1}$, while the results for $\pdv$ are
systematically different. Since the Gaussian FoG model cannot well
describe the FoG effect seen in simulations as implied in
Fig.~\ref{fig:re_dm}, the inaccuracy of the Gaussian FoG model causes a
systematic underestimation in the band powers of $\pdv$ if including the
higher-$k$ modes. However, the two results for $\pdd$ and $\pdv$ agree
over an overlapping range of $k$, up to $k=0.16~h{\rm Mpc}^{-1}$, within
the error bars.  For comparison, we also show the results obtained
assuming $k_{\rm max}=0.16~h{\rm Mpc}^{-1}$ and the Taylor ($\sigma$)
FoG model (see Eq.~[\ref{eq:FOG}]). The Taylor ($\sigma$) model is found
to give a less biased reconstruction of $\pdv$, implying the importance
of the assumed FoG model even around $k\simeq 0.16~h{\rm Mpc}^{-1}$.

 \begin{table}
  \begin{center}
   \begin{tabular}{|l|l|l|}
     \hline
        FoG model  &  $k_{\rm max}=0.3\hompc$ & $k_{\rm max}=0.16\hompc$ \\
     \hline
           Gaussian ($\sigma$) & $155^{+24.5}_{-30.5}$ &  $215^{+118}_{-137}$ \\
           Taylor ($\sigma$) & $105^{+23.1}_{34.4}$  &  $170^{+96.2}_{-109}$ \\
           Taylor ($\sigma+\tau$) & ($287^{+17.9}_{-17.0}$, $356^{+73.1}_{-76.2}$) & ---\\
     \hline
   \end{tabular}
    \caption{The best-fit FoG parameters 
assuming different $k_{\rm max}$ for the results of $z=0$ simulations. 
The units of the numbers
   shown here are $h{\rm Mpc}^{-1}$. 
For the Taylor
   ($\sigma+\tau$) model the parameters $\sigma$ and $\tau$ are not well
   constrained if using the redshift-space spectrum information up to
   $k_{\rm max}=0.16~\hompc$, and therefor are not shown here.}
\label{tb:FoG_002}
  \end{center}
\end{table}

\begin{figure}
\begin{center}
\includegraphics[width=8.cm,angle=0]{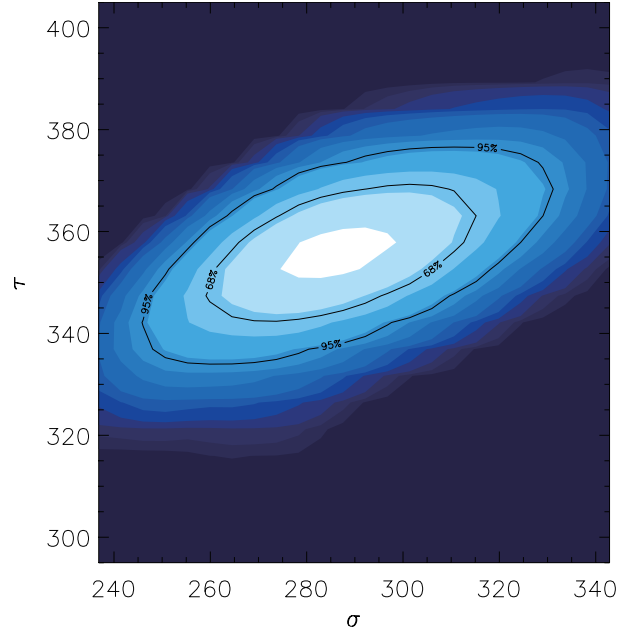}
\end{center}
\caption{The contours represents the marginalized 2d probability
 distribution in the parameter space ($\sigma, \tau$) of the Taylor
 ($\sigma+\tau$) FoG model (for the results shown in
 Figs.~\ref{fig:re_dm} and \ref{fig:re_diff_dm}). 
The two contours levels represent the
 confidence levels 68\% and 95\%, respectively. }
 \label{fig:sigma_tau_002}
\end{figure}

Table~\ref{tb:FoG_002} summarizes the best-fit FoG parameters and the
marginalized confidence ranges that are obtained for the reconstructions
at $z=0$ assuming different $k_{\rm max}$: $k_{\rm max}=0.3$ and
$0.16\hompc$, respectively.  For the Taylor ($\sigma+\tau$) FoG model,
the error of $\tau$ parameter is smaller than that of $\sigma$, because
the $\tau$ parameter has a stronger dependence on $k\mu$ as $
(\tau k\mu)^4$, than the $\sigma$-term does. Fig.~\ref{fig:sigma_tau_002}
shows the 2D posterior distribution in the ($\sigma$,$\tau$) sub-space.

\begin{figure*}
\begin{center}
\begin{minipage}[c]{1.00\textwidth}
\centering
\includegraphics[width=8.0cm,angle=0]{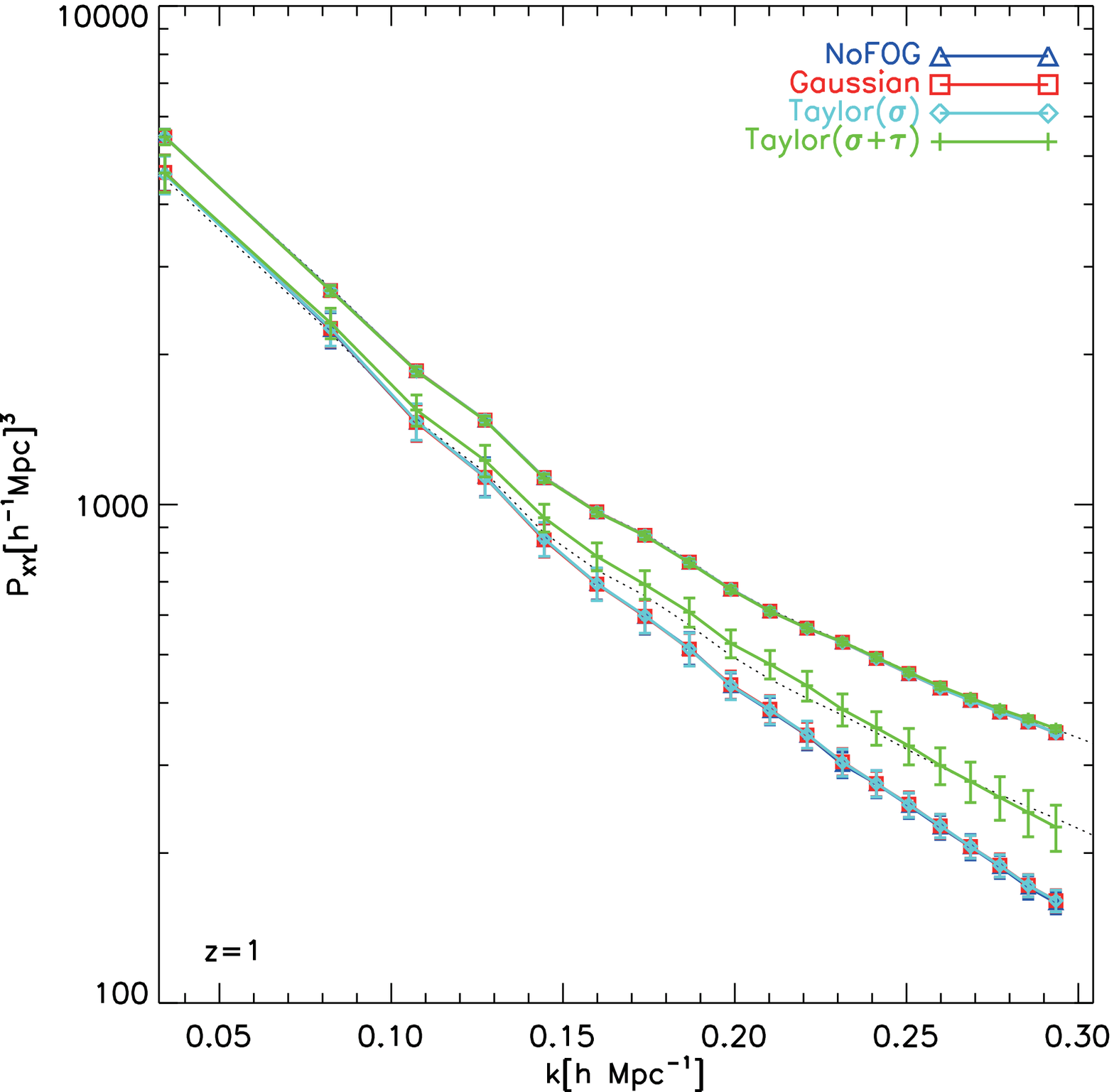}
\includegraphics[width=7.6cm,angle=0]{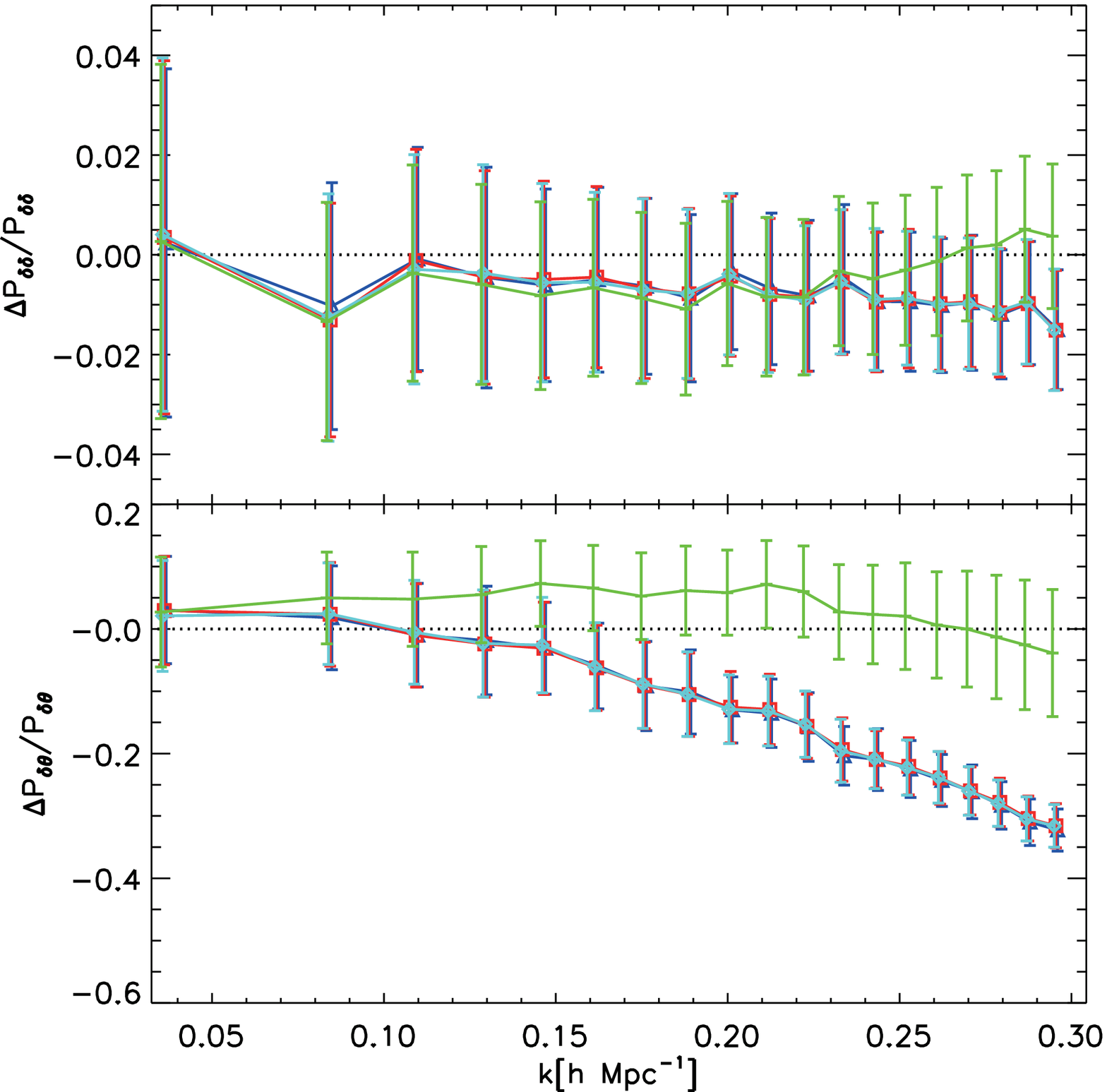}\\
\includegraphics[width=5cm,angle=0]{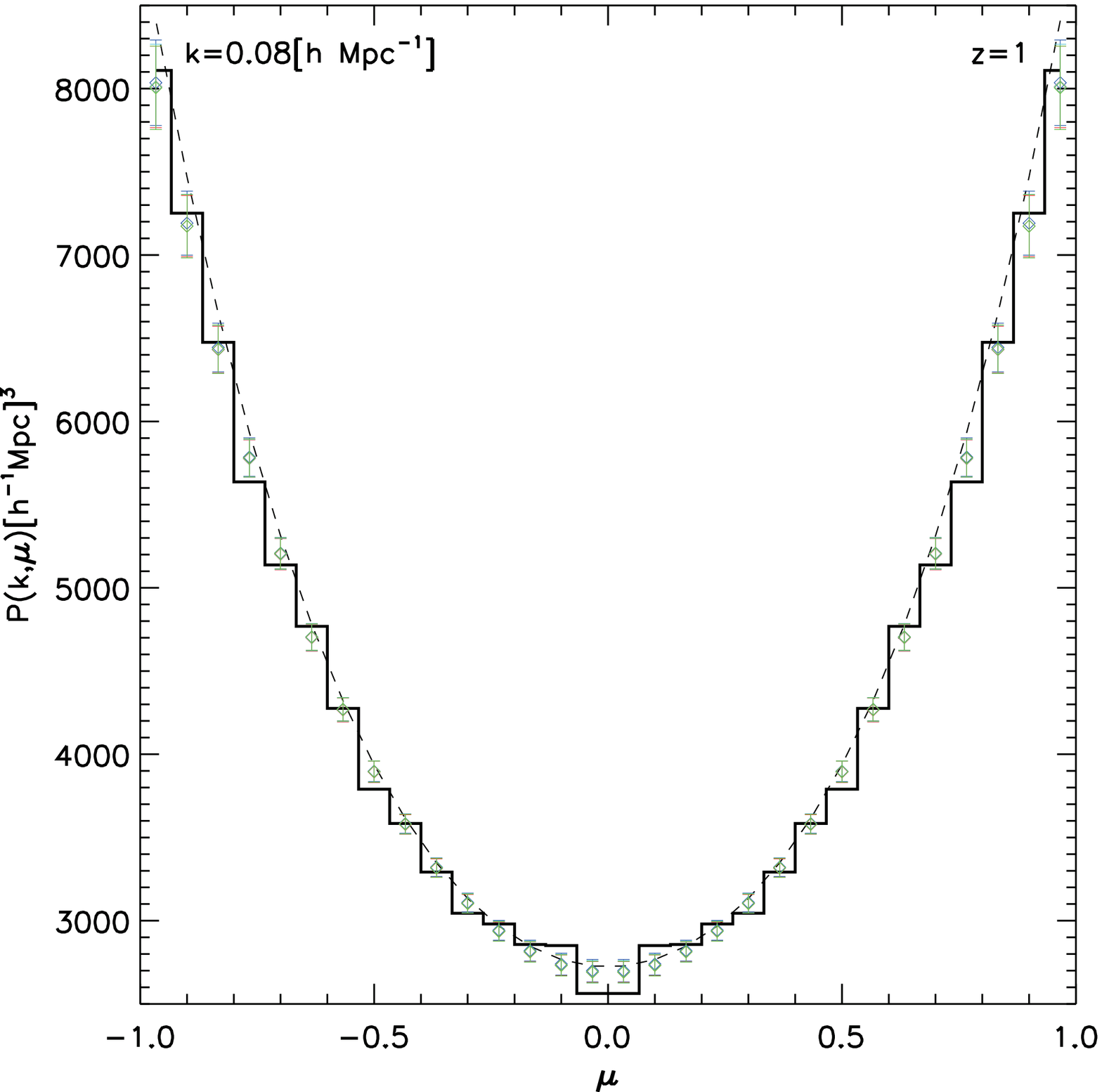}
\includegraphics[width=5cm,angle=0]{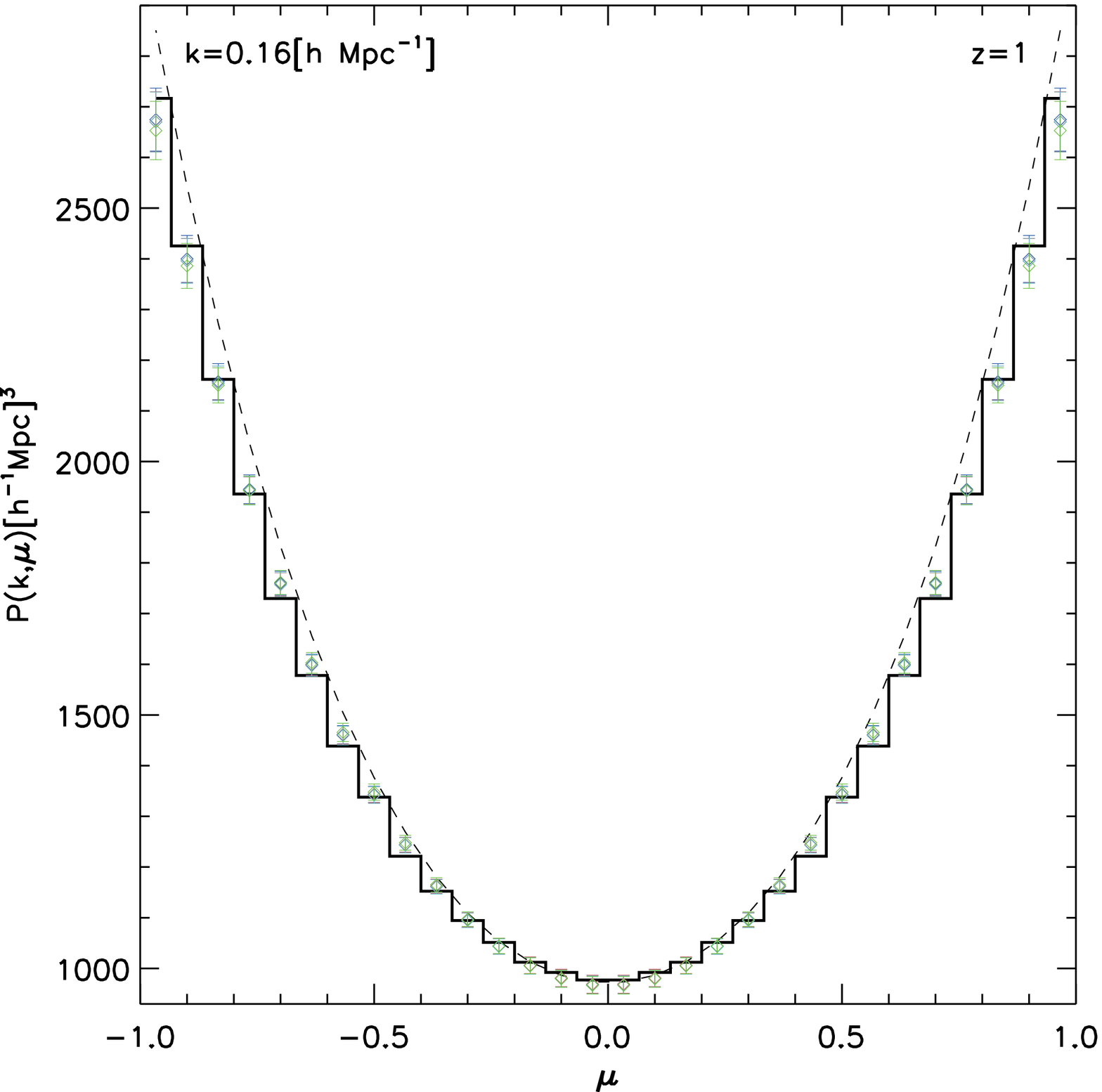}
\includegraphics[width=5cm,angle=0]{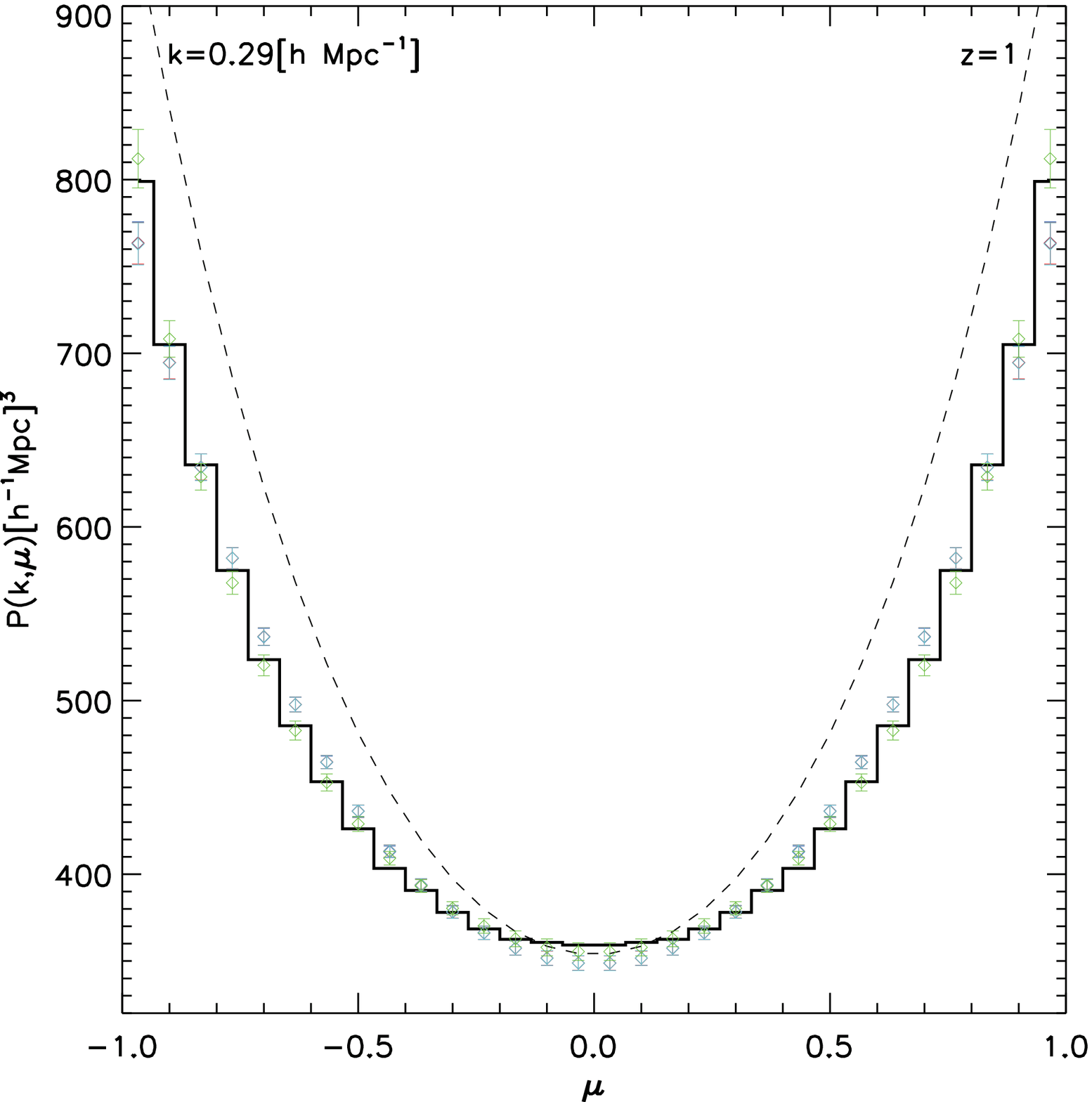}
\end{minipage}
\caption{Same as in Figs.~\ref{fig:re_dm}, \ref{fig:re_diff_dm} and
\ref{fig:2d_slices_z=0}, but for redshift $z=1$. 
\label{fig:2d_all_methods_001}}
\end{center}
\end{figure*}

 \begin{table}
  \begin{center}
   \begin{tabular}{|l|l|l|}
     \hline
        FoG model  &  $k_{\rm max}=0.3\hompc$ & $k_{\rm max}=0.16\hompc$ \\
     \hline
           Gaussian ($\sigma$) & $17.4^{+18.5}_{-12.2}$ &  $82.6^{+82.4}_{-57.6}$ \\
           Taylor ($\sigma$) & $18.6^{+18.8}_{-12.8}$  &  $82.4^{+76.4}_{-57.1}$ \\
           Taylor ($\sigma+\tau$) & ($240^{+18.4}_{-20.2}$, $336^{+12.6}_{-14.4}$) & ---\\
     \hline
   \end{tabular}
    \caption{The best fit FoG parameters assuming different $k_{\rm
   max}$ for $z=1$, as in Table~\ref{tb:FoG_002}. }
\label{tb:FoG_001}
  \end{center}
\end{table}

\begin{figure}
\begin{center}
\includegraphics[width=8.5cm,angle=0]{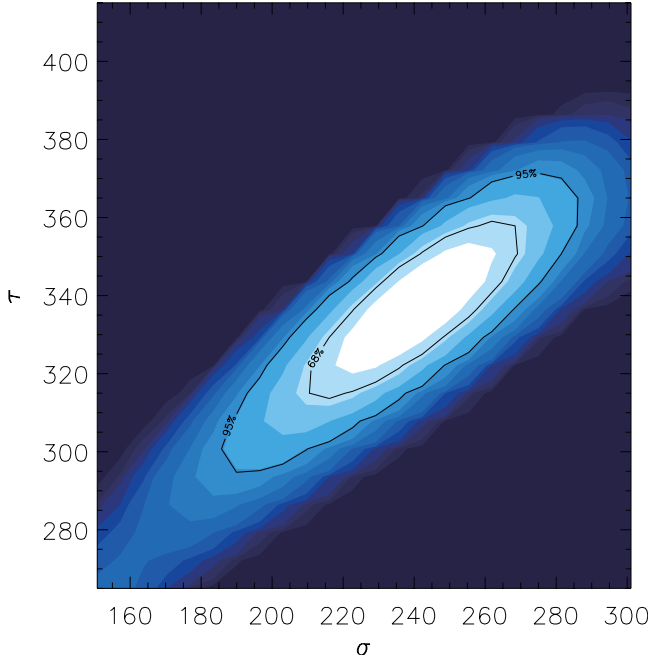}
\end{center}
\caption{The posterior distribution in the ($\sigma,\tau$) parameter
 space for the Taylor ($\sigma+\tau$) FoG model at $z=1$, as in
 Fig.~\ref{fig:exmp_contour}.
}
\label{fig:sigma_tau_001}
\end{figure}

Nonlinearities in matter clustering are less significant at higher
redshifts. Hence the likelihood reconstruction of power spectrum we are
studying may work better for higher
redshifts. Fig.~\ref{fig:2d_all_methods_001} shows the results using
simulations at $z=1$, similarly to Figs.~\ref{fig:re_dm},
\ref{fig:re_diff_dm}, and \ref{fig:2d_slices_z=0}. Note that we assumed
$k_{\rm max}=0.3~h{\rm Mpc}^{-1}$ as done in the $z=0$ reconstruction.
In fact the FoG effect is smaller at $z=1$ than $z=0$, e.g. as seen from
the bottom-right panel of Fig.~\ref{fig:2d_all_methods_001} compared to
Fig.~\ref{fig:2d_slices_z=0}.  However, the Taylor ($\sigma+\tau$) FoG
model seems still needed in order to better recover the simulation
spectra. The bottom-right panel clearly shows that, although the FoG effect is
smaller around $\mu=0$ compared to the $z=0$ result
(Fig.~\ref{fig:2d_slices_z=0}), the Taylor ($\sigma+\tau$) model better
captures the simulation results around $\mu=\pm 1$, showing a stronger
dependence of $\mu$ than the Gaussian or Taylor ($\sigma$) 
(or Lorentzian) models do. 
This implies that it is important to properly include the
scale-dependent FoG effect for the power spectrum reconstruction, at
least up to $z\simeq 1$ we have studied. As given in
Table~\ref{tb:FoG_001}, a non-zero $\tau$ parameter is favored to
capture the FoG effect seen in simulations. Fig.~\ref{fig:sigma_tau_001}
shows the 2D posterior distribution in the ($\sigma$,$\tau$) sub-space,
displaying a strong correlation between the two parameters.

\subsection{Reconstruction of halo power spectra} 
\label{sec:results_halos}

Now let us move on to the reconstruction of halo power spectra, which
are more relevant for a galaxy survey, using the halo catalogs
constructed from 70 simulation realizations (see \S~\ref{sec:dm_catalog}
for details). The halo clustering in redshift space is least affected by
the FoG effect, because the halos are treated as points and the redshift
distortion effect on halo clustering is caused only by their bulk
motions in large-scale structure, not by the internal virial motion
within one halo. Therefore we can naively expect a more accurate
reconstruction of the density and velocity power spectra for halos based
on the maximum likelihood method we have developed in this
paper. However, unfortunately, this is not that simple as shown below.

For halo power spectrum we need to take into account the effect of shot
noise arising from an imperfect sampling of the density fluctuation
field
due to the finite number of halos.  In this case the maximum
likelihood for power spectrum reconstruction needs to be modified as
\begin{eqnarray}
-2\ln{\cal L}&=&\sum_{k_i, \mu_a}N(k_i,\mu_a)\left[
 \frac{\hat{P^s}(k_i,\mu_a)-\pshotnoise}{P^s(k_i,\mu_a)}\right.\nonumber\\
&&\hspace{2em}
+\left.\ln \frac{P^s(k_i,\mu_a)}{\hat{P^s}(k_i,\mu_a)-\pshotnoise}-1
\right], \label{eq:2dmethod_halo}
\end{eqnarray}
where $\pshotnoise=1/\bar{n}$ is the shot noise contamination, and
$\bar{n}$ is the mean number density of halos.  In the following we
simply assume that the shot noise is not a free parameter and 
given by the mean number density of halos we use for the power spectrum
reconstruction: $\pshotnoise=1/\bar{n}$ \citep[see][for a promising
method to further suppress the shot noise
contamination]{Seljaketal:09}. The shot noise expression is not accurate
for the actually measured power spectrum \citep[e.g.][]{Smithetal:07},
but the residual shot noise, even if exists, primarily contaminates to
the spectrum that is proportional to $\mu^0$, i.e. the density-density
spectrum $\pdd$.

However the main obstacle we have faced is that we cannot reliably
measure the velocity field of halos (see \S~\ref{sec:halo_spectra} for
details) and cannot therefore have the velocity-related power spectra,
which are needed to assess the performance of our reconstruction method
by comparing with the reconstructed spectra $\pdd$ and $\pdv$.  Rather
we decided to use the dark matter (N-body particles) velocity field
instead of estimating the halo velocity field, assuming that the halo
bulk-velocity field is unbiased from the matter velocity field, which is
often assumed in the literature.

Hence, before going to the halo spectrum reconstruction, we make a
simple test to study whether or not the power spectrum reconstruction is
affected by the shot noise contamination. This test can be done
by applying our reconstruction method to the catalogs with reduced
N-body particles.
To be more precise we randomly select N-body particles from
each simulation realization ($z=0$) until the number density of
particles selected becomes the same to the density of halo catalogs, 
$\bar{n}\simeq 3.8\times10^{-4} h^{3}{\rm
Mpc}^{-3}$.
 Then by using the reduced N-body particles in each
simulation we compute the redshift-space power spectrum taking into
account redshift modulation due to the velocity field of each
particle. These procedures preserve the underlying spectra of $\pdd$ and
$\pdv$. Thus we can compare the spectra with the spectra reconstructed
by applying the maximum likelihood method to the redshift-space spectrum
of reduced N-body particles, where the shot noise contamination is
subtracted from the measured spectrum according to
Eq.~(\ref{eq:2dmethod_halo}).

\begin{figure*}
\begin{center}
\begin{minipage}[c]{1.00\textwidth}
\centering
      \includegraphics[width=5.8cm,angle=0]{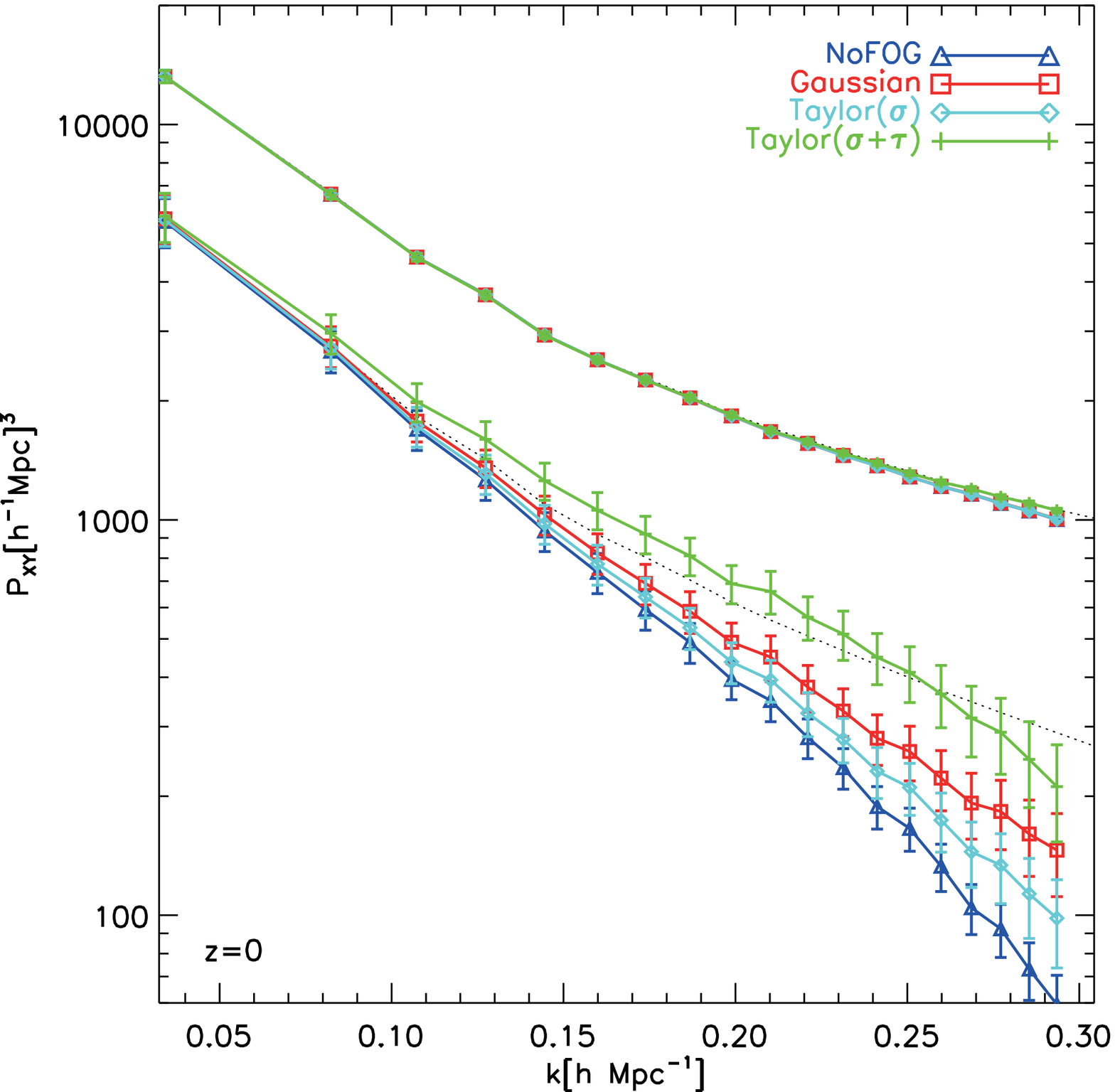}
      \includegraphics[width=5.8cm,angle=0]{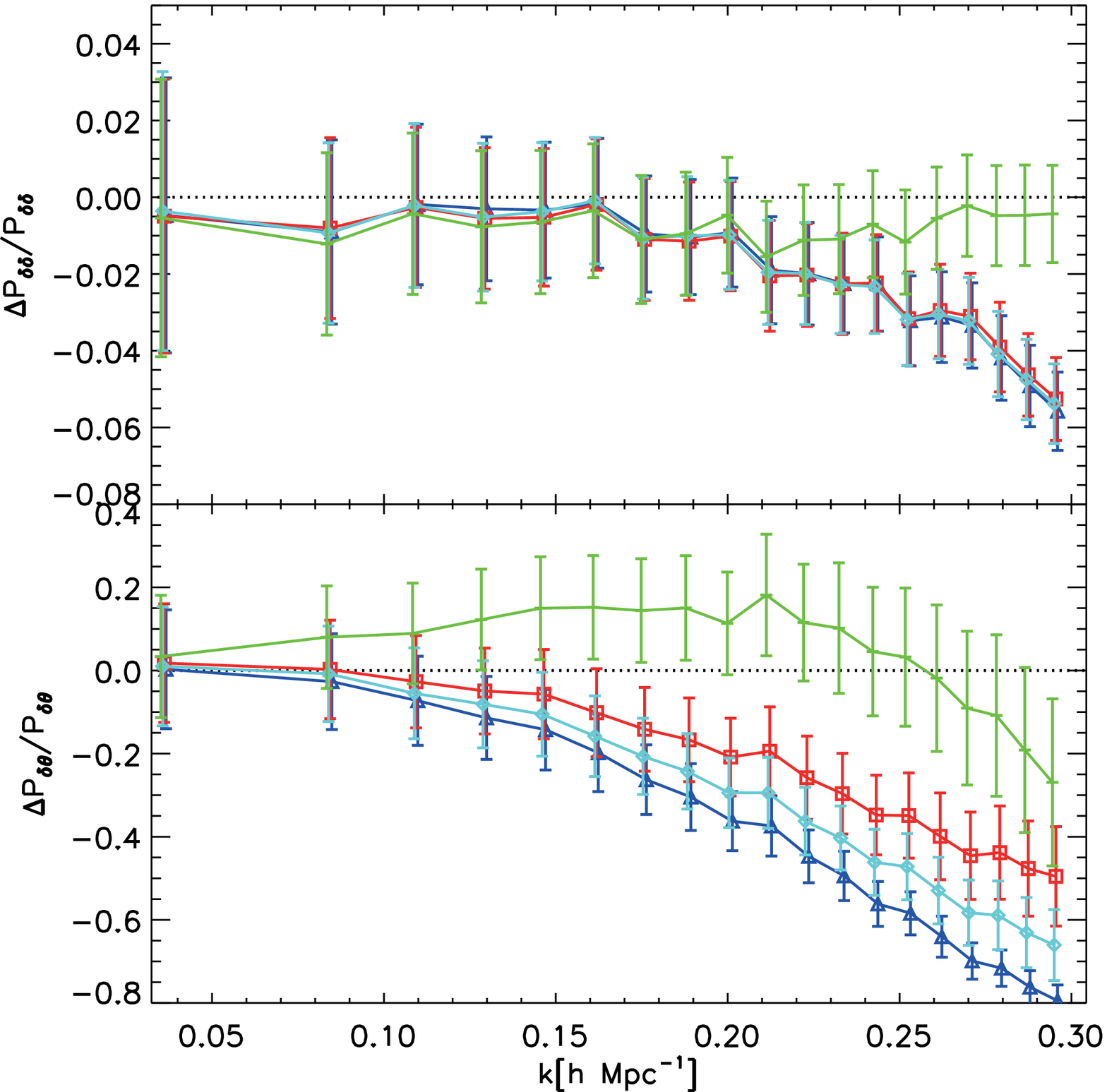}
     \includegraphics[width=5.8cm,angle=0]{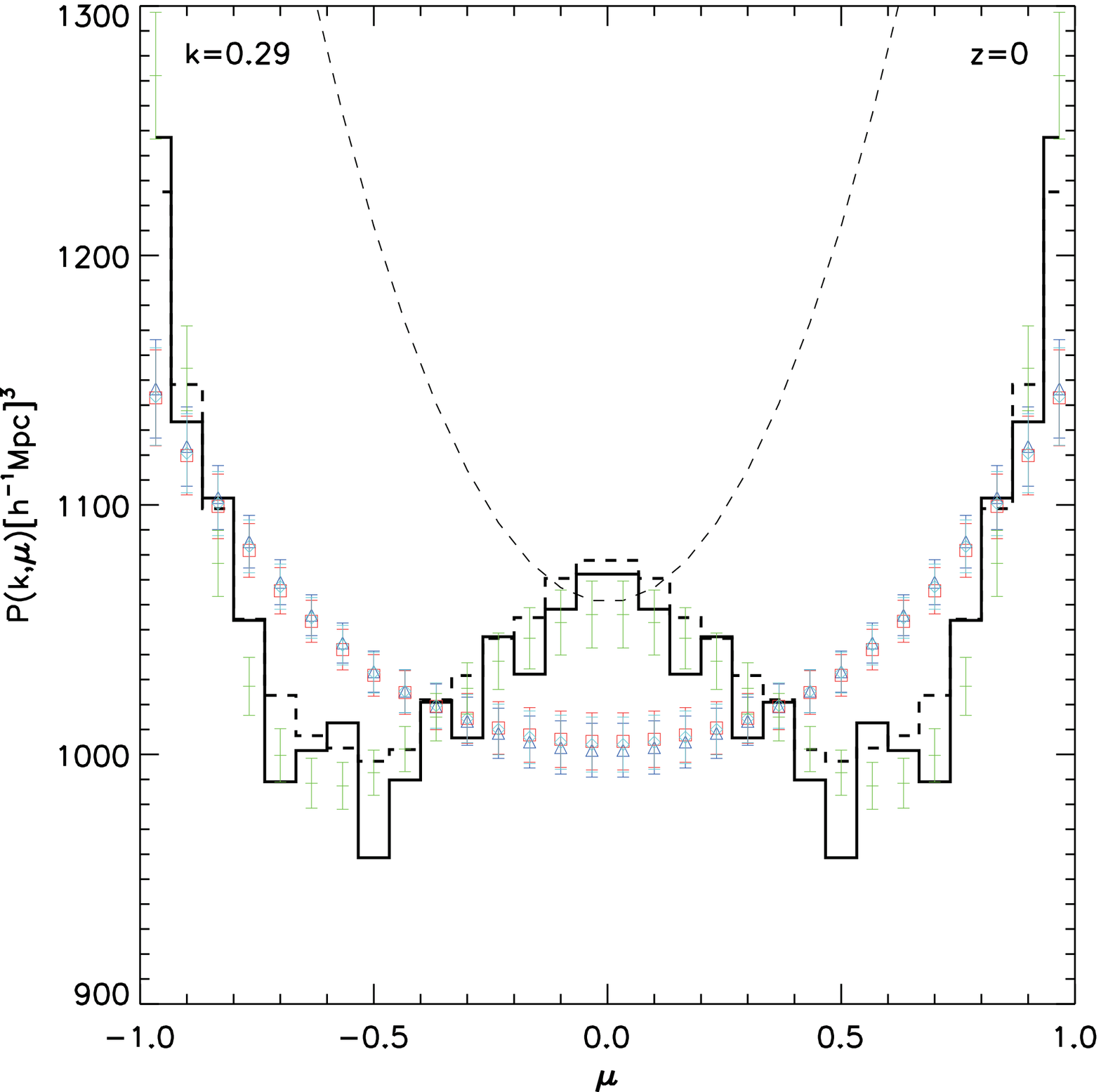}
\end{minipage}
\caption{Reconstruction results of the power spectra for the reduced
N-body particle distribution as in Fig.~\ref{fig:2d_all_methods_001},
where a smaller number of N-body particles are randomly selected in each
simulation realization in such a way that the mean number density of the
resulting particles becomes comparable to that of halo catalogs we will
use below (see \S~\ref{sec:dm_catalog}): $\bar{n}=3.8\times
 10^{-4}~h^{3}{\rm Mpc}^{-3}$.
In this case the shot noise term arising from
a finite number of the sampled N-body particles affects the
redshift-space power spectrum measurement.  
We applied the power spectrum reconstruction method
(Eq.~[\ref{eq:2dmethod}]) to the
redshift-space power spectrum after simply subtracting the expected shot noise
term $1/\bar{n}$ from the measured spectrum. 
The dotted curves in the left and right panels and the spectrum in
denominator in the middle panel are the input power spectrum, which are
 the same as the spectra of original N-body particles in
 Figs.~\ref{fig:re_dm}, \ref{fig:re_diff_dm} and \ref{fig:2d_slices_z=0}.
It is found that our method nicely recovers the input power spectra
 even in the presence of shot noise contamination. 
\label{fig:2d_all_methods_002_reduce}}
\end{center}
\end{figure*}

Fig.~\ref{fig:2d_all_methods_002_reduce} shows the reconstruction
results (symbols in each panels) for the catalogs of reduced N-body
particles.  The directly measured spectra in the left panel are similar
to the curves in Fig.~\ref{fig:re_dm}, although we found a small
difference in the directly measured $\pdd(k)$ at $k\simgt 0.2~h{\rm
Mpc}^{-1}$  due to the residual shot noise effect
\citep{Smithetal:07}.  Fig.~\ref{fig:2d_all_methods_002_reduce}
clearly shows that, even in the presence
of shot noise, our reconstruction method recovers the power spectra to a
similar precision to the results in Figs.~\ref{fig:re_dm},
\ref{fig:re_diff_dm} and \ref{fig:2d_slices_z=0}. Hence we conclude that
the shot noise is not a serious source of systematics for our method.

\begin{figure*}
\begin{center}
\begin{minipage}[c]{1.00\textwidth}
\centering
\includegraphics[width=8.0cm,angle=0]{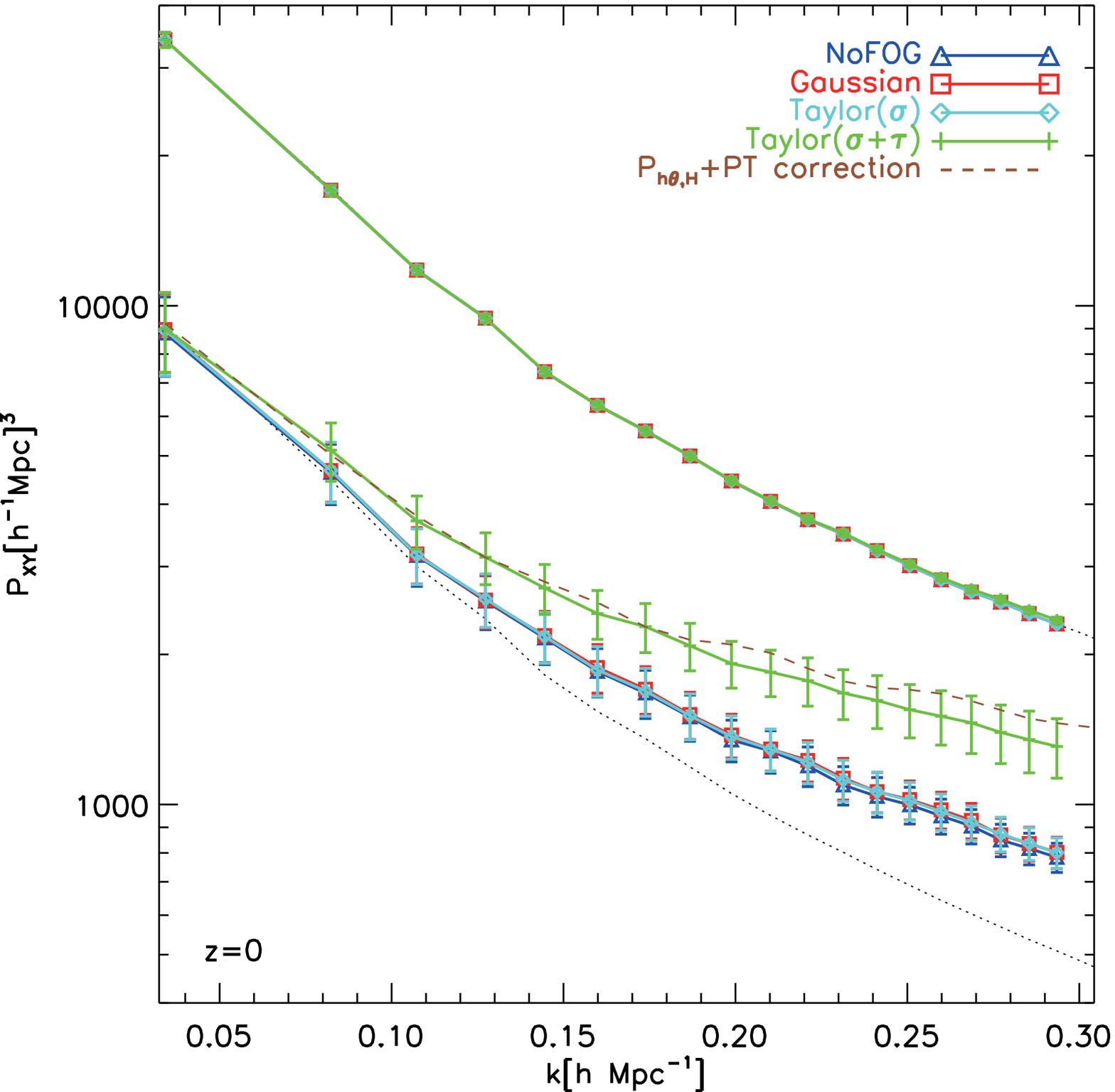}
\includegraphics[width=7.6cm,angle=0]{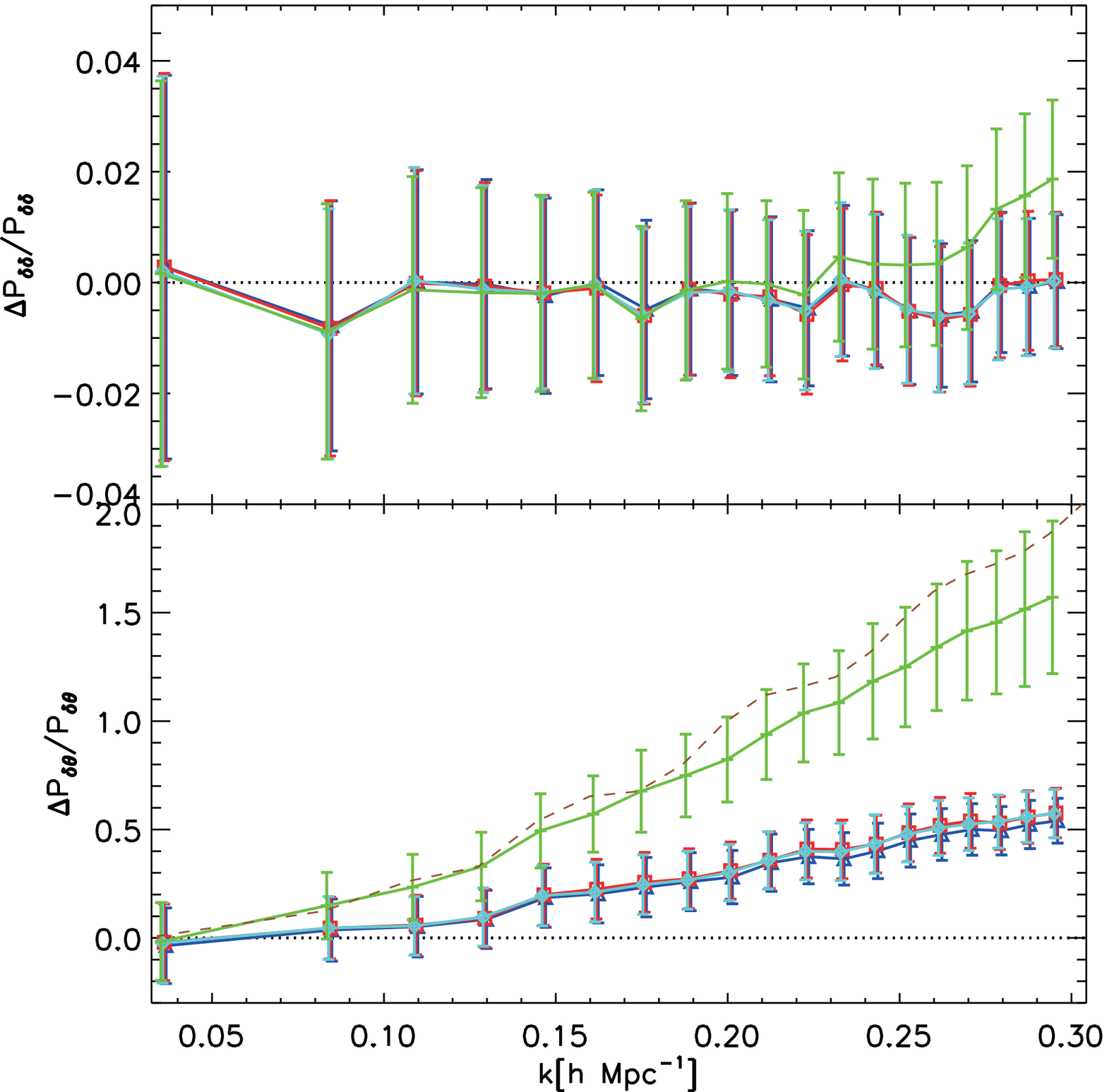}
\includegraphics[width=5cm,angle=0]{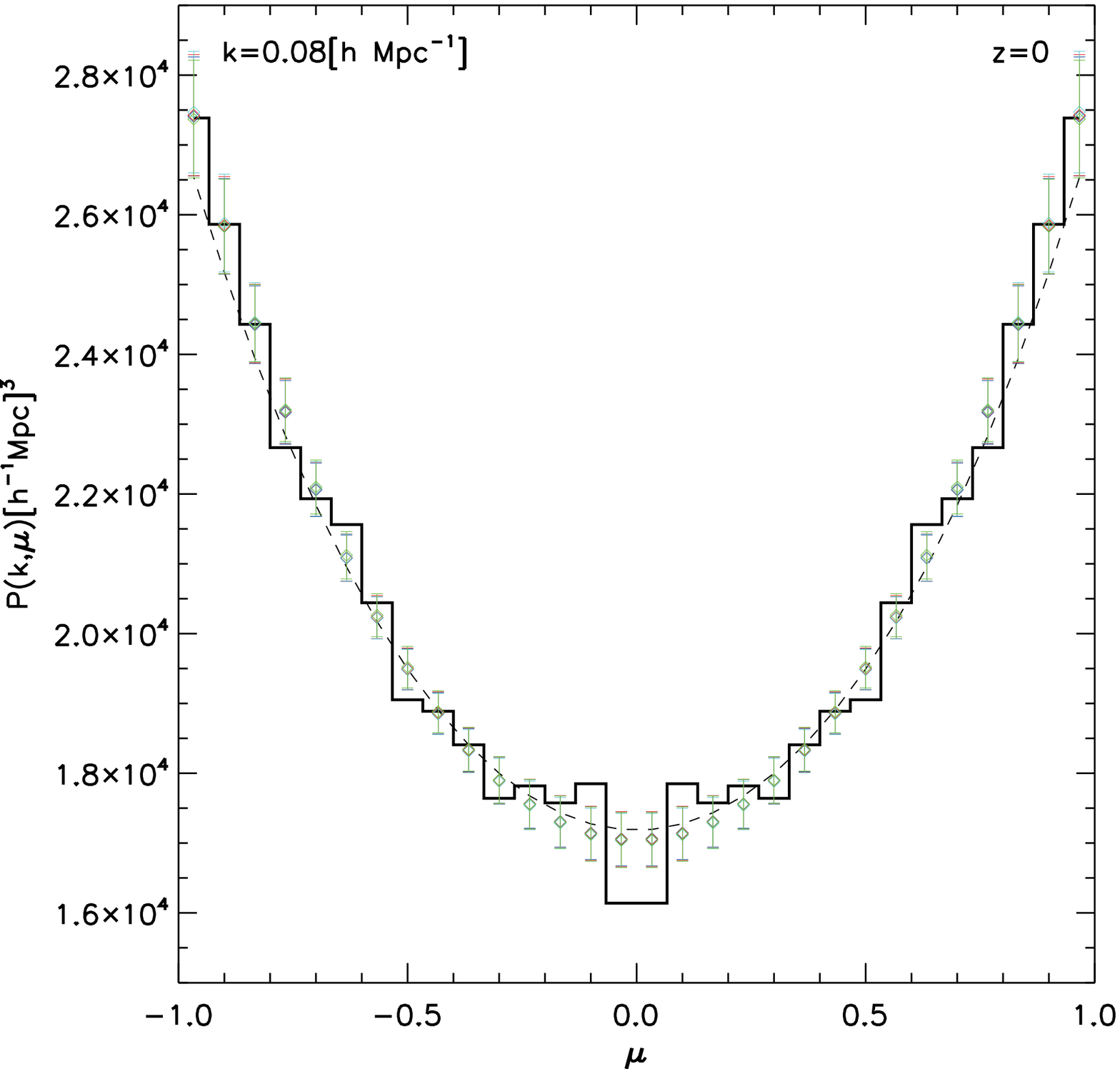}
\includegraphics[width=5cm,angle=0]{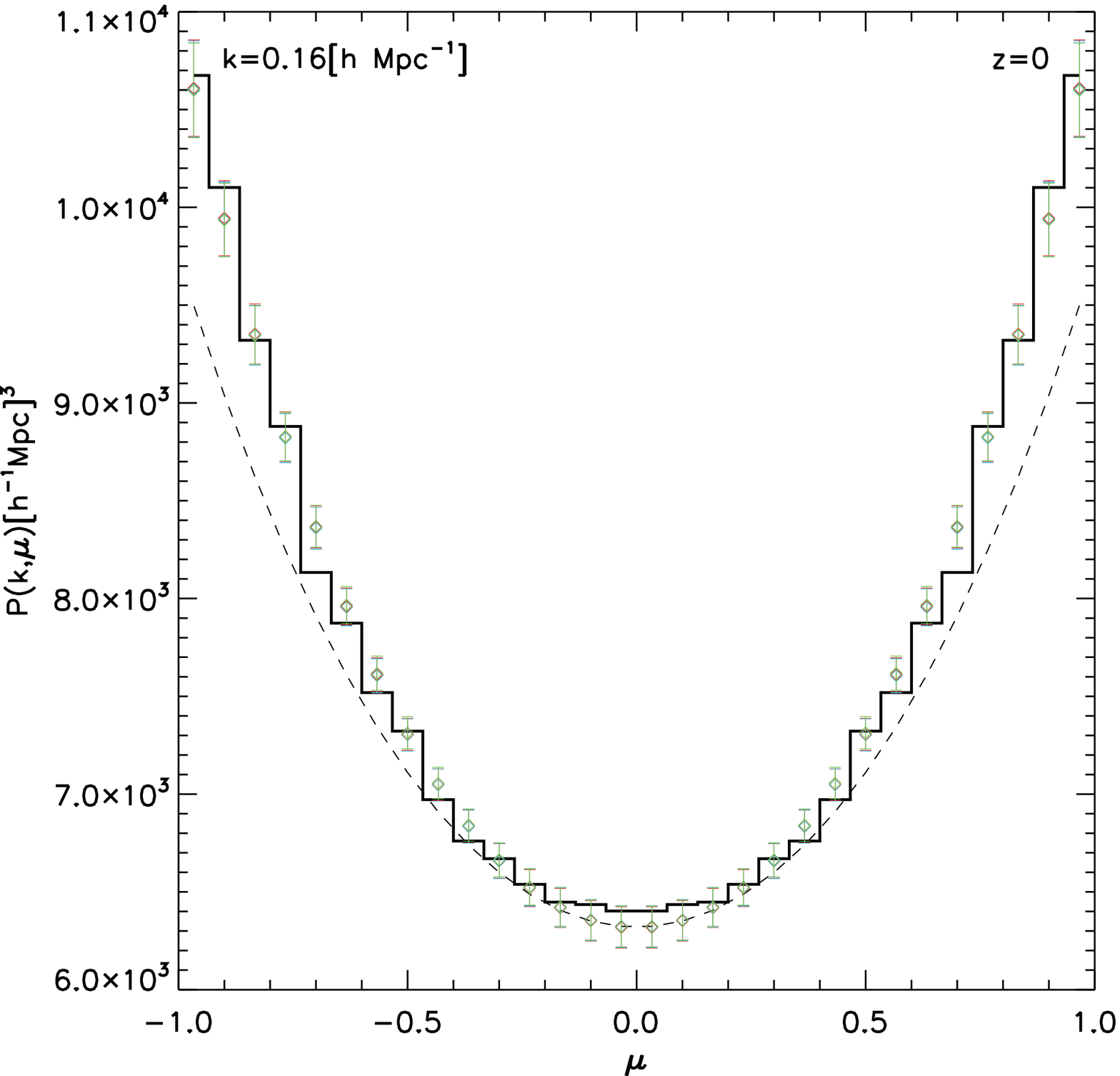}
\includegraphics[width=5cm,angle=0]{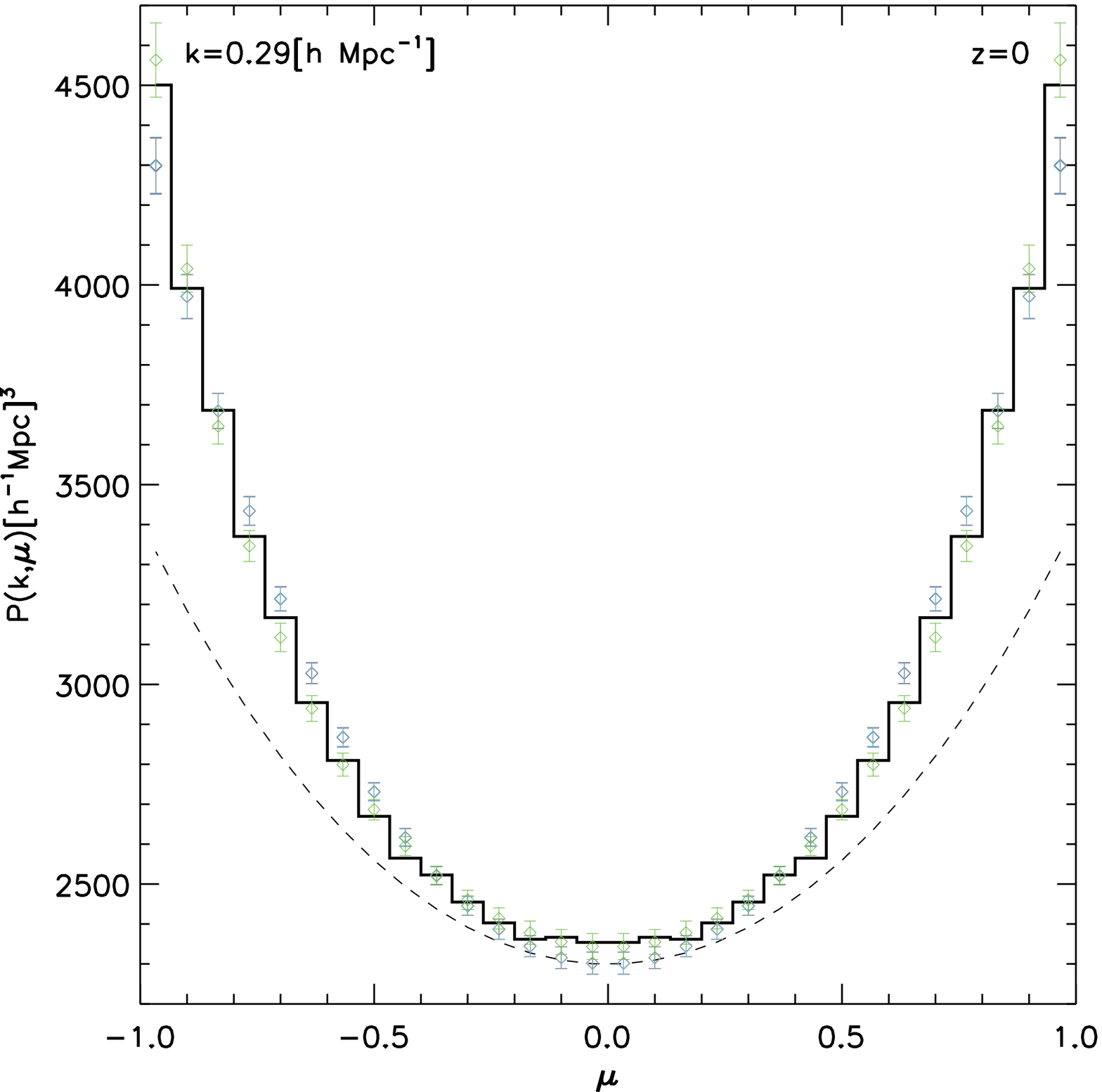}
\end{minipage}
\caption{Same as in Figs.~\ref{fig:re_dm}, \ref{fig:re_diff_dm} and
\ref{fig:2d_slices_z=0}, but for halo spectra at $z=0$. As described in
\S~\ref{sec:halo_spectra}, the input density-velocity spectra $\pdv$,
which are used to compare with the reconstructed spectra, we used the
spectra between the halo density field and the N-body particle velocity
field, because the halo velocity field is hard to construct due to a
coarse sampling of the velocity field. As can be clearly seen from the
lower three panels (especially two lower-right panels), the measured
redshift-space power spectra show greater amplitudes 
than the redshift-space 
spectrum inferred from the simulation, without the FoG effect
$F=1$ (see text for discussion). The dashed curves in the upper-left and
-right panels show the results obtained by adding the perturbation
theory prediction of the nonlinearity correction term
(Eq.~[\ref{eq:higher_Ps_halo_text}]) to the directly measured
 $\pdv(k)$. In the predictions, we assumed the halo bias parameters
 $b_1=1.6$ and $b_2=0$, where the linear bias parameter is estimated by
 comparing the density power spectra ($\pdd$) of dark matter and halos
 at small $k$. 
 \label{fig:2d_all_methods_halo002}}
\end{center}
\end{figure*}

\begin{figure}
\includegraphics[width=8.7cm,angle=0]{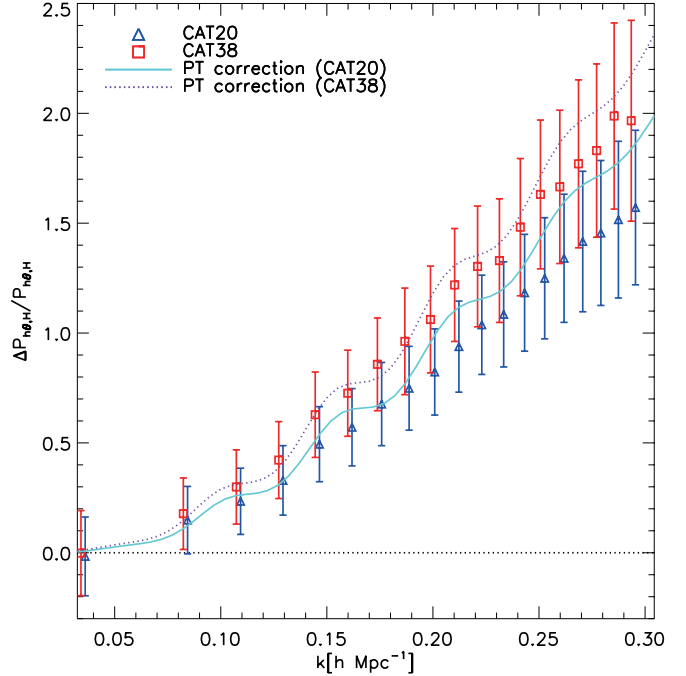}
\caption{Ratios of the reconstructed spectra of $\mu^2$
 to the directly measured $\pdv(k)$
(as in
Fig.~\ref{fig:2d_all_methods_halo002}), for different halo catalogs.
The two results are slightly shifted in a horizontal direction for
 illustrative purpose. The two different halo catalogs are defined from
 halos with masses greater than $9.8\times 10^{12}~h^{-1}M_\odot$
 (triangles) and $1.86\times 10^{13}~h^{-1}M_\odot$ (squares), which
 contain at least 20 and 38 N-body particles as members,
 respectively. For more massive halos, the reconstructed spectrum of
 $\mu^2$ shows greater amplitudes. The solid and dashed curves show the
 results including the nonlinearity correction term to $\pdv(k)$ as done
 in Fig.~\ref{fig:2d_all_methods_halo002}: $\pdv(k)+\delta
 P_{\mu^2}(k)$. To compute the nonlinearity corrections, 
we assumed $b_1=1.6$
 and 1.84 for the less and more massive halo catalogs, respectively, but
 assumed $b_2=0$ for both the catalogs. 
\label{fig:halo_biastest}}
\end{figure}

Now we move to the reconstruction of halo power spectra.
Fig~\ref{fig:2d_all_methods_halo002} shows the results for halo catalogs
at $z=0$. 
First of all, the reconstruction
can successfully recover the density power spectrum $\pdd(k)$ over a
range of wavenumbers we consider, as a result of properly correcting for
the shot noise and the redshift distortion.
The accurate reconstruction of $\pdd$ is relevant for the BAO
experiments, and the results imply that our method may allow us to
further use the broad-band shape of $\pdd(k)$ to improve cosmological
constraints. However, in contrast to the results for N-body particles
shown in Figs.~\ref{fig:re_dm} and \ref{fig:2d_all_methods_002_reduce},
the reconstruction fails to recover the density-velocity power spectrum
$\pdv(k)$. The reconstructed $\pdv$ for halos gives higher amplitudes
than the directly measured power spectrum, irrespective of the different
FoG models. 
In fact, as explicitly shown
in the right panel, the measured redshift-space power spectrum shows
{\em greater} amplitudes than
predicted by the Kaiser formula
(Eq.~[\ref{eq:Pgg}] with no FoG effect, i.e. $F=1$). The enhancement in
the power spectrum amplitudes is opposite to the FoG effect, which
always suppresses the amplitudes. 

We argue below that the results in Fig.~\ref{fig:2d_all_methods_halo002}
can be understood by the nonlinearity effect on the redshift-space power
spectrum. As we briefly discussed around Eq.~(\ref{eq:higher_Ps}), the
nonlinear clustering causes a correction to the Kaiser formula of
redshift-space power spectrum \citep[also see][for a more extensive
discussion]{Scoccimarro04,Taruyaetal:10}. Assuming that the density
perturbation is greater than the velocity field, which can be even more
validated for highly biased halos with $b>1$, the leading-order
correction term is found to arise from the
cross-bispectrum $\langle
\tilde{\delta}\tilde{\delta}\tilde{\theta}\rangle$ (see
Eq.~[\ref{eq:higher_Ps}]):
\begin{equation}
\delta P_s(k,\mu)\leftarrow
k_\parallel\left\langle\tilde{\delta}(\bmf{k}')
\int\!\frac{d^3\bmf{q}}{(2\pi)^3}\frac{q_\parallel}{q^2}\tilde{\theta}(\bmf{q})
\tilde{\delta}(\bmf{k}-\bmf{q})
\right
\rangle.
\end{equation}
We tried to measure this correction term from the simulations, but could
not obtain the reliable results as the bispectrum measurement is very
noisy. Instead we here use the perturbation theory prediction assuming a
$\Lambda$CDM cosmology (or equivalently Einstein gravity).  In
Appendix~\ref{sec:higher_Kaiser} we explicitly derive the leading-order
correction term given as a function of the linear mass power
spectrum. We find that the leading-order correction only contributes to
the redshift-space power spectrum at the power of $\mu^2$:
\begin{equation}
P_{\rm halo}^s(k,\mu)=\pdd(k)+2\mu^2\left[\pdv(k) + \delta P_{\mu^2}(k)\right]
+\mu^4\pvv(k),
\label{eq:phh}
\end{equation}
where $\delta P_{\mu^2}(k)$ is the correction term. Including the halo
bias parameters, the correction term is expressed as
\begin{eqnarray}
&&\hspace{-2em}\delta P_{\mu^2}(k) =\frac{fb_1^2k^3}{(2\pi)^2}\left[
\int_0^\infty\!dr
\int_{-1}^{1}dx~ x\right.
\nonumber\\
&&\hspace{0em}
\times \left\{ \frac{r^3}{7}(-1+7rx-6x^2)
P_{\delta\delta}^L(k)
\right.\nonumber\\
&&\hspace{1em}
\left.
+\frac{1}{7}(7x+3r-10rx^2)P_{\delta\delta}^L(kr)
\right\}
\frac{P^L_{\delta\delta}(k\sqrt{1+r^2-2rx})}{(1+r^2-2rx)}
\nonumber\\
&&\hspace{1em}
\left.
-P_{\rm \delta\delta}^L(k)\int_0^\infty\!dr
\frac{2}{3}(1+r^2)P_{\delta\delta}^L(kr)
\right]\nonumber\\
&&+
\frac{fb_1b_2k^3}{(2\pi)^2}\int_0^\infty\!dr\int_{-1}^{1}
\!dx~ xr P_{\delta\delta}^L(kr)P_{\delta\delta}^L(k\sqrt{1+r^2-2rx}), 
\nonumber\\
\label{eq:higher_Ps_halo_text}
\end{eqnarray}
where $f\equiv d\ln D/d\ln a$ ($D$ is the growth rate) and $b_1$ and
$b_2$ are the linear and nonlinear bias parameters. Note that we here
assumed that the velocity field of halos is unbiased with respect to the
velocity field of dark matter. 
Eq.~(\ref{eq:higher_Ps_halo_text}) clearly shows that the nonlinearity
correction term depends on halo bias parameters. The first term depends
on the linear bias parameter as $\propto b_1^2$. Compared to the
density-velocity power spectrum $\pdv(k)$, which depends on $b_1$ as
$\pdv \propto b_1$, the nonlinearity correction term can be more
important for more biased halos, or equivalently more massive halos.
\cite{Taruyaetal:10} derived more comprehensive equations for dark
matter including other nonlinear terms which arise from the
cross-bispectra such as
$\langle\tilde{\delta}\tilde{\theta}\tilde{\theta} \rangle$. The other
terms are found to have the contributions of $\mu^{2n}$ ($n=1,2,\cdots
4$), but depend on halo bias as $\propto b_1$. For highly biased halos
with $b_1>1$ as halos we are studying, the term given by
Eq.~(\ref{eq:higher_Ps_halo_text}) has most dominant
contribution. However, the perturbation theory is known to be less
accurate for lower redshifts such as $z=0$, due to the stronger
nonlinear clustering effects. Hence the results shown here still need to
be more carefully studied.  

The dashed curves in the upper-left and -right panels of
Fig.~\ref{fig:2d_all_methods_halo002} show the results where we added
the nonlinearity correction term to $\pdv(k)$ measured from the
simulation assuming the Taylor $(\sigma+\tau)$ FoG model:
$\pdv(k)+\delta P_{\mu^2}(k)$. To obtain the theory prediction of the
nonlinearity correction term, we assumed
$b_1=1.6$ and $b_2=0$, where $b_1$ is estimated by comparing the density
power spectra $\pdd$ for dark matter and halos at small $k$. More
exactly we computed the correction power spectrum of $\mu^2$ using the
full expression in \cite{Taruyaetal:10}, also taking into account the
halo bias dependences on the different terms that are either
proportional to $b_1$ or $b_1^2$. Eq.~(\ref{eq:higher_Ps_halo_text})
gives the similar shape, but about 10\% higher amplitudes than the
curves in Fig.~\ref{fig:2d_all_methods_halo002}.  Interestingly, the
nonlinearity correction term increases, rather than suppresses, the
amplitudes of real-space power spectrum that is proportional to $\mu^2$
in the redshift-space power spectrum. The enhancement increases with
increasing $k$. We should also emphasize that such a nice agreement
including the nonlinearity correction can be found only if using the
Taylor-($\sigma+\tau$) FoG model. Given the fact that the redshift-space
power spectrum of halos is least affected by the FoG effect, our results
imply that the Taylor ($\sigma+\tau$) model has more degrees of freedom
than other one-parameter FoG models and can effectively capture
higher-order contributions of $\mu^{2n}$ ($n=1,2,\cdots$) that arise
from nonlinearity effects as studied in \cite{Taruyaetal:10}.

We also studied the halo power spectrum reconstruction using the halo
catalogs constructed from $z=1$ simulations. We similarly found that the
reconstructed power spectra of $\mu^2$ show greater amplitudes than
expected from the measured $\pdv(k)$. The nonlinearity correction term
is found to similarly explain the reconstructed power spectrum, in
slightly less agreement, if using the reconstructed power spectrum
obtained from the Taylor $(\sigma+\tau)$ FoG model. We
have also found a subtle contamination of the residual shot noise for
the $z=1$ results, and
therefore we here show the results for $z=0$ for illustrative clarity.

To obtain more insights on the halo power spectrum results, in
Fig.~\ref{fig:halo_biastest} we study the reconstruction results for
$\pdv$ using different halo catalogs where halos are selected with
different mass thresholds.
To be more precise, we made the new catalogs by employing higher mass
threshold, $1.86\times10^{13}h^{-1}M_{\sun}$ (including more than 38
N-body member particles), rather than the threshold
$9.8\times10^{12}h^{-1}M_{\sun}$ (20 particles) we have so far
used. Note that the number density for more massive halos is
$\bar{n}\simeq 1.9\times 10^{-4}~h^3{\rm Mpc}^{-3}$, compared to
$3.8\times 10^{-4}~h^3{\rm Mpc}^{-3}$ for our fiducial halo
catalogs. The estimated bias parameter is $b_1=1.84$ compared to
$b_1=1.6$. Fig.~\ref{fig:halo_biastest} shows the reconstruction power
spectrum of $\mu^2$ for 
the different halo catalogs, compared to the density-velocity
spectrum.
The figure shows that the reconstructed spectrum for more massive halos
has higher amplitudes than for less massive halos. The solid and dotted
curves show the predictions obtained by adding the nonlinearity
correction term (Eq.~[\ref{eq:higher_Ps_halo_text}]) to the directly
measured $\pdv(k)$, where we used in the computation the linear bias
parameters above and assumed $b_2=0$ for simplicity. 
The nonlinearity correction, which depends on the halo bias,
fairly well reproduces the reconstruction
results.  In summary such nonlinearity correction terms need to be
included when interpreting the reconstructed power spectra for halos, or
more generally galaxies. We again emphasize that our method reconstructs
band powers of the real-space power spectrum, which are proportional to $\mu^2$ in the
redshift-space power spectrum, rather than the band powers of $\pdv$
alone.

Finally we comment on the impact of nonlinearity correction on the
reconstructions results for dark matter (N-body particles), which we
showed in the preceding section. For the dark matter spectrum, which
has $b=1$ by definition, the nonlinearity correction
(Eq.~[\ref{eq:higher_Ps_halo_text}]) is smaller compared to the
results of halos. However, using the perturbation theory predictions, we
found that the nonlinearity correction is not negligible. Including the
nonlinearity correction improves agreement with the input power
spectrum over a range of wavenumbers up to $ k \simeq  0.2~h$Mpc$^{-1}$ 
for the
Taylor-($\sigma+\tau$) FoG results shown in the middle panel of
Fig.~\ref{fig:re_diff_dm}. However, the nonlinearity correction
increases the disagreement at the larger $k$. In summary we conclude
that our reconstruction method can well recover the real-space power
spectrum, which is proportional to $\mu^2$ in the redshift-space power
spectrum, up to $k\simeq 0.2~h$Mpc$^{-1}$ for both dark matter and
halos, {\em if} including the nonlinearity corrections.

\section{Summary and Discussion}
\label{sec:summary}

In this paper we have developed a maximum likelihood based method of
reconstructing the real-space power spectra of density and velocity
fields, from the two-dimensional, redshift-space clustering of dark
matter and halos (supposedly galaxies). This method is developed in
analogy with the CMB power spectrum reconstruction method
\citep{Verde03}.

By assuming the form of redshift-space power spectrum given by 
Eq.~(\ref{eq:2dmethod}), we developed a method of reconstructing the
band powers of $\pdd$ and $\pdv$ at each $k$ bins, being marginalized
over uncertainties in the band powers at different $k$ bins and the
parameters to model the FoG effect, in such a way that the 
likelihood of the redshift-space power spectrum measured becomes
maximized. One assumption we have employed for the method is the
functional form of  redshift-space power spectrum
(Eq.~[\ref{eq:2dmethod}]), where the Kaiser formula and the FoG effect
is given by multiplicative functions. In fact this form is expected
based on the halo model picture \citep[][also see Hikage et al. in
preparation]{White:01,Seljak:01}. 
The real-space power spectra, especially at such large length scales
($k\simlt 0.3~\hompc$), contains cleaner cosmological information in the
linear or quasi-nonlinear regimes, and are relatively easier to develop
a sufficiently accurate model by using a suit of
simulations and/or refined perturbation theory. 
Furthermore, by measuring the velocity-related power
spectra in a model-independent way,
we
can open up a new window of testing gravity on cosmological scales. That
is, we can address whether or not the velocity field inferred is
consistent with the gravity field inferred from the density field,
because the density and velocity fields are related to each other via
gravity theory.

We have carefully tested our method by comparing the reconstructed
real-space power spectra with the spectra directly measured from
simulations of 70 realizations, for dark matter (N-body particles) as
well as halos. For matter power spectra (i.e. N-body particles), we
showed our method nicely recovers the power spectra $\pdd$ and $\pdv$
over a range of scales $k\simlt 0.3~\hompc$ and at redshifts $z=0$ and
$z=1$ (see Figs.~\ref{fig:re_dm}, \ref{fig:re_diff_dm} and
\ref{fig:2d_all_methods_001}), to accuracies within the statistical
errors, {\em if} we use the Taylor ($\sigma+\tau$) FoG model (see
Eq.~[\ref{eq:FOG}]), which has more degrees of freedom (2 parameters)
than the other models, Gaussian, Lorentzian and single-parameter Taylor
FoG models.  Our results imply that the FoG effect seen in simulations
has a complex scale-dependence, and is important to take into account
the scale dependence in order to obtain an unbiased reconstruction of the
band powers of $\pdd$ and $\pdv$ at scales down to
$k\simeq0.3\hompc$. In other words, the FoG effect affects the
redshift-space power spectrum over a wide range of wavenumbers. Hence  an
inaccurate modeling of the FoG effect causes a biased estimate of the
power spectra. 
It is also worth noting that the reconstruction causes
correlations between the band powers of different power spectra and at
different $k$-bins and the FoG model parameters (see
Figs.~\ref{fig:exmp_contour} and \ref{fig:2d_all_methods_001}).

For the halo power spectrum, we showed that our method again nicely
recovers the density power spectrum $\pdd$ over a wide range of
wavelengths, up to $k\simeq 0.3~h{\rm Mpc}^{-1}$
 (see Fig.~\ref{fig:2d_all_methods_halo002}). Such an
accurate reconstruction of $\pdd$ is very promising, because the shape
and amplitude information of $\pdd$ are sensitive to cosmological
parameters such as the tilt and running index of the primordial power
spectrum and neutrino masses
\citep[e.g.][]{Takadaetal:06,Saitoetal:09,Saitoetal:10}. On a
measurement side, the halo power spectrum can be estimated from actual
galaxy redshift survey, e.g. based on the method developed in
\cite{Reidetal:10} where galaxy pairs with small spatial separations are
clipped out. Although the halo power spectrum is supposed to be less
contaminated by the FoG effect, it is very important to minimize the
residual FoG contamination in order to extract unbiased cosmological
information from the measured halo power spectrum. For example, since
the FoG effect causes a suppression in the power spectrum amplitude, a
residual FoG contamination would cause a bias in neutrino mass
constraints because the main effect of massive neutrinos is also the
suppression on power spectrum amplitude (Hikage et al. in preparation).
Our method can give a robust way of measuring the density power
spectrum, minimizing the FoG contamination. 

For the halo velocity power spectrum $\pdv$, we found some
difficulty. First of all, we could not reliably reconstruct the velocity
field of halos from simulations, due to too sparse sampling of halos'
velocities. Hence we instead used the velocity field of dark matter
(N-body particles) assuming that the large-scale bulk motions of halos
are unbiased from the velocities of dark matter, which has been often
assumed in the literature. Note that the {\em real-space} velocity field
we considered here contains only the large-scale information at $k\simlt
0.3~h{\rm Mpc}^{-1}$, and therefore is not affected by any virial
motions within halos. As a result, we found that the reconstructed power
spectrum of $\mu^2$ systematically differs from $\pdv(k)$ directly
measured from the simulations
(Fig.~\ref{fig:2d_all_methods_halo002}). In fact, the measured
redshift-space halo power spectrum shows greater amplitudes than the
spectrum inferred from the Kaiser formula of redshift-space spectrum,
without the FoG effect that causes a suppression in the redshift-space
power spectrum amplitudes. 

Therefore we argued that the halo power spectrum is affected by the
nonlinearity effect.
In Appendix~\ref{sec:higher_Kaiser}, assuming a $\Lambda$CDM cosmology
or Einstein gravity, we derived the leading-order nonlinearity correction
to the Kaiser formula of redshift-space power spectrum, which arises
from the cross-bispectrum 
 between the density  and velocity
perturbations. We meant by the leading-order term that the term appears
to have
the largest contribution, assuming that the density perturbation is
greater than the velocity field at length scales of interest. 
We found
that the leading-order contribution is proportional to $\mu^2$ and has
greater amplitudes for more biased halos, i.e. more massive halos. We
showed that adding the perturbation theory prediction to the simulation
$\pdv(k)$ better matches the reconstructed power spectrum (see
Fig.~\ref{fig:2d_all_methods_halo002}). We also found that, by using the
different halo catalogs defined with different mass thresholds,  
the halo bias
dependence of the nonlinearity correction is seen in the
reconstructed power spectra of $\mu^2$ 
(see
Fig.~\ref{fig:halo_biastest}).

Hence a more appropriate statement for our maximum likelihood method is
that the method can recover the real-space power spectra, which are
proportional to $\mu^0$ and $\mu^2$, respectively, in the measured
redshift-space power spectrum, including marginalization over
uncertainties in the FoG effect. In other words the reconstructed
spectrum of $\mu^2$ is not necessarily the same as the density-velocity
power spectrum $\pdv(k)$, which we have used in our comparison. The
nonlinearity effect on the real-space power spectrum of $\mu^2$ needs to
be included if we want to use the reconstructed power spectrum to
constrain cosmological parameters as well as to test
 gravity theory. On the other hand, we 
found that the power spectrum of $\mu^4$ is very noisy to reconstruct,
for the ranges of wavenumbers and redshifts we have considered in this
paper. 

Recently \cite{Taruyaetal:10} studied the redshift-space power spectrum
including nonlinearity effects, based on the extended perturbation
theory. They found that the nonlinear correction terms including the
higher-order terms of $O(\theta^2)$ have the contributions that are
proportional to $\mu^{2n}$ ($n=1,2,\dots, 4$) in the redshift-space
power spectrum. The comparison of the theoretical prediction with the
reconstructed power spectrum based on our method is very interesting,
and will be studied elsewhere. 

One encouraging result is our method can unbiasedly recover the
real-space density power spectrum $\pdd(k)$ even in the presence of
redshift distortion effect. Given this result our method may offer even
a new means of obtaining geometrical constraints on the Hubble expansion
rate and the angular diameter distances beyond the usual BAO
constraints. We have assumed throughout this paper that the underlying
cosmology is known. However, this is obviously not true for an actual
observation. In reality, we have to assume a reference cosmological
model to perform the clustering analysis of galaxies, and the assumed
cosmology generally differs from the underlying true cosmology.  An
imperfect cosmological model causes additional angular anisotropies in
the measured redshift-space power spectrum -- the so-called cosmological
distortion. In terms of Eq.~(\ref{eq:2dmethod}) an incorrect cosmology
leads some power of the density power spectrum $\pdd(k)$ of $\mu^0$ to
leak into the power spectrum with powers higher than $\mu^2$ in the
redshift-space power spectrum. Contrary, if we seek the reconstructed
$\pdd$ of maximum amplitudes with varying reference cosmological models,
we may be able to obtain the cosmological constraints \citep[also
see][for a similar discussion]{Padmanabhan:08}. This method looks
similar to the Alcock-Paczynski test
\citep{AP,MatsubaraSuto:96,Ballingeretal:96}, but our method may have
practical advantages: our method allows us to measure the real-space
power spectrum of $\mu^0$ in a model-independent way as well as to
derive cosmological constraints being marginalized over uncertainties in
the FoG effect. The feasibility of this method is our future work and
will be presented elsewhere.

Our reconstruction method is done in the two-dimensional Fourier space
of $(k,\mu)$. In practice one may want to exclude the Fourier modes
around $\mu\simeq \pm 1$, which are more affected by the FoG effect. In
our method it is straightforward to include a masking of the modes
around $\mu\pm 1$; that is, the real-space power spectra are
reconstructed by using the Fourier modes in redshift space, excluding
the modes around $\mu\pm 1$. We have tried several masking methods of
$\mu$, 
but
could not find any significant differences from the results shown in
this paper. 

In this paper we have ignored some observational effects for
simplicity. For example, to apply our method to actual data, we need to
include effects such as survey window function and the curvature of the
sky. These effects have been well studied \citep[e.g.][]{Tegmarketal:04},
and would be rather straightforward to include, although a further
careful study needs to be done.

\section*{Acknowledgments}
We thank E. Komatsu, B. Jain, T. Nishimichi, N. Padmanabhan, R. Sheth,
D. Spergel and A. Taruya for useful discussion and valuable comments.
This work is in part supported in part by JSPS Core-to-Core Program
``International Research Network for Dark Energy'', by Grant-in-Aid for
Scientific Research from the JSPS Promotion of Science, by Grant-in-Aid
for Scientific Research on Priority Areas No. 467 ``Probing the Dark
Energy through an Extremely Wide \& Deep Survey with Subaru Telescope'',
by World Premier International Research Center Initiative (WPI
Initiative), MEXT, Japan, and by the FIRST program ``Subaru Measurements
of Images and Redshifts (SuMRe)''.

\bibliography{aamnem99,pgvpaper_arXiv}

\appendix

\section{Likelihood function of redshift-space power spectrum}
\label{sec:like}

In this appendix we derive the likelihood function of redshift-space
power spectrum. To do this, we assume that the density fluctuation field
of large-scale structure tracers (matter or galaxies) is Gaussian and
obeys the following Gaussian likelihood function:
\begin{equation}
{\cal L}\propto \frac{1}{\sqrt{{\rm det}{\bmf{C}}}} \exp\left[-\frac{1}{2} \int\!\frac{d^3\bmf{x}_i}{V} \int\!\frac{d^3\bmf{x}_j}{V}
\delta(\bmf{x}_i)\bmf{C}^{-1}_{ij}
\delta(\bmf{x}_j)
\right], \label{eqn:likelihood}
\end{equation}
where $\delta(\bmf{x})$ is the density fluctuation field (of matter or
galaxies), $\bmf{C}_{ij}$ is the correlation matrix between the fields
$\delta(\bmf{x}_i)$ and $\delta(\bmf{x}_j)$, $\bmf{C}^{-1}_{ij}$ is
the inverse matrix, and ${\rm det}\bmf{C}$ is the determinant of the
matrix $\bmf{C}$. 
The covariance matrix or the correlation function, $\bmf{C}_{ij}$, can be
expressed in terms of the power spectrum $P(k)$ as
\begin{eqnarray}
\bmf{C}_{ij}&\equiv& \kaco{\delta(\bmf{x}_i) \delta(\bmf{x}_j)
}\nonumber\\
&=&\frac{1}{V_s^2}\sum_{\bmf{k}}\sum_{\bmf{k}'} \kaco{\tilde{\delta}_{\bmf{k}} \tilde{\delta}_{\bmf{k}}
}e^{i\bmf{k}\cdot\bmf{x}_i}e^{i\bmf{k}'\cdot\bmf{x}_j}
\nonumber\\
&=&\frac{1}{V_s}\sum_{\bmf{k}}P(k)e^{\bmf{k}\cdot(\bmf{x}_i-\bmf{x}_j)},
\end{eqnarray}
where $V_s$ is the survey volume.  Given a finite-volume survey, we
introduced the discrete Fourier transformation of the density
field~\citep[e.g. see][]{TakadaBridle:07}, where the fundamental Fourier
mode is given by the survey size; $k_f=2\pi/L$ ($V_s=L^3$). Here we
ignored survey geometry and boundary effects for simplicity.  Therefore
the inverse of the covariance matrix can be given as
\begin{equation}
{\bmf{C}}^{-1}_{ij}=\sum_{\bmf{k}}\frac{V_s}{P(k)} e^{i\bmf{k}\cdot(\bmf{x}_i-\bmf{x}_j)},
\end{equation}
This can be proved because the product of the covariance matrix and its
inverse matrix satisfies the orthogonal relation, which is formally
computed as
\begin{eqnarray}
\bmf{C}_{ij}\bmf{C}^{-1}_{jk} &\equiv &\int\!\frac{d^3\bmf{x}_j}{V_s}
\bmf{C}_{ij}\bmf{C}^{-1}_{jk}
\nonumber\\
&=&\int\!\frac{d^3\bmf{x}_j}{V_s} \sum_{\bmf{k}}\sum_{\bmf{k}'} P(k)\frac{1}{P(k')}e^{i\bmf{k}\cdot(\bmf{x}_i-\bmf{x}_j)}
e^{i\bmf{k}'\cdot(\bmf{x}_j-\bmf{x}_k)}
\nonumber\\
&=& \sum_{\bmf{k}}\sum_{\bmf{k}'} P(k)\frac{1}{P(k')}e^{i\bmf{k}\cdot\bmf{x}_i} e^{i\bmf{k}'\cdot\bmf{x}_k}\delta^K_{\bmf{k}-\bmf{k}'}
\nonumber\\
&=& \sum_{\bmf{k}}e^{i\bmf{k}\cdot(\bmf{x}_i-\bmf{x}_k)}
\nonumber\\
&=&V_s\delta_D^3(\bmf{x}_i-\bmf{x}_k),
\end{eqnarray}
where $\delta^K(\bmf{k}-\bmf{k}')$ is the Kronecker-type delta function; 
$\delta^K(\bmf{k}-\bmf{k}')=1$ if $\bmf{k}=\bmf{k}'$ within the bin
width, otherwise $\delta^K(\bmf{k}-\bmf{k}')=0$ (see
\cite{TakadaBridle:07}).  

Using the equations above, the argument in the exponential of
Eq.~(\ref{eqn:likelihood}) can be reduced to the following form in
Fourier space: 
\begin{eqnarray}
\int\!\frac{d^3\bmf{x}_i}{V_s} 
\int\!\frac{d^3\bmf{x}_j}{V_s} 
\delta(\bmf{x}_i)\bmf{C}^{-1}_{ij}\delta(\bmf{x}_j)&=&
\int\!\frac{d^3\bmf{x}_i}{V_s} \int\!\frac{d^3\bmf{x}_j}{V_s}
\nonumber\\
&&\hspace{-13em}
\times\frac{1}{V_s^2}\sum_{\bmf{k}}\sum_{\bmf{k}'}\sum_{\bmf{k}''
}\tilde{\delta}_{\bmf{k}}e^{i\bmf{k}\cdot\bmf{x}_i} \frac{V_s}{P(k')}e^{i\bmf{k}\cdot(\bmf{x}_i-\bmf{x}_j)}
\tilde{\delta}_{\bmf{k}''}e^{i\bmf{k}''\cdot\bmf{x}_j}
\nonumber\\
&&\hspace{-10em}=\frac{1}{V_s}\sum_{\bmf{k}}\sum_{\bmf{k}'}\sum_{\bmf{k}'' }\tilde{\delta}_{\bmf{k}} \frac{1}{P(k')} \tilde{\delta}_{\bmf{k}''}
\delta^K_{\bmf{k}+\bmf{k}'} \delta^K_{-\bmf{k}'+\bmf{k}''}
\nonumber\\
&&\hspace{-10em}=\frac{1}{V_s} \sum_{\bmf{k}}\frac{\tilde{\delta}_{\bmf{k}}\tilde{\delta}_{-\bmf{k}}
}{P(k)}\nonumber\\
&&\hspace{-10em}=\frac{1}{V_s} \sum_{\bmf{k}}\frac{|\tilde{\delta}_{\bmf{k}}|^2 }{P(k)}. \label{eqn:exp}
\end{eqnarray}
Note that, in equation above, I assumed that $|\tilde{\delta}_{\bmf{k}}|^2$ and $P(k)$ have same dimensions such that the combination
$|\tilde{\delta}_{\bmf{k}}|^2/P(k)$ becomes dimension-less.

Therefore
the log-likelihood function of density fluctuation field
(Eq.~[\ref{eqn:likelihood}]) is reduced to the log-likelihood function
of the power spectrum: 
\begin{equation}
-2\ln{\cal L}
= \sum_{\bmf{k}}\left[ \frac{|\tilde{\delta}_{\bmf{k}}|^2}{P(k)}+\ln P(k)
\right], \label{eqn:loglikelihood}
\end{equation}
where we have ignored the constant additive term.  The log-likelihood
above is rewritten in terms of the power spectrum estimator as
\begin{equation}
-2\ln{\cal L}=\sum_{k_i,\mu_a} N(k_i,\mu_a)\left[ \frac{\hat{P}(k_i,\mu_a)}{P(k_i,\mu_a)} +\ln P(k_i,\mu_a) \right].
\end{equation}
Here $\hat{P}$ is the power spectrum estimator 
\begin{equation}
\hat{P}(k_i,\mu_a)\equiv \frac{1}{N(k_i,\mu_a)}\sum_{\bmf{k}\in k_i, \mu_a} |\tilde{\delta}_{\bmf{k}}|^2, \label{eqn:est_P}
\end{equation}
where the summation $\sum_{\bmf{k}\in (k_i,\mu_a)}$ is over Fourier
modes $\bmf{k}$ satisfying the condition that the wavevector $\bmf{k}$
lies in the bin labeled by the length and the azimuthal angle between the
line-of-sight direction and the wavevector, $(k_i,\mu_a)$, and
$N(k_i,\mu_a)$ is the number of independent Fourier modes:
 $N(k_i,\mu_a)\equiv \sum_{\bmf{k}\in(k_i,\mu_a)}\approx 2\pi
k_i^2\Delta k\Delta\mu/(2\pi/L)^3$ for $k_i\gg 2\pi/L$. The
$\mu_a$-dependence of power spectrum accounts for redshift distortion
effect. 

Adding the constant term into the equation above such that the log-likelihood function can be maximized if the theory power spectrum
$P(k_i,\mu_a )$ is equal to the estimated one, we can arrive, as in the
CMB case \citep{Verde03}, at the expression:
\begin{equation}
-2\ln{\cal L}=\sum_{k_i,\mu_a} N(k_i,\mu_a)\left[ \frac{\hat{P}(k_i,\mu_a)}{P(k_i,\mu_a)} +\ln \frac{P(k_i,\mu_a)}{\hat{P}(k_i,\mu_a)}-1
\right].
\end{equation}
Hence we can use this log-likelihood to estimate the underlying power
spectrum $P(k_i,\mu_a)$ at each bin, given the observed power spectrum
$\hat{P}(k_i,\mu_a)$. 

\section{Nonlinear correction terms of the Kaiser formula}
\label{sec:higher_Kaiser}

In this appendix, following the method developed in \cite{Scoccimarro04}
and \cite{Taruyaetal:10}, we derive the leading-order correction term of
higher-order perturbations to the Kaiser formula for the redshift-space
power spectrum, assuming a $\Lambda$CDM cosmological model based on the
Einstein gravity. 

Let us begin our discussion with recalling that the redshift distortion
effect on a given tracer of the large-scale structure is recognized as a
mapping between redshift- and real-space positional vectors:
\begin{equation}
\bmf{s}=\bmf{x}+u_\chi \hat{z}, 
\end{equation}
where $\bmf{s}$ and $\bmf{x}$ are the positional vectors in redshift-
and real-spaces, respectively, $u_\chi$ is the line-of-sight component
of the normalized peculiar velocity (see around Eq.~[\ref{eq:u_chi}])
and $\hat{z}$ is the unit vector of the line-of-side direction in the
real-space coordinate system. The mass or number conservation law gives
the relation between the density perturbations in redshift- and
real-spaces: 
\begin{equation}
\delta_s(\bmf{s})=\left[1+\delta(\bmf{x})\right]
\left[
1+\frac{\partial u_\chi}{\partial \chi}(\bmf{x})\right]^{-1}-1.
\end{equation}

Therefore the Fourier transform of the redshift-space density
perturbation can be expressed in terms of the real-space density and
velocity perturbation fields as
\begin{eqnarray}
\tilde{\delta}_s(\bmf{k})&\equiv& \int\!d^3\bmf{s}~
 \delta_s(\bmf{s})e^{-i\bmf{k}\cdot\bmf{s}}\nonumber\\
&&\hspace{-4em}=\int\!d^3\bmf{s}~
\left[
(1+\delta)\left(
1+\frac{\partial u_\chi}{\partial \chi}\right)^{-1}
-1
\right]e^{-i\bmf{k}\cdot\bmf{s}}
\nonumber\\
&&\hspace{-4em}=\int\!d^3\bmf{x}~
\left[
1+\delta - \left(
1+\frac{\partial u_\chi}{\partial \chi}\right)
\right]e^{-i\bmf{k}\cdot\bmf{x}-ik_\parallel u_\chi}\nonumber\\
&&\hspace{-4em}\approx \int\!d^3\bmf{x}~\left(
\delta - \frac{\partial u_\chi}{\partial \chi}\right)
\left(1-ik_\parallel u_\chi \right)
e^{-i\bmf{k}\cdot\bmf{x}} + O(u^2)\nonumber\\
&&\hspace{-4em}= \tilde{\delta}(\bmf{k})+\mu^2\tilde{\theta}(\bmf{k})
+k\mu
\int\!\frac{d^3\bmf{q}}{(2\pi)^3}\frac{q_\parallel}{q^2}\tilde{\delta}(\bmf{k}-\bmf{q})\tilde{\theta}(\bmf{q})+O(\theta^2), \nonumber\\
\end{eqnarray}
where we have expressed the peculiar velocity field as the
velocity-divergence field as $u_\chi=i(k_\parallel/k)\theta$ (see around
Eq.~[\ref{eq:delta_s_app}]), and $\mu\equiv k_\parallel/k$. On the third
equality on the r.h.s. of the equation above, we have used the Jacobian,
$|\partial \bmf{x}/\partial \bmf{s}|$ to make the integration variable
change, $\bmf{s}\rightarrow \bmf{x}$. Here, given the fact that the
density perturbation is greater than the velocity perturbation, we kept
the leading-order nonlinear correction term which has the order of
$O(\delta\theta) $ and ignored the higher-order terms than
$O(\theta^2)$. 

Therefore, the nonlinear correction term to the redshift-space power
spectrum is found to be 
\begin{equation}
\delta P_s(k,\mu)=2(k\mu)\int\!\frac{d^3\bmf{q}}{(2\pi)^3}
\frac{q_\parallel}{q^2}B_{\delta\delta\theta}(-\bmf{k},\bmf{k}-\bmf{q},\bmf{q}), 
\end{equation}
where the bispectrum is defined as
\begin{equation}
\langle
\tilde{\delta}(\bmf{k}_1)\tilde{\delta}(\bmf{k}_2)
\tilde{\theta}(\bmf{k}_3)
\rangle\equiv
B_{\delta\delta\theta}(\bmf{k}_1,\bmf{k}_2,\bmf{k}_3)
(2\pi)^3\delta_D^3(\bmf{k}_1+\bmf{k}_2+\bmf{k}_3).
\end{equation}
Using the perturbation theory of structure formation
\citep[e.g.][]{JainBertschinger:94} we can express the bispectrum in
terms of the linear power spectrum as
\begin{eqnarray}
B_{\delta\delta\theta}(\bmf{k}_1,\bmf{k}_2,\bmf{k}_3)&=&2f\left[
F_2(\bmf{k}_2,\bmf{k}_3)P_{\delta\delta}^L(k_2)
P_{\delta\delta}^L(k_3)
\right.\nonumber\\
&&\hspace{-8em}\left.+
F_2(\bmf{k}_1,\bmf{k}_3)P_{\delta\delta}^L(k_1)
P_{\delta\delta}^L(k_3)
+
G_2(\bmf{k}_1,\bmf{k}_2)P_{\delta\delta}^L(k_1)
P_{\delta\delta}^L(k_2)
\right],\nonumber\\
\end{eqnarray}
where $f\equiv d\ln D/d\ln a$, $\tilde{\theta}
=f\tilde{\delta}$ at the linear order for our notation 
and the kernels $F_2$ and $G_2$ are
defined as 
\begin{eqnarray}
F_2(\bmf{k}_1,\bmf{k}_2)&\equiv &
\frac{5}{7}+\frac{1}{2}
\left(
\frac{1}{k_1^2}+\frac{1}{k_2^2}
\right)\left(\bmf{k}_1\cdot\bmf{k}_2\right)
+\frac{2}{7}\frac{\left(\bmf{k}_1\cdot\bmf{k}_2\right)^2}{k_1^2k_2^2},
\nonumber\\
G_2(\bmf{k}_1,\bmf{k}_2)&\equiv&
\frac{3}{7}+\frac{1}{2}
\left(
\frac{1}{k_1^2}+\frac{1}{k_2^2}
\right)\left(\bmf{k}_1\cdot\bmf{k}_2\right)
+\frac{4}{7}\frac{\left(\bmf{k}_1\cdot\bmf{k}_2\right)^2}{k_1^2k_2^2}.
\nonumber\\
\end{eqnarray}

For halo clustering, we need to further take into account halo
bias. Here we simply assume that halo bias is deterministically given as
a function of the underlying mass density field
\citep[e.g.][]{FryGaztanaga:94}: 
%
\begin{equation}
\delta_h(\bmf{x})=b_1\delta_m(\bmf{x})+\frac{b_2}{2}\delta_m^2(\bmf{x}),
\end{equation}
where $b_1$ and $b_2$ are the linear and nonlinear bias parameters. On
the other hand we assume that the velocity field of halos is unbiased to
the velocity field of dark matter.  Then we can similarly compute the
correction term to the redshift-space power spectrum of halos:
\begin{eqnarray}
\delta P_{s,h}(k,\mu;z)&=& 
2b_1^2(k\mu)\int\!\frac{d^3\bmf{q}}{(2\pi)^3}
\frac{q_\parallel}{q^2}
B_{\delta\delta\theta}(-\bmf{k},\bmf{k}-\bmf{q},\bmf{q})\nonumber\\
&&\hspace{-6em}+2 fb_1b_2\left[
\int\!\frac{d^3\bmf{q}}{(2\pi)^3}\hat{q}_z^2P_{\delta\delta}^L(q)
P_{\delta\delta}^L(\bmf{k}-\bmf{q})\right. \nonumber\\
&&+P_{\delta\delta}^L(k)
\int\!\frac{d^3\bmf{q}}{(2\pi)^3}\hat{q}_z^2P_{\delta\delta}^L(q)\nonumber\\
&&\left.+\frac{1}{2}\hat{k}_z^2P_{\delta\delta}^L(k)\int\!\frac{d^3
 \bmf{q}}{(2\pi)^2}P_{\delta\delta}^L(q)\right].
\label{eq:higher_Ps_halo}
\end{eqnarray}
%

We can further simplify the equation (\ref{eq:higher_Ps_halo}) by using
the usual formula developed in Appendix A and B of \cite{Taruyaetal:10}
(also see \cite{Matsubara:08a}). The formula tells that the following
equation holds for an arbitrary scalar function $f(\bmf{q},\bmf{k})$:
\begin{equation}
\int\!\!\frac{d^3\bmf{q}}{(2\pi)^3}\frac{q_\parallel}{q^2}
f(\bmf{k},\bmf{q})
=\mu Q(\bmf{k},\bmf{q}), 
\label{eq:q_formula}
\end{equation}
where $\mu=k_z/k $ and 
\begin{eqnarray}
Q(k)&\equiv &\frac{k^2}{(2\pi)^2}\int_0^\infty\!\!dr
 \int_{-1}^{1}\!\!dx~ rx f(k,r,x). 
\end{eqnarray}
Here we have introduced the integration variable transformations as 
$q=kr$ and $\bmf{q}\cdot\bmf{k}=k^2rx$. 

Therefore, by comparing Eqs.~(\ref{eq:higher_Ps_halo}) and
(\ref{eq:q_formula}), we can find that the nonlinear correction term to
the Kaiser formula of the redshift-space power spectrum is proportional
to $\mu^2$. That is, the leading-order nonlinear correction term only
contaminates to the density-velocity power spectrum in the Kaiser
formula. If we define the correction term as $\delta P_s^{\rm
Kaiser}(k,\mu)\equiv 2\mu^2 \delta P_{\mu^2}(k)$, we find that the
real-space power spectrum $\delta P_{\mu^2}(k)$ is given as
\begin{eqnarray}
&&\hspace{-2em}\delta P_{\mu^2}(k) =\frac{fb_1^2k^3}{(2\pi)^2}\left[
\int_0^\infty\!dr
\int_{-1}^{1}dx~ x\right.
\nonumber\\
&&\hspace{0em}
\times \left\{ \frac{r^3}{7}(-1+7rx-6x^2)
P_{\delta\delta}^L(k)
\right.\nonumber\\
&&\hspace{1em}
\left.
+\frac{1}{7}(7x+3r-10rx^2)P_{\delta\delta}^L(kr)
\right\}
\frac{P^L_{\delta\delta}(k\sqrt{1+r^2-2rx})}{(1+r^2-2rx)}
\nonumber\\
&&\hspace{1em}
\left.
-P_{\rm \delta\delta}^L(k)\int_0^\infty\!dr
\frac{2}{3}(1+r^2)P_{\delta\delta}^L(kr)
\right]\nonumber\\
&&+
\frac{fb_1b_2k^3}{(2\pi)^2}\int_0^\infty\!dr\int_{-1}^{1}
\!dx~ xr P_{\delta\delta}^L(kr)P_{\delta\delta}^L(k\sqrt{1+r^2-2rx}).
\nonumber\\
\label{eq:higher_Ps_halo_2}
\end{eqnarray}

Eq.~(\ref{eq:higher_Ps_halo_2}) has several interesting
implications. First of all, the nonlinear correction term scale with
halo bias as $\delta P_{\mu^2}\propto b_1^2$.  Since the
density-velocity power spectrum for halos scales as
$P_{\delta\theta}\propto b_1$, the correction term can be more important
for more biased halos. Secondly, once the gravity theory is
assumed (here the Einstein gravity), the correction term can be computed
as a function of cosmological models. \cite{Taruyaetal:10} further
derived other correction terms arising from the higher-order
perturbations than $O(\theta^2)$, and then showed that the nonlinear
correction term can show remarkable nice agreement with the simulation
results. In the main text (see Sec.~\ref{sec:results_halos}) we use
Eq.~(\ref{eq:higher_Ps_halo_2}) to explain the power spectrum
reconstruction results for halos, where we found that the reconstructed
power spectrum, which is proportional to $\mu^2$, shows sizable
difference from the input $P_{\delta\theta}(k)$.

\label{lastpage}
\end{document}